\documentclass[12pt]{iopart}
\expandafter\let\csname equation*\endcsname\undefined
\expandafter\let\csname endequation*\endcsname\undefined
 
\usepackage{amsmath}
\usepackage{mathtools}
\usepackage[colorlinks=true,linkcolor=black,citecolor=black,urlcolor=black,filecolor=black,
bookmarksnumbered=true,unicode]{hyperref}
\usepackage{cite}
\makeatletter
\renewcommand{\@biblabel}[1]{[#1]\hspace{2em}} 
\makeatother
\usepackage{cuted}  
\usepackage{ragged2e}

\providecommand{\keywords}[1]
{
  \small	
  \textbf{\textit{Keywords---}} #1
}

\usepackage[english]{babel}
\usepackage {cancel}
\usepackage{ascmac}
\usepackage{threeparttable}
\usepackage{caption}
\usepackage{comment}
\usepackage{mathdots}
\usepackage[utf8]{inputenc}
\usepackage{times}
\usepackage{newtxtext}
\usepackage{mathptmx}
\usepackage{amsthm}
\makeatletter
\renewcommand\subsubsection{%
  \@startsection{subsubsection}{3}{\z@}%
    {-3.25ex\@plus -1ex \@minus -.2ex}
    {1.5ex \@plus .2ex}
    {\reset@font\normalsize\itshape}%
}%

\makeatother
\makeatletter
\renewcommand{\thefootnote}{\arabic{footnote}}
\renewcommand{\@makefnmark}{\textsuperscript{\thefootnote}} 
\long\def\@makefntext#1{\noindent\makebox[1.8em][r]{\@thefnmark.}\,#1}
\makeatother

\newcommand{\YNedit}[1]{#1}
\newcommand{\YNblueedit}[1]{#1}
\usepackage[top=30mm, bottom=30mm, left=25mm, right=25mm]{geometry}
\usepackage{color}
\usepackage{graphicx}
\usepackage{braket}
\usepackage[subrefformat=parens]{subcaption}
\usepackage{txfonts}
\usepackage{enumerate}
\usepackage{bm}
 
\definecolor{gray}{gray}{0.5}
\definecolor{emerald}{cmyk}{1,0,0.5,0}
\definecolor{bluegreen}{cmyk}{0.85,0,0.33,0}
\definecolor{violet}{cmyk}{0.79,0.88,0,0}	
\renewcommand{\eqref}[1]{\textup{\ensuremath{(\ref{#1})}}}

\newcommand{\1}{\mbox{1}\hspace{-0.23em}\mbox{l}}
\begin{document}
\title{Lindbladian $\mathcal{PT}$ phase transitions}
\author{Yuma Nakanishi$^1$ and Tomohiro Sasamoto$^2$}
\address{$^1$ Institute for Physics of Intelligence, University of Tokyo, 7-3-1 Hongo, Bunkyo-ku, Tokyo 113-0033, JAPAN}
\address{$^2$ Department of Physics, Institute of Science Tokyo, 2-12-1 Ookayama, Meguro-ku, Tokyo 152-8551, JAPAN}
\ead{nakanishi.y@phys.s.u-tokyo.ac.jp}
\begin{abstract}
A parity-time (PT) transition is a spectral transition characteristic of non-Hermitian generators; it typically occurs at an exceptional point, where multiple eigenvectors coalesce.
The concept of a PT transition has been extended to Markovian open quantum systems, which are described by the Gorini-Kossakowski-Sudarshan-Lindblad (GKSL) equation.
Interestingly, the PT transition in many-body Markovian open quantum systems, the so-called \textit{Lindbladian $\mathcal{PT}$ (L-$\mathcal{PT}$) phase transition}, is closely related to two classes of exotic nonequilibrium many-body phenomena: \textit{continuous-time crystals} and \textit{non-reciprocal phase transitions}. 
In this review, we describe the recent advances in the study of L-$\mathcal{PT}$ phase transitions.
First, we define PT symmetry in three distinct contexts: non-Hermitian systems, nonlinear dynamical systems, and Markovian open quantum systems, highlighting the interconnections between these frameworks.
Second, we develop mean-field theories of L-$\mathcal{PT}$ phase transitions for collective-spin systems and for bipartite bosonic systems with particle-number conservation.
Within these classes of models, we show that L-$\mathcal{PT}$ symmetry can induce a breaking of continuous time-translation symmetry down to a discrete one, leading to persistent periodic dynamics.
We further demonstrate that the L-$\mathcal{PT}$ phase transition point is typically \textit{a critical exceptional point}, where multiple collective excitation modes with zero excitation spectrum coalesce. 
These findings establish an explicit connection to continuous-time crystals and non-reciprocal phase transitions. 
Third, going beyond the mean-field theory, we analyze statistical and quantum properties, such as purity and quantum entanglement indicators of time-independent steady states for several specific models with the L-$\mathcal{PT}$ symmetry. 
Finally, we discuss future research directions for L-$\mathcal{PT}$ phase transitions.
\end{abstract}

\noindent\keywords{Parity-Time symmetry, Dissipative phase transition, Time crystal, Non-reciprocal phase transition, Critical exceptional point,  Collective spin system}
\maketitle
\pagestyle{plain}
\tableofcontents
\twocolumn
\small
\section{Introduction}
\renewcommand{\theequation}
{1.\arabic{equation} }
\setcounter{equation}{0}
\label{sec1}
Phase transition is a remarkable phenomenon in which small changes in external conditions, such as temperature, pressure, or chemical potential, can result in significant changes in the macroscopic properties of a system. Water freezing and metals becoming superconducting are just a few striking examples. 
Phase transitions arise not only in thermal equilibrium and quantum ground states, but also in far-from-equilibrium settings driven externally or subject to dissipation. These systems represent a diverse class of phase transitions with rich dynamical behaviors.

In this review, we focus on a notable class of nonequilibrium quantum phase transitions: those involving the breaking of parity-time (PT) symmetry under Lindbladian dynamics. We refer to these as \textit{Lindbladian $\mathcal{PT}$ phase transitions} and review their theoretical foundations and recent developments, highlighting the central roles of dissipation and symmetry in generating novel nonequilibrium quantum phases.

\subsection{Thermal, Quantum, and Dissipative phase transitions}
\label{sec11}
\subsubsection{Thermal phase transitions }
\label{sec111}
In closed quantum systems, the time evolution is governed by the Schr$\ddot{\rm{o}}$dinger equation, 
\begin{align}
\label{Shc}
i\partial_{t}\ket{\psi}=H\ket{\psi}
\end{align}
where the state of the system $\ket{\psi}$ evolves unitarily under the action of the Hermitian Hamiltonian $H$.
In thermal equilibrium, the systems are described by Gibbs states $\rho_{\rm Gibbs}:=\exp(-\beta H)/Z$ with inverse temperature $\beta$ and partition function $Z:=\textrm{Tr}[\exp(-\beta H)]$. 
In the thermodynamic limit, the parameter space spanned by the model’s control variables splits into sharply distinct regions, referred to as phases of matter; the transition between such phases is \textit{a thermal phase transition}~\cite{landau1980statphys1, pathria2011statmech,goldenfeld1992lectures, Nishimori}.
Such thermal phase transitions emerge from the competition between
thermal disorder and interaction-driven ordering and are signaled by the non-analytic behavior of the free energy $F:=-\ln Z/\beta$ or its derivatives (Table $\ref{phasetra}$).

In continuous phase transitions, the order parameter changes continuously from zero in the disordered phase to a finite value in the ordered phase,
and critical phenomena appear: susceptibility, specific heat, correlation length, and other observables diverge as power laws with universal exponents that are largely insensitive to microscopic details.
At the mean-field level, this behavior can be captured by Landau theory. 
For example, for a real order parameter \(\phi({\bf x})\) with a $Z_2$ symmetry \(\phi\!\to\!-\phi\), the free-energy functional $\mathcal{F}[\phi]$ can be expanded as a power series 
\begin{align}
\label{free}
   \mathcal{F}[\phi]=\int d{\bf x}\ f_0+\tfrac{r}{2}\phi^2+\tfrac{u}{4}\phi^4+\mathcal{O}(\phi^6), 
\end{align}
with $u>0$. In the spatially uniform case, for $r>0$, the disordered state with \(\phi=0\) is stable. As $r$ approaches zero from above, this state loses stability, triggering \emph{spontaneous symmetry breaking (SSB)}:  
for \(r<0\) the system relaxes to one of the degenerate minima at \(\phi\neq0\), thereby choosing a specific orientation and breaking the symmetry~\cite{landau1980statphys1, pathria2011statmech,goldenfeld1992lectures, Nishimori}.

To move beyond the uniform mean-field description and capture critical
fluctuations, \emph{Landau–Ginzburg–Wilson} (LGW) framework promotes
\(\phi\) to a coarse-grained field \(\phi(\mathbf{x})\) and augments the
functional with a gradient term \(|\nabla \phi(\mathbf{x})|^{2}\).
Wilson’s renormalization group analysis then integrates short-wavelength modes, capturing critical fluctuations and producing non-mean-field exponents that characterize the universality class in low dimensions~\cite{pathria2011statmech,goldenfeld1992lectures, Nishimori,Amit}.

\subsubsection{Quantum phase transition}
In the zero-temperature limit, the relevant state is the \emph{ground state} of the Hamiltonian $H$. In this case, phase transitions are driven by quantum fluctuations generated by the Heisenberg uncertainty principle instead
of thermal fluctuations, and closing the many-body energy gap at the critical point is their defining signature
\cite{carr2010understanding,Sachdev2011,Tasaki}. These are called \textit{quantum phase transitions}.

When a \emph{local} order parameter can still be identified, the LGW philosophy remains almost unchanged.
Quantum fluctuations add an imaginary-time direction, yielding an effective dimension $d_{\mathrm{eff}}=d+z$, with spatial dimension $d$ and dynamical exponent $z$ that relates energy and momentum, $\omega\!\sim\!k^{\,z}$. The quantum critical point then maps to a classical transition in \(d_{\mathrm{eff}}\) dimensions; for example, the one-dimensional transverse-field Ising chain with $z=1$ falls into the two-dimensional classical Ising universality class.
Similar to their thermal counterparts, continuous quantum phase transitions are often accompanied by the \emph{spontaneous breaking of the unitary symmetry}.

\footnotetext[1]{
Let \(\mathcal{H}\) be a Hilbert space describing the state space of a quantum system, and let \(|\psi\rangle,\ |\phi\rangle \in \mathcal{H}\) be arbitrary quantum states.
An operator \(U: \mathcal{H} \to \mathcal{H}\) is called a \emph{unitary} if it preserves the inner product:
\[
  \langle U\psi \mid U\phi \rangle = \langle \psi \mid \phi \rangle \quad \text{for all } |\psi\rangle, |\phi\rangle \in \mathcal{H}.
\]
This condition implies \(U^\dagger U = \mathbb{I}\), where \(U^\dagger\) is the Hermitian adjoint of \(U\). Moreover, \(U\) is a \emph{linear} operator:
\[
  U(\alpha|\psi\rangle + \beta|\phi\rangle) = \alpha\,U|\psi\rangle + \beta\,U|\phi\rangle
\]
for all complex scalars \(\alpha, \beta\).

\ \ In contrast, operator \(A\) is \emph{anti-unitary} if it preserves the inner products
up to complex conjugation and acts \emph{anti-linearly} on state superpositions. 
Formally, an anti-unitary \(A\) satisfies
\[
  \langle \psi \mid \phi \rangle = 
  \bigl\langle A\phi \mid A\psi \bigr\rangle^*,
\]
and for complex coefficients \(a,b\),
\[
  A\bigl(a\,\lvert\psi\rangle + b\,\lvert\phi\rangle\bigr)
  = a^*\,A\lvert\psi\rangle + b^*\,A\lvert\phi\rangle.
\]
}

\begin{table*}[htbp]
   \vspace*{0cm}
     \hspace*{0cm}
\scalebox{0.9}{\small
  \begin{tabular}{|c||c|c|c|}  \hline
     &Thermal phase transition & Quantum phase transition & Dissipative phase transition \\  \hline\hline
System & Hamiltonian  & Hamiltonian  & Liouvillian (e.g. Lindbladian) \\ \hline
    State & Gibbs state & Ground state & Steady-state \\ \hline
    Transition point & \begin{tabular}{c} Nonanalyticity of the free energy  \\ or its derivatives\end{tabular} & Energy gap 
    closing & Liouvillian gap closing \\ \hline
 \end{tabular}}
\caption{\footnotesize Comparison of thermal, quantum and dissipative phase transitions~\cite{Kessler}.}
\label{phasetra}
 \end{table*}

\subsubsection{Unitary symmetry and spontaneous symmetry breaking}
\label{secSSB}

To make precise what we mean by \emph{spontaneous breaking of unitary symmetry},
we briefly review the basic notions of symmetry and its breaking in ground states~\cite{carr2010understanding,Sachdev2011,Tasaki}.
A Hamiltonian $H$ is said to possess unitary symmetry if it commutes with a unitary operator $U$ \footnotemark, 
\begin{align}
\label{eq:symmetry}
[H,U]=0 .
\end{align}
Here $[\cdot,\cdot]$ denotes the commutator, $[A,B]:=AB-BA$.
Equation ($\ref{eq:symmetry}$) guarantees \(U\)-invariant dynamics, yet in the \emph{thermodynamic limit} the ground state may violate the symmetry.

A standard lattice formulation proceeds by specifying the order of limits: one introduces a symmetry-breaking field \(\epsilon\), takes the thermodynamic limit, and only then sends \(\epsilon\to 0\).
Let \(\Lambda\) be a finite lattice of volume \(|\Lambda|=V\) and \(o_{\bf x}\) be a local observable at a discrete lattice site $\bf x$ with \([o_{\bf x},U]\neq0\). Couple a symmetry-breaking field \(\epsilon>0\) to \(o_{\bf x}\), with \(O:=\sum_{{\bf x}\in\Lambda} o_{\bf x}\),
\begin{equation}
\label{epshion}
H_\epsilon \;=\; H -\epsilon O.
\end{equation}
Let \(\langle\cdot\rangle_{\epsilon,V}\) denote the ground-state expectation for \(H_\epsilon\)
at volume \(V\). The SSB is said to occur if
\begin{equation}
\label{eq:quasiavg}
m \;=\; \lim_{\epsilon\to 0}\;\lim_{V\to\infty}\;
\frac{1}{V}\langle O\rangle_{\epsilon,V} \;\neq\; 0 .
\end{equation}
The quantity $m$ is the order parameter.
If the limits are reversed, the expectation vanishes at any finite \(V\).

Applying an explicit symmetry-breaking field modifies the Hamiltonian \(H\) and its ground state. To diagnose SSB without altering the original Hamiltonian or ground state, we characterize the long-range order (LRO) via a ground-state correlator. Specifically, we define
\begin{align}
\label{lro}
\sigma^{2}
:= \lim_{|{\bf x-y}|\to\infty}\,\lim_{V\to\infty}
\braket{ o_{\bf x} o_{\bf y}}_{0,V}.
\end{align}
Equivalently,
$\sigma^{2}
= \lim_{V\to\infty}\braket{  O^{2}}_{0,V} /V^{2},$   
and we take \(\sigma\ge 0\) by definition. If \(\sigma>0\), the system has an LRO.


It has been shown quite generally that a nonzero \(\sigma\)~\eqref{lro} guarantees the corresponding SSB. In other words, the expectation value of the order parameter 
\(m\)~\eqref{eq:quasiavg} remains nonvanishing by first taking the thermodynamic limit and then taking the symmetry breaking field $\epsilon$ to zero~\cite{Kaplan1989OrderParameter, Tasaki, KomaTasaki1993CMP}.
In Appendix~\ref{transising}, a simple example of the transverse Ising model is briefly discussed.

\subsubsection{Dissipative phase transitions}
In contrast to the equilibrium setting discussed above, we now turn to open quantum systems: quantum systems coupled to external degrees of freedom such that information and energy are continuously exchanged with the environment, leading to decoherence, relaxation, and excitation.
The competition between coherent dynamics and dissipation can induce nonequilibrium non-analytic changes in steady-state properties, termed \emph{dissipative phase transitions (DPTs)}~\cite{Kessler,fazio2024many,kamenev2023field}.

The dynamics of an open quantum system is generally described by the superoperator $\hat{\mathcal{E}}_{(t,t_0)}$ (a linear map acting on the space of linear operators), which maps a density matrix at $t_0$ to another density matrix at $t$ as $\rho(t)=\hat{\mathcal{E}}_{(t,t_0)}\rho(t_0)$~\cite{Breuer, ARivas}.
(In this review, superoperators are denoted by hats.)
To ensure the physical consistency of the reduced dynamics, we shall assume that, for all
$t\ge t_0$, the dynamical map $\hat{\mathcal{E}}_{(t,t_0)}$ is completely positive and
trace preserving (CPTP).
If, in addition, the dynamics is Markovian and time-homogeneous, the density matrix evolves according to a time-independent master equation of 
the Gorini–Kossakowski–Sudarshan–Lindblad (GKSL) form~\cite{Lindblad, GKS}:
\begin{align}
\label{dm}
\dot{\rho}=\hat{\mathcal{L}}\rho,
\end{align}
where generator $\hat{\mathcal{L}}$ is called the \emph{Lindbladian} (Liouvillian of the GKSL form) \footnote{
In this review, we use the term \emph{Liouville space} for the state operator space $K$, taken as a subspace of trace-class operators on the system Hilbert space $\mathcal H$ and containing physical density operators.
To accommodate unbounded observables in infinite-dimensional systems, we specify in parallel a space of observables $K'$ and choose $K$ such that $O\rho$ is trace class for all $O\in K'$ and $\rho\in K$.
}
and is explicitly given by 
\begin{align}
\label{lindblad}
\hat{\mathcal{L}}\rho&=\hat{\mathcal{L}}_0[H]\rho+\sum_{\mu}\mathcal{\hat{D}}[L_{\mu}]\rho\nonumber
\\&:=-i[H,\rho]+\sum_{\mu}(2L_{\mu}\rho L_{\mu}^{\dagger}-\{L_{\mu}^{\dagger}L_{\mu},\rho\}).
\end{align}
Here, $H$ is the Hermitian Hamiltonian responsible for coherent dynamics, and $L_\mu$ denotes a Lindblad operator characterizing a dissipative process. The superoperator $\hat{\mathcal{L}}_0[H]$ and $\mathcal{\hat{D}}[L]$ are the coherent evolution and the dissipative effects due to the coupling to the environment, respectively. The first and second dissipative terms represent the quantum jumps and continuous non-unitary dissipation, respectively.
In the context of DPTs, the Lindbladian spectrum provides essential information~\cite{Minganti,Kessler}. Given that the steady state is unique, it coincides with an eigenstate with a zero eigenvalue. The Lindbladian gap (Liouvillian gap), defined as the difference between zero eigenvalue and the maximum real part of any other eigenvalue in the spectrum (also known as the asymptotic decay rate), typically determines the system's relaxation rate.
Typically, a DPT is accompanied by the closing of the Lindbladian gap, which leads to a divergent relaxation time. Similarly to closed systems, the spontaneous breaking of unitary symmetry is one of the fundamental processes underlying the emergence of continuous DPTs~\cite{Minganti}. 

Even within the Markovian setting, open quantum systems host a wide variety of intrinsically nonequilibrium DPTs, whose phases and critical behavior are shaped by the interplay of coherent drive, interactions, and dissipation. Examples include laser transitions~\cite{minganti2021liouvillian}, superradiance transitions~\cite{KirtonKeeling2017}, absorbing-state transitions~\cite{marcuzzi2016absorbing}, dissipative topological phase transitions~\cite{nie2021dissipative, nava2023lindblad}, non-reciprocal phase transitions~\cite{nadolny2025nonreciprocal, chiacchio2023nonreciprocal}, time-crystal transitions~\cite{Iemini}, and strong-to-weak phase transitions~\cite{gu2024spontaneous}.

\YNedit{Experimentally, signatures of DPTs, including finite-size manifestations of dissipative critical behavior, have now been reported across a variety of experimental platforms.}
\YNedit{Collectively coupled atomic ensembles in cavity-QED and free-space settings provide controlled access to collective spin dynamics, long-range light-mediated interactions, and collective decay. These platforms have revealed critical slowing down~\cite{Brennecke2013PNAS}, dynamical hysteresis~\cite{Klinder2015PNAS}, and superradiant phase transitions~\cite{Zhiqiang2017Optica,Ferri2021PRX,Ferioli,Song2025SciAdv}. In a driven--dissipative atomic Bose--Hubbard array, bistability between a superfluid branch and a resistive branch, together with critical slowing down within the bistable regime, provided evidence of a DPT~\cite{Labouvie2016PRL}. In superconducting circuits, experiments observed bistable switching in a circuit-QED lattice and during photon-blockade breakdown~\cite{Fitzpatrick2017PRX,Fink2017PhotonBlockadeBreakdown}, critical slowing down in a bistable circuit-QED system~\cite{Brookes2021SciAdv}, and long-lived metastability with a strongly reduced Lindbladian gap at a first-order DPT in a Duffing oscillator~\cite{Chen2023NatCommun}. Semiconductor microcavities hosting exciton-polariton modes provide driven--dissipative optical platforms in which dynamical hysteresis and photon-correlation measurements probe dissipative critical behavior~\cite{Rodriguez2017PRL,Fink2018DPTPhotonCorrelations}. In ultracold atomic Bose gases in optical lattices, time-resolved measurements revealed stochastic switching between metastable normal and superfluid states, an effective Lindbladian gap, and dynamical hysteresis, providing evidence for a first-order DPT~\cite{Benary2022NJP,Roehrle2024SciPost}. A two-photon-driven Kerr resonator exhibited finite-size signatures of both first- and second-order DPTs: hysteresis at the first-order transition, spontaneous symmetry breaking at the second-order transition~\cite{Beaulieu2025NatCommun}.}

\YNedit{Related experiments have also demonstrated engineered open many-body dynamics in settings closely connected to dissipative critical phenomena. Localized dissipation was engineered in an atomic Bose--Einstein condensate, revealing nonmonotonic atom loss and dissipation-induced Zeno dynamics~\cite{Barontini2013PRL}. Controllable two-body loss was used to investigate the Mott-insulator-to-superfluid crossover in a driven--dissipative Bose--Hubbard system~\cite{Tomita2017SciAdv}. A superconducting Jaynes--Cummings dimer exhibited a dissipation-driven localization transition from classical Josephson oscillations to a macroscopically self-trapped state~\cite{Raftery2014PRX}.}

\subsection{Parity-Time symmetry}
\label{sec12}

In closed quantum systems, anti-unitary symmetries such as time reversal are fundamental in the representation of symmetries. According to the CPT theorem, any relativistic quantum field theory that is microcausal and Poincaré covariant and satisfies the standard spectral and stability assumptions is invariant under the combined action of charge conjugation (C), parity (P), and time reversal (T)~\cite{StreaterWightman}.


By contrast, in open systems, anti-unitary symmetries play an important role in the reality of the spectrum. The dynamical generator for open systems is generically non-Hermitian, and thus its spectrum is complex. The presence of an anti-unitary symmetry (or, more generally, pseudo-Hermiticity) allows for a parameter regime in which all eigenvalues are real~\cite{MostafazadehA1}.
This possibility was highlighted in the context of PT-symmetric quantum mechanics, introduced by Bender and Boettcher in 1998~\cite{Bender}. 

The parity operator $P$ implements spatial inversion, transforming position and momentum as $x\to-x$, $p\to-p$. The time-reversal operator $T$, being anti-unitary, reverses the direction of time and momentum as $t\to-t$, $p\to-p$, and complex conjugate scalar quantities such that $i\to-i$. The (non-Hermitian) Hamiltonian $H$ is said to be PT-symmetric if it commutes with the PT operator~\cite{Bender} 
\begin{align}
\label{PT2}
[H, PT]=0,
\end{align}
where the Hamiltonian $H$ generates the Schr$\ddot{\rm{o}}$dinger dynamics ($\ref{Shc}$).

A PT-symmetric Hamiltonian can exhibit a spectral and dynamical transition, known as the \textit{PT transition}~\cite{MostafazadehA1, MostafazadehA2, MostafazadehA3}. Specifically, the dynamics of the system change from oscillatory behavior to exponential decay or growth as the eigenvalues undergo a real-to-complex transition. This transition is associated with PT symmetry breaking of the eigenvector. Moreover, it is typically marked by the emergence of an \textit{exceptional point (EP)}~\cite{Kato, Heiss}, which is the parameter value of a non-Hermitian operator at which the algebraic multiplicity exceeds the geometric multiplicity. For simple cases, two eigenvalues and their associated eigenvectors coalesce, rendering the Hamiltonian non-diagonalizable and producing a Jordan block. \YNedit{The branch-point topology of EPs was observed experimentally in a microwave-cavity system~\cite{Dembowski2001PRL}.}


\YNedit{Early optical experiments demonstrated both \textit{passive} PT-symmetry breaking\footnote{Here, \textit{passive} PT symmetry refers to loss-only systems mapped to a balanced gain--loss form by an overall imaginary energy shift or state normalization, whereas \textit{active} PT symmetry refers to systems with physical gain and loss.} and a PT-symmetry-breaking transition in a waveguide with balanced gain and loss~\cite{guo2009observation,RoterC}. Later experiments explored PT symmetry in synthetic photonic lattices and coupled microresonators~\cite{Regensburger2012Nature,Chang2014NaturePhotonics}, as well as in electronic circuits and mechanical oscillators~\cite{Schindler,Bender2}. Beyond spectral PT transitions, experiments have also revealed a range of phenomena associated with EPs, including asymmetric mode switching under dynamical encircling~\cite{Doppler2016Nature}, enhanced sensing~\cite{hodaei2017enhanced}, higher-order EP responses~\cite{wang2019arbitrary}, exotic lasing phenomena~\cite{feng2014single,hodaei2014parity}, non-reciprocal propagation~\cite{peng2014parity}, and topological energy transfer~\cite{xu2016topological}.}

\YNedit{Subsequent experiments have also brought PT-symmetric and EP-related physics into quantum platforms. PT-symmetry breaking was observed in a single-spin system~\cite{Wu1}, and information-flow critical behavior near an EP was characterized in single-photon interferometric simulations of PT-symmetric non-unitary dynamics~\cite{Kawabata2}. Quantum state tomography across an EP was carried out in a post-selected dissipative superconducting qubit~\cite{Naghiloo}. In atomic platforms, PT transitions were observed in a dissipative Floquet system of ultracold atoms with state-dependent loss~\cite{Li2019NatCommun}. Experiments have also moved toward many-body settings: a PT-symmetric non-Hermitian quantum many-body system was realized in an optical lattice with controlled one-body loss~\cite{Takasu2020PTEP}, and non-Hermitian PT-symmetry breaking was reported in an interacting Rydberg-atom array~\cite{Zhang2025RydbergPT}. Thus, PT transitions have been observed across a broad range of platforms, from classical-wave systems to quantum and many-body settings.}


PT transitions can resemble conventional phase transitions through symmetry breaking and qualitative changes in dynamics. However, they are characterized by the transition of eigenvalues from real values to complex-conjugate pairs in a non-Hermitian spectrum and do not inherently require collective many-body behavior. They therefore need not constitute phase transitions in the thermodynamic sense. In this review, we refer to the real-to-complex transition associated with PT-symmetry breaking as \textit{spectral} PT-symmetry breaking, to distinguish it from other notions.
\footnote{Anti-unitary symmetries can also be spontaneously broken in ordinary Hermitian many-body systems; for example, a time-reversal-invariant Hamiltonian may admit a ferromagnetic ground state that breaks time-reversal symmetry. This is conceptually distinct from the spectral PT transition discussed here: for Hermitian generators the spectrum remains real and no real-to-complex transition occurs.
}


\subsection{Lindbladian $\mathcal{PT}$ phase transitions}
\label{sec13}
The spontaneous breaking of unitary symmetries often underlies the occurrence of continuous phase transitions. In contrast, spectral PT-symmetry breaking yields a nonequilibrium dynamical transition. These facts raise the following central question:

\begin{quote}
\textit{Can the spectral breaking of PT symmetry, an anti-unitary symmetry, characterize phase transitions in nonequilibrium steady states of open quantum systems? 
}
\end{quote}



To answer this question, we begin by defining $\mathcal{PT}$ symmetry in the context of the Lindbladian. (We use the calligraphic symbol $\mathcal{PT}$ to denote parity–time symmetry at the superoperator level.) Comparing the Schr$\ddot{\rm{o}}$dinger equation ($\ref{Shc}$) and the master equation ($\ref{dm}$), the superoperator $i\hat{\mathcal{L}}$ plays a role analogous to that of a (non-Hermitian) Hamiltonian.
Consequently, a naive definition of $\mathcal{PT}$ symmetry in Lindbladian is to demand that it commutes with a $\mathcal{PT}$ superoperator: 
\begin{align}
\label{firstilpt}
[i\hat{\mathcal{L}},\hat{\mathcal{P}}\hat{\mathcal{T}}]=0,
\end{align}
where $\hat{\mathcal{P}}$ and $\hat{\mathcal{T}}$ are parity and time-reversal superoperators, respectively. \footnote{If one is interested in spectral breaking of an anti-unitary symmetry, one may instead impose
\(\{i\hat{\mathcal{L}},\hat{\mathcal{P}}\hat{\mathcal{T}}\}=0\); 
physically, however, this corresponds to \emph{anti-}\(\mathcal{PT}\) symmetry. 
In this review, we focus on physical PT-symmetric models. 
For the relation between anti-\(\mathcal{PT}\)-symmetric Lindbladians and L-\(\mathcal{PT}\) symmetry, see Sec.~\ref{sec42} and~\ref{sec53}.
}
Under this definition, the Lindbladian eigenvalue $\lambda$ is restricted to purely imaginary eigenvalues or pairs symmetric about the imaginary axis, $\{\lambda,-\lambda^*\}$. However, owing to the trace-norm contractivity of the CPTP semigroups, every eigenvalue satisfies Re$[\lambda]\leq0$.
Hence, a Lindbladian with dissipation cannot satisfy Eq.~\eqref{firstilpt}. 

A straightforward way to impose $\mathcal{PT}$ symmetry is to redefine the Lindbladian via a uniform spectral shift,
$\hat{\mathcal{L}}' := \hat{\mathcal{L}} - \gamma \hat{1},$ with $\gamma\in\mathbb{R}$~\cite{Prosen}.
In such a construction, a $\mathcal{PT}$ transition can appear in the spectrum of the shifted Lindbladian $\hat{\mathcal{L}}'$, but it is irrelevant to steady-state properties and therefore does not induce a DPT. 
We note that, likewise, for the effective non-Hermitian Hamiltonian obtained by neglecting quantum-jump terms, PT symmetry can be restored only after an overall imaginary energy shift is applied~\cite{ashida2017parity,Li2019NatCommun,Zhang2025RydbergPT,jaramillo2020pt, Takasu2020PTEP}. The subsequent PT-symmetry breaking does not give rise to a DPT (see Sec.~\ref{sec311} for details).

As an alternative approach, Huber et al. proposed another definition of $\mathcal{PT}$ symmetry for Lindbladians by interpreting the time-reversal operation as an exchange between the gain and loss~\cite{Huber2}. This is given by the following (slightly modified) condition~\cite{Nakanishi2,Nakanishi3}:
\begin{align}
\label{HuberPT2}
\hat{\mathcal{L}}[\mathbb{PT}(H); \mathbb{PT}(L_\mu),\mu=1,2,...]=\hat{\mathcal{L}}[H;L_\mu,\mu=1,2,...],
\end{align}
with $\mathbb{PT}(O)=PTO^{\dagger}(PT)^{-1}$. This ensures that the Lindbladian remains invariant under the combined PT transformation applied to both the Hamiltonian and set of Lindblad operators.

With this symmetry ($\ref{HuberPT2}$) in place, Huber et al. demonstrated that a DPT occurs in a collective spin system with alternating gain and loss without spontaneous breaking of unitary symmetries~\cite{Huber2, Huber1}. In addition, they implied that the PT symmetry breaking of the time-independent steady-state (TISS), which is an eigenmode with zero eigenvalues, triggers that DPT. They further obtained closed-form expressions in the thermodynamic limit for both the purity and the entanglement negativity of the TISS, and showed that these quantities, which are zero in the disordered phase, acquire finite values once the transition point is crossed~\cite{Huber1}.

Following these studies, we have shown for the same model that both \textit{spectral} and steady-state dynamical transitions occur at the PT-symmetry breaking point in the thermodynamic limit~\cite{Nakanishi}. This DPT is referred to as the \textit{Lindbladian $\mathcal{PT}$ (L-$\mathcal{PT}$) phase transition}.


However, the underlying reasons for this specific symmetry requirement and the exact role of the $\mathcal{PT}$ symmetry ($\ref{HuberPT2}$) in facilitating such a DPT were not initially well understood. 
One of the main challenges in understanding these transitions arises from the fact that extracting the physical consequences of symmetries is not always straightforward. This difficulty is primarily due to the technical issue that the L-$\mathcal{PT}$ symmetry ($\ref{HuberPT2}$) cannot be rewritten as a commutator with the Lindbladian and an arbitrary superoperator, unlike conventional unitary or anti-unitary symmetries. Consequently, the Lindbladian is not block-diagonalized into symmetry sectors, 
so standard symmetry-breaking techniques and their usual physical interpretations become much less effective.

\begin{figure}[t]
   \vspace*{-0.1cm}
     \hspace*{0.1cm}
\includegraphics[bb=0mm 0mm 90mm 150mm,width=0.33\linewidth]{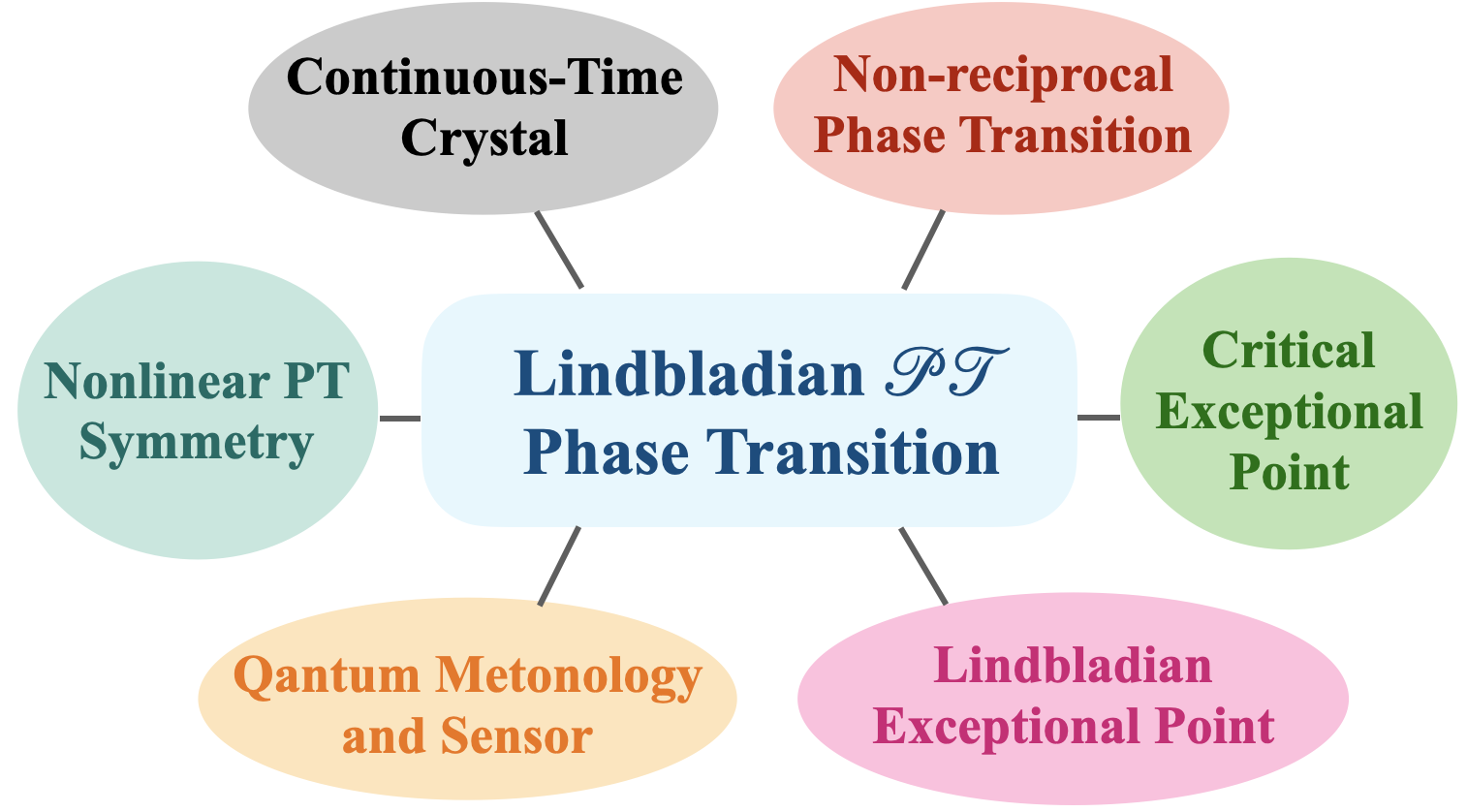}    \caption{\footnotesize\justifying Connections between L-$\mathcal{PT}$ phase transitions and other physical phenomena, symmetry and applications.}
    \label{figProsenPT1}
\end{figure}

\begin{table*}[t]
\centering
\renewcommand{\arraystretch}{1.1}
{\footnotesize
\begin{tabular}{ll|ll}
\hline
\textbf{Abbr.} & \textbf{Meaning} & \textbf{Abbr.} & \textbf{Meaning} \\
\hline
BTC   & Boundary Time Crystal             & L-$\mathcal{PT}$ & Lindbladian Parity--Time \\
CEP   & Critical Exceptional Point        & LEP   & Lindbladian (Liouvillian) Exceptional Point \\
CTC   & Continuous-Time Crystal           & LMG   & Lipkin--Meshkov--Glick \\
DCTC  & Dissipative Continuous-Time Crystal & LRO  & Long-Range Order \\
DDM   & Driven Dicke Model                & n-PT  & Nonlinear Parity--Time \\
DPT   & Dissipative Phase Transition      & PIE   & Purely Imaginary Eigenvalue \\
EP    & Exceptional Point                 & PT    & Parity--Time \\
GKSL  & Gorini--Kossakowski--Sudarshan--Lindblad & SSB   & Spontaneous Symmetry Breaking \\
HP    & Holstein--Primakoff               & TISS  & Time-Independent Steady State \\
\hline
\end{tabular}
\caption{\footnotesize Alphabetical list of frequently used abbreviations.}
\label{tab:abbreviations}
}
\end{table*}

\subsection{Nonlinear PT symmetry, Continuous-time crystals, and Quantum effects}
Despite these challenges, it has been shown analytically that L-$\mathcal{PT}$ symmetric models can exhibit novel DPTs across a broad class of systems 
by connecting two different notions of PT symmetries for microscopic and macroscopic dynamics developed in different contexts. Specifically, it has been proven that when the Lindbladian exhibits an L-$\mathcal{PT}$ symmetry at the microscopic level, its corresponding mean-field equation has a \textit{nonlinear PT (n-PT) symmetry}, and a DPT associated with \textit{spontaneous} n-PT symmetry breaking occurs. 
\footnote{
Symmetry breaking at a bifurcation with a finite number of degrees of freedom is usually not referred to as ``spontaneous symmetry breaking.'' Nevertheless, for a broad class of models, there is a clear correspondence between the symmetry breaking induced by such a bifurcation and the order-parameter spontaneous symmetry breaking that characterizes a phase transition (e.g., Landau theory). Whenever this correspondence holds, we will refer to the symmetry as \emph{spontaneously} broken. Appendix~\ref{phasebifur} provides illustrative examples that clarify the correspondence between phase transitions and bifurcations.} 
This result applies to a broad class of systems, including spatially extended bosonic models with conserved particle numbers and single collective-spin systems.


Furthermore, by linear stability analysis, it has been shown that the stable PT-symmetric solution of such systems generically displays persistent periodic oscillations. In addition, the transition point is characterized with a critical exceptional point (CEP). At this point, multiple collective excitation modes coalesce into a collective zero mode (e.g., the Nambu-Goldstone mode)~\cite{Fruchart, Hanai2, zelle2024universal, suchanek2023irreversible, belyansky2025phase, shmakov2025field, liu2025universal, weis2022exceptional}.
These findings reveal the deep connection between L-$\mathcal{PT}$ phase transitions and other nonequilibrium many-body phenomena~\cite{Nakanishi2, Nakanishi3}, such as \textit{continuous-time crystals (CTCs)}~\cite{Wilczek, Iemini}, nonequilibrium phases of matter that spontaneously break the continuous time-translation symmetry into a discrete one, and \textit{non-reciprocal phase transitions}~\cite{Fruchart, Hanai2, zelle2024universal, Hanai, you2020nonreciprocity, saha2020scalar,  suchanek2023entropy, nadolny2025nonreciprocal, chiacchio2023nonreciprocal}, where new forms of order or dynamics emerge due to non-reciprocal interactions. The standard continuous non-reciprocal phase transitions are associated with CEPs~\cite{Fruchart}. 
Their behavior is unique to nonequilibrium systems and cannot be captured within a Landau description.

\begin{table*}[t]
{ \centering
  \footnotesize
  \begin{tabular}{lll}
    \hline\hline
    Symbol & Acts on & Definition \\ 
    \hline
    $PT$ 
      & State (vector) $|\psi\rangle$ in the Hilbert space $\mathcal{H}$
      & $P:\,|\psi\rangle \mapsto P|\psi\rangle,\quad
         T:\,|\psi\rangle \mapsto T|\psi\rangle$ \\[3pt]
    $\hat{\mathcal{P}}\hat{\mathcal{T}}$
      & Observables $O\in K'$ and state operators $\rho\in K$ 
      & $\hat{\mathcal{P}}[X]=PXP^{-1},\quad
         \hat{\mathcal{T}}[X]=TXT^{-1},\quad (X=O\ \text{or}\ \rho)$ \\[3pt]
    $\tilde{P}\tilde{T}$
      & State vector $\mathbf{q}(t)\in\mathbb{C}^{n}$ in nonlinear dynamical systems
      & $\tilde{P}:\,\mathbf{q}(t)\mapsto \tilde{P}\mathbf{q}(t),\quad
         \tilde{T}:\,\mathbf{q}(t)\mapsto \mathbf{q}^{*}(-t)$ \\[3pt]
    $\mathbb{PT}$
      & Hamiltonian and Lindblad operators in Lindbladians
      & $\mathbb{PT}(O)=PT\,O^{\dagger}(PT)^{-1}$ \\
    \hline\hline
  \end{tabular}
    \caption{\footnotesize Summary of the notation for PT transformations used in this review.}
    \label{tapt}}
\end{table*}

While the above semiclassical analysis already highlights the exotic nature of L-$\mathcal{PT}$ phase transitions, calculations beyond the mean-field have shown that they also have a pronounced impact on quantum properties, in particular on entanglement. Several driven–dissipative models with L-$\mathcal{PT}$-symmetry exhibit transition points accompanied by strong enhancements of spin squeezing~\cite{lee2014dissipative, Hannukainen}, negativity~\cite{Huber1, Hannukainen}, genuine multipartite correlations~\cite{AC} and Quantum Fisher information (QFI)~\cite{Cabot, Montenegro, AC, cabot2025quantum} (although these studies are not originally formulated in the language of L-$\mathcal{PT}$ symmetry). 
These results suggest a close connection between L-$\mathcal{PT}$ phase transitions and the generation of entanglement: near such points, quantum correlations and metrological resources can be strongly enhanced, potentially opening up new possibilities for quantum sensing and precision metrology~\cite{Cabot, Montenegro, cabot2025quantum}.

In this paper, we review recent advances in L-$\mathcal{PT}$ phase transitions, mainly based on Refs.\cite{Nakanishi, Nakanishi2, Nakanishi3, Huber2, Huber1}, and provide additional data that reinforce these results. Section~\ref{sec2} provides a brief overview of PT symmetry and transitions in non-Hermitian systems and nonlinear dynamical systems. In Section~\ref{sec3}, we introduce two distinct notions of unitary symmetries in Lindbladians and DPTs with their breaking. In Section~\ref{sec4}, we explain the properties of L-$\mathcal{PT}$ symmetry and L-$\mathcal{PT}$ phase transitions, drawing comparisons with PT symmetry and transitions in non-Hermitian systems and shifted Lindbladians. In Section~\ref{sec5}, we analyze the mean-field theory of L-$\mathcal{PT}$ phase transitions, highlighting that L-$\mathcal{PT}$ symmetry can generally induce persistent oscillations in single-collective spin systems. Importantly, the transition point is associated with a spontaneous n-PT symmetry breaking and typically corresponds to a CEP. Additionally, in the PT-symmetry broken phase, a pair of stable and unstable fixed points emerges, replacing the steady-state degeneracy characteristic of the conventional SSB. Furthermore, we reveal a deep connection with non-reciprocal phase
transitions.
In Section~\ref{sec6}, we organize
the definition of CTCs in Markovian open quantum systems and explore the interplay between the L-$\mathcal{PT}$ symmetry and CTCs. Furthermore, we analyze a two–collective spin model with L-$\mathcal{PT}$ symmetry for which exact Lindbladian eigenvalues (around steady-state in the thermodynamic limit) can be obtained, thus deepening its connection to CTC behavior.
Section~\ref{sec7} delves into the statistical and quantum properties of L-$\mathcal{PT}$ phase transitions for specific collective spin models. Finally, Section~\ref{sec8} concludes the review, summarizing key findings and outlining potential future directions.
For convenience, we summarize the frequently used abbreviations and the notation for PT transformations in Table~\ref{tab:abbreviations} and Table~\ref{tapt}, respectively.

\section{PT symmetry and its breaking in non-Hermitian and nonlinear dynamical systems}
\renewcommand{\theequation}{2.\arabic{equation} }
\setcounter{equation}{0}
\label{sec2}

As mentioned in Sec.~\ref{sec12}, PT symmetry has played a central role in the study of non-Hermitian systems.
While the earliest studies of PT symmetry concentrated on systems described by the linear Schr$\ddot{\rm{o}}$dinger equation, many experimental platforms used to explore non-Hermitian physics inherently incorporate nonlinear effects, for example, Kerr nonlinearity in optics or mean-field interactions in BECs described by the Gross–Pitaevskii equation. 

These nonlinear effects naturally enrich the dynamics of the system and have led to a variety of exotic phenomena related to PT symmetry, such as the formation of stable solitons~\cite{abdullaev2011solitons, yang2014symmetry, sarma2014continuous, nixon2012stability, zezyulin2011stability}, PT-symmetric vortices~\cite{chen2014discrete}, transmission suppression~\cite{miroshnichenko2011nonlinearly}, unidirectional wave propagation~\cite{ramezani2010unidirectional, lin2011unidirectional}, nonlinear PT-symmetric lasers~\cite{ge2016nonlinear}, and PT-symmetry breaking chaos~\cite{lu2015pt}.
\YNedit{Related nonlinear PT physics has also been explored experimentally in several platforms. In electronic circuits, gain saturation has been used to realize robust wireless power transfer based on PT-symmetric dynamics~\cite{Assawaworrarit2017Nature}. In photonic systems, experiments have observed optical solitons in PT-symmetric lattices~\cite{Wimmer2015NatCommun}, power-dependent control of PT symmetry~\cite{xia2021nonlinear}, and nonlinear localization and soliton formation in two-dimensional PT-symmetric photonic lattices~\cite{Muniz2019PRL}.}

Sections~\ref{sec21} and~\ref{sec22} provide a concise review of the PT symmetry for systems described by a discrete Schr$\ddot{\rm{o}}$dinger-type equation and nonlinear dynamical systems, respectively. 
Readers familiar with PT symmetry in non-Hermitian systems may skip Section~\ref{sec21}. For a more detailed review, we refer the reader to Refs.\cite{bender1999, bender2005introduction, bender2007making, el2018non, ozdemir2019parity, ashida2020non} for the linear (non-Hermitian) case and Ref.\cite{Konotop} for the nonlinear case. 

\subsection{Non-Hermitian system}
\label{sec21}

In Section~\ref{sec1}, we introduced the non-Hermitian PT symmetry ($\ref{PT2}$) as defined by the commutation relation between the Hamiltonian (the generator of the Schr$\ddot{\rm{o}}$dinger equation), and the combined operator of the physical parity and time-reversal operators. We now extend this definition to a finite dimensional matrix $H$ in a Schr$\ddot{\rm{o}}$dinger-type equation and a generalized parity-time operation. 
This generalization is not limited to quantum systems, and operators $P$ and $T$ are not required to be physical parity or time-reversal operations.

Let us consider a Schr$\ddot{\rm{o}}$dinger-type equation, $i\partial_{t}\ket{\psi}=H\ket{\psi}$ with an $n$-component vector $\ket{\psi}\in\mathbb{C}^{n}$ and an $n\times n$ complex matrix $H\in\mathbb{C}^{n\times n}$. We call the matrix $H$ a Hamiltonian even if it is non-Hermitian, i.e. $H\neq H^\dagger$.
The Hamiltonian $H$ is said to be PT-symmetric~\cite{Bender, MostafazadehA1, MostafazadehA2, MostafazadehA3} if it commutes with the anti-unitary operator $PT$
\begin{align}
\label{PT}
[H, PT]=0,
\end{align}
where $P:\mathbb{C}^{n}\to \mathbb{C}^{n}$ is a unitary (generalized parity) operator and $T:\mathbb{C}^{n}\to \mathbb{C}^{n}$ is an anti-unitary (generalized time-reversal) operator,
Throughout this review, we restrict to the commonly used setting
\begin{align}
\label{opPT}
P=P^{\dagger}, \ \ P^{2}=T^{2}=1, \ \ [P, T]=0,
\end{align}
for simplicity.


When the Hamiltonian $H$ has PT symmetry ($\ref{PT}$), a PT transformation of the eigenvector $\ket{\psi_i}$ with eigenvalue $E_i$ is also the eigenvector with eigenvalue $E_i^*$ as
\begin{align}
H(PT\ket{\psi_i})=PTH\ket{\psi_i}=PTE_i\ket{\psi_i}=E_i^*(PT\ket{\psi_i}),
\end{align}
where in the last equality we used the anti-unitarity of $T$, which yields $T(E_i\ket{\psi_i})=E_i^*T\ket{\psi_i}$.
PT symmetry is said to be unbroken if all eigenvectors $\ket{\psi_i}$ are also eigenvectors of the PT operator ($PT \ket{\psi_i}\propto \ket{\psi_i}$), in which case all eigenvalues are real ($E_i=E_i^*$). Meanwhile, PT symmetry is said to be broken if some eigenvectors $\ket{\psi_i}$ are not those of the PT operator ($PT \ket{\psi_i}\ \cancel{\propto}\ \ket{\psi_i}$). 
 In the PT-broken case, complex conjugate eigenvalue pairs appear ($E_i\neq E_i^*$). This spectral transition with PT symmetry breaking is referred to as \textit{the PT transition}. 
 
\begin{figure}[t]
   \vspace*{-1.6cm}
     \hspace*{0.7cm}
\includegraphics[bb=0mm 0mm 90mm 150mm,width=0.37\linewidth]{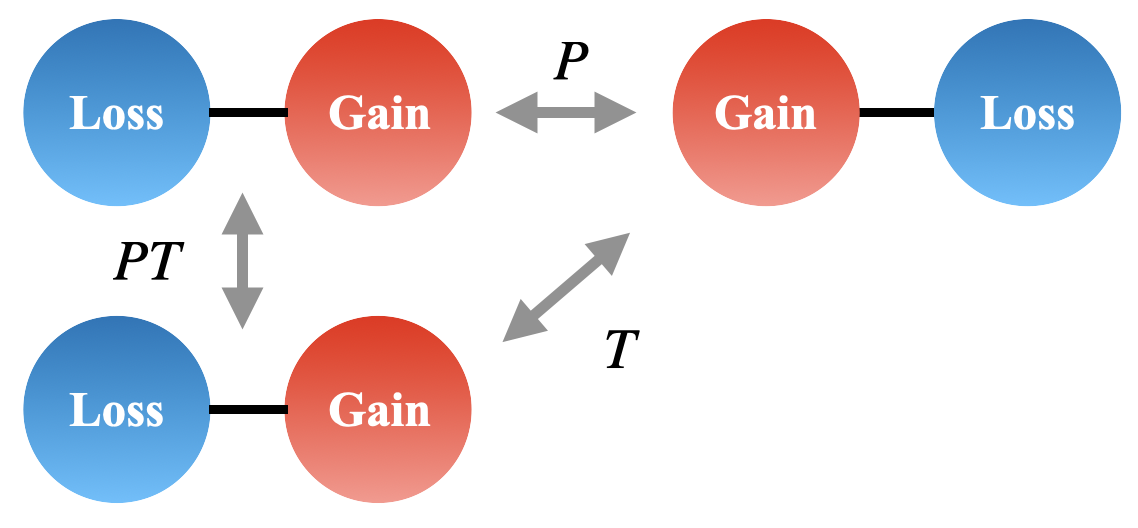}
\caption{\footnotesize\justifying A basic PT-symmetric example with balanced gain and loss: the parity operator exchanges two particles, while the time-reversal operator swaps gain and loss. [Inspired by Ref.\cite{ozdemir2019parity}]}
    \label{figPT4}
\end{figure}

To illustrate the above general statement, we consider the simplest PT-symmetric system with balanced gain and loss at a rate $\Gamma>0$ (Figure $\ref{figPT4}$). Its effective non-Hermitian Hamiltonian is described by
\begin{align}
\label{2times2}
H=\left(
   \begin{array}{cc}
     -i\Gamma& g \\
     g  & i\Gamma
   \end{array}
  \right),
\end{align}
with coherent interaction strength $g>0$.
One readily verifies that the non-Hermitian Hamiltonian in Eq.~($\ref{2times2}$) is PT-symmetric under the parity operator 
\begin{align}
P=\left(\begin{array}{cc}
     0& 1 \\
     1 & 0
   \end{array}\right),
\end{align}
and the time-reversal operator $T$ is taken as $T = K$, where $K$ denotes the complex conjugation ($i \mapsto -i$).
Throughout what follows, we take $T=K$.
The eigenvalues $E_{\pm}$ and eigenvectors $\ket{\psi_{\pm}}$ are given by
\begin{align}
E_{\pm}=\pm\sqrt{g^{2}-\Gamma^{2}},\ \ \
\ket{\psi_{\pm}}=\left(
   \begin{array}{c}
   -i\Gamma\pm\sqrt{g^{2}-\Gamma^{2}}\\
     g
        \end{array}
  \right),
\end{align}
respectively. For $\Gamma<g$, both eigenvalues are real, and the eigenvectors are PT-symmetric (i.e., $PT\ket{\psi_{\pm}}\propto\ket{\psi_{\pm}}$), indicating oscillating solutions. For $\Gamma>g$, on the other hand, the eigenvalues are complex conjugate pairs, and the PT symmetry of the eigenvectors is broken (i.e., $PT\ket{\psi_{\pm}}\ \cancel{\propto}\ \ket{\psi_{\pm}}$), that is, one exponentially amplifying (unstable) and one exponentially decaying (stable) mode. Moreover, the two eigenvectors coalesce at the transition point, and the Hamiltonian becomes non-diagonalizable. This point is called \textit{the exceptional point (EP)}~\cite{Kato, Heiss}.

\subsection{Nonlinear dynamical system}
\label{sec22}
We consider a finite-dimensional nonlinear dynamical system on $\mathbb{C}^n$ of the form
\begin{equation}
\label{nldy}
i\,\partial_t \mathbf{q}={\bf{f}}({\bf{q}}),
\end{equation}
with $\mathbf{q}:=(q_1,q_2,..,q_n)\in\mathbb{C}^n$ and ${\bf{f}}:=(f_1,f_2,...,f_n)$ with component functions $f_j:\mathbb{C}^{n}\to\mathbb{C}$ that are in general nonlinear.

\subsubsection{Nonlinear PT symmetry }

The concept of PT symmetry has been generalized to nonlinear dynamical systems~\cite{li2011pt, zezyulin2012nonlinear, konotop2012discrete, sarma2014continuous, leykam2013discrete, kevrekidis2013nonlinear, Konotop}. 
We define nonlinear PT (n-PT) symmetry as 
\begin{align}
\label{nonlinearPT}
\tilde{P}\tilde{T}{\bf{f}}({\bf{q}})={\bf{f}}({\tilde{P}\tilde{T}\bf{q}}),
\end{align}
where $\tilde{P}$ is a parity matrix, and $\tilde{T}$ acts as the complex conjugation $i\to-i$ together with time reversal $t\to-t$. (Symmetry operations in nonlinear dynamical systems will be denoted with tildes.) Note that the n-PT symmetry ($\ref{nonlinearPT}$) reduces to the conventional PT symmetry ($\ref{PT}$) if ${\bf{f}}({\bf{q}})$ is linear, that is, ${\bf f}({\bf{q}})=H{\bf{q}}$. Upon satisfying condition ($\ref{nonlinearPT}$), if ${\bf{q}}(t)$ is a solution, then its PT-transformed one $\tilde{P}\tilde{T}{\bf{q}}(t)=\tilde{P}{\bf{q}}^*(-t)$ is also a solution.

Ref.~\cite{Konotop} adopts a slightly different definition of nonlinear PT symmetry.
The vector field is factorized as ${\bf{f}}({\bf{q}})=F({\bf{q}}){\bf{q}}$ and the matrix-valued function $F({\bf{q}})$ is required to satisfy
\begin{align}
\label{nonlinearPTmat}
\tilde{P}\tilde{T}F({\bf{q}})(\tilde{P}\tilde{T})^{-1}=F(\tilde{P}\tilde{T}{\bf{q}})\ \ \ \ \ \textrm{for all {\bf{q}}}.
\end{align}
Condition~\eqref{nonlinearPTmat} is more restrictive than Eq.~\eqref{nonlinearPT} and, more importantly, depends on the particular choice of the factorization ${\bf f}({\bf{q}})=F({\bf{q}})\,{\bf{q}}$.
Since such a factorization is not unique, the same nonlinear dynamical system may admit both choices of $F({\bf{q}})$ that satisfy Eq.~\eqref{nonlinearPTmat} and choices that do not \footnote{For example, we consider a nonlinear dynamical system of the form 
$i\partial_t{\bf{q}}=\left(\begin{array}{c}
    \alpha|\beta|^{2} \\
    \beta|\alpha|^{2}
   \end{array}\right)
  =F({\bf{q}}){\bf{q}}$
   with ${\bf{q}}:=(\alpha,\beta)^T$ ($\alpha,\beta\in\mathbb{C}$). Then, matrix $F_{1}({\bf{q}}):=\left(\begin{array}{cc}
    |\beta|^{2}&0 \\
    0&|\alpha|^{2}
   \end{array}\right)$ satisfies definition ($\ref{nonlinearPTmat}$) with $\tilde{P}=\left(\begin{array}{cc}
    0&1 \\
    1&0
   \end{array}\right)$, but matrix $F_{2}({\bf{q}}):=\left(\begin{array}{cc}
    0&\alpha\beta^{*} \\
    0&|\alpha|^{2}
   \end{array}\right)$ does not.}.
For this reason, in this review we adopt the definition~\eqref{nonlinearPT}, which directly constrains the nonlinear vector field ${\bf f}({\bf{q}})$ and is independent of any auxiliary decomposition.

\subsubsection{Nonlinear PT symmetry breaking}
\label{sec222}
We now clarify the breaking of n-PT symmetry in the steady-state. We first consider a time-independent steady state, i.e., a fixed point ${\bf{q}}_0$. The PT symmetry is said to be unbroken when ${\bf{q}}_0=\tilde{P}\tilde{T}{\bf{q}}_0$. If this equality fails, the PT symmetry is said to be broken. 

Next, we consider a time-periodic solution \(\mathbf{q}(t)\) with period \(T\), i.e. \(\mathbf{q}(t)=\mathbf{q}(t+T)\). In the linear limit, one of these solutions takes the form \(\mathbf{q}(t)=e^{-iE_i t}\ket{\psi_i}\), where \(PT\,\ket{\psi_i}=e^{i\theta_i}\ket{\psi_i}\) with \(\theta_i\in\mathbb{R}\) in PT-unbroken phase. Assuming \(E_i\) to be nonzero, we obtain
\begin{align}
PT\,\mathbf{q}(t)
= e^{-iE_i\bigl(t-\theta_i/E_i\bigr)}\ket{\psi_i}
= \mathbf{q}\bigl(t-\theta_i/E_i\bigr).
\end{align}
Thus, in the linear case the PT transformation acts as a time translation by \(\theta_i/E_i\). Motivated by this, in the nonlinear case we allow for a global time shift \(t_{0}\in\mathbb{R}\), and say that the motion is n-PT symmetric if there exists \(t_{0}\) such that
\begin{align}
\mathbf{q}(t)=\tilde{P}\tilde{T}\,\mathbf{q}(t-t_{0}),
\end{align}
and otherwise n-PT symmetry is broken.



In non-Hermitian systems, the spectral PT symmetry breaking signals a dynamical transition. For nonlinear dynamical systems, a similar link has been explored in specific models both analytically and numerically~\cite{barashenkov2014exactly, cuevas2013pt, li2011pt, li2013revisiting, achilleos2012dark, kevrekidis2013nonlinear}, but a general correspondence between n-PT symmetry breaking of solutions and dynamical transitions has not been established.
By contrast, we find that the dynamical transitions from oscillatory to exponential decay occur at the \textit{spontaneous} n-PT symmetry breaking point for general single-collective spin models or bipartite bosonic systems with particle-number conservation.
This statement will be explained in detail in Section~\ref{sec5}.

\subsubsection{Nonlinear anti-PT symmetry}
It is also useful to introduce a closely related notion, which we refer to as nonlinear anti-PT (n-anti-PT) symmetry. We say that the nonlinear dynamical system has the n-anti-PT symmetry if
\begin{align}
\label{nonlinearantiPT}
\tilde{P}\,{\bf f}^*({\bf q})=-{\bf f}\bigl(\tilde{P}{\bf q}^*\bigr),
\end{align}
where the asterisk denotes complex conjugation. When ${\bf f}({\bf{q}})=H{\bf{q}}$, it corresponds to a conventional anti-PT symmetry, $\{H,PT\}=0$ with $T=K$. Here $\{\cdot,\cdot\}$ denotes the anti-commutator, $\{A,B\}:=AB+BA$. If condition~\eqref{nonlinearantiPT} is satisfied and
${\bf q}(t)$ is a solution, then its anti-PT–transformed trajectory
$\tilde{P}{\bf q}^*(t)$ is also a solution. In contrast to the n-PT symmetry, this transformation does not involve an explicit time reversal $t\to -t$.

\section{Lindbladian dynamics, Unitary symmetries, and Dissipative Phase Transitions}
\renewcommand{\theequation}{3.\arabic{equation} }
\setcounter{equation}{0}
\label{sec3}
Symmetries and their spontaneous breaking provide insights into the presence of conserved quantities and the occurrence of phase transitions. Here, we summarize unitary symmetries in Lindbladians and DPTs associated with their spontaneous breaking. 
This section is designed to be useful to readers who are not familiar with phase transitions in open quantum systems. For those interested in a quicker overview of the L-$\mathcal{PT}$ phase transitions, we recommend skipping this section.  
In Section~\ref{sec31}, we describe the general properties of open quantum systems governed by the GKSL equation. In Section~\ref{sec32}, we explain the relation between persistent oscillations and purely imaginary eigenvalues in open quantum systems. In Section~\ref{sec33}, we introduce two types of unitary symmetries for Lindbladians. Finally, in Section~\ref{sec34}, we define DPTs and briefly review DPTs involving spontaneous breaking of unitary symmetries.

\subsection{Fundamental properties of the GKSL equation}
\label{sec31}
\subsubsection{Effective non-Hermitian Hamiltonian}
\label{sec311}
The GKSL equation ($\ref{lindblad}$) can be rewritten as
\begin{align}
\label{lindblad2}
\frac{d\rho}{dt}=-i(H_{\rm{eff}}\rho-\rho H_{\rm{eff}}^{\dagger})+\sum_{\mu}2L_{\mu}\rho L_{\mu}^{\dagger},
\end{align}
with the effective non-Hermitian Hamiltonian 
\begin{align}
\label{effective}
H_{\rm{eff}}:=H-i\sum_{\mu}L_{\mu}^{\dagger}L_{\mu}.
\end{align}
The quantum jump terms $2L_{\mu}\rho L_{\mu}^{\dagger}$ can be neglected in the short‑time (rare‑jump) regime or under post‑selection of no‑jump trajectories, indicating that the dynamics is well approximated by the effective non‑Hermitian Hamiltonian~\cite{WisemanMilburn2010, Daley2014}. 

Assuming \(H_{\mathrm{eff}}\)~\eqref{effective}
is diagonalizable, for a right eigenvector $H_{\mathrm{eff}}|\psi_j\rangle=E_j|\psi_j\rangle$,
\begin{align}
E_j={\langle\psi_j|H|\psi_j\rangle}
-i{\langle\psi_j|\sum_\mu L_\mu^\dagger L_\mu|\psi_j\rangle},
\end{align}
so that ${\rm Im}\,[E_j]\le 0$.
A PT-symmetric operator has a spectrum symmetric with respect to the real axis
($E_j$ implies $E_j^*$). Therefore, if dissipation is present,
$H_{\mathrm{eff}}$ cannot be PT-symmetric, i.e., $[H_{\mathrm{eff}},PT]\neq 0$.



Note that, in postselected (no-jump) dynamics, an overall imaginary constant in the eigenvalue spectrum can be removed by normalizing the quantum state; equivalently, one may re-center the spectrum by a \emph{purely imaginary energy shift}
\(H_{\mathrm{eff}}' := H_{\mathrm{eff}} + i a I\) with \(a\in\mathbb{R}\)~\cite{ashida2017parity,jaramillo2020pt}.
Such a shift translates all eigenvalues without changing eigenvectors, and is often useful in passive-PT settings where the effective non-Hermiticity contains a uniform loss offset.
For the shifted Hamiltonian \(H_{\mathrm{eff}}'\), spectral PT-symmetry breaking can be discussed and may be accompanied by non-trivial critical behavior~\cite{ashida2017parity}.
However, this critical behavior concerns ground-state properties of \(H_{\mathrm{eff}}'\) rather than the full GKSL dynamics.
Thus, these spectral PT transitions are distinct from the steady-state phase transitions that are the focus of this review.

\subsubsection{Time evolution for observables}
We define the Hilbert--Schmidt inner product by
\begin{align}
\langle \sigma,\rho\rangle_{\mathrm{HS}}:=\operatorname{Tr}(\sigma^\dagger \rho).
\end{align}
The adjoint generator $\hat{\mathcal{L}}^\dagger$ is defined through this duality,
\begin{align}
\label{eq:HS-duality}
\braket{O,\hat{\mathcal{L}}(\rho)}_{\mathrm{HS}}
=
\braket{(\hat{\mathcal{L}}^\dagger O),\rho}_{\mathrm{HS}}.
\end{align}
The adjoint generator for Lindbladian in Eq.~\eqref{lindblad} acts on any operator $O$ as
\begin{align}
\hat{\mathcal{L}}^\dagger(O)
&=
\hat{\mathcal{L}}^\dagger[H](O)
+\sum_{\mu}\hat{\mathcal{D}}^\dagger[L_\mu](O)\nonumber\\
&:=
i[H,O]
+\sum_{\mu}[L^\dagger_\mu,O]L_\mu-L^\dagger_\mu[L_\mu,O].
\label{eq:adjoint}
\end{align}
Hence, the equations of motion for an observable $O$ are written as
\begin{align}
\frac{d}{dt}\langle O\rangle(t)
&=
\operatorname{Tr}\!\left[\,O\,\dot{\rho}(t)\,\right]
=
\operatorname{Tr}\!\left[\,({\hat{\mathcal{L}}^\dagger }O)\,\rho\,\right],
\end{align}
with $\braket{O}:=\textrm{Tr}[O\rho]$.
The trace preservation of the dynamics generated by $\hat{\mathcal{L}}$ implies unitality in the Heisenberg picture, $\hat{\mathcal{L}}^\dagger[\1]=0$, where $\1$ is the identity operator.

\subsubsection{Lindbladian eigenvalues and eigenmodes}
\label{sec313}
The Lindbladian spectrum encodes key information about DPTs. The eigenvalue problem for the Lindbladian is defined at the level of operators as
\begin{equation}
\label{eigen}
  \hat{\mathcal L}\,\rho_i = \lambda_i\,\rho_i ,
\end{equation}
where $\rho_i$ is the right eigenmode, and $\lambda_i\in\mathbb C$ are the corresponding eigenvalues. 
 It is generally known that the real part of the eigenvalue $\textrm{Re}[\lambda_i]$ of the Lindbladian $\hat{\mathcal L}$ is always non-positive
${\rm Re}[\lambda_{i}$]$\leq0, \forall i$. If Eq.~\eqref{eigen} holds, then $\hat{\mathcal L}\,\rho_i^\dagger = \lambda_i^*\,\rho_i^\dagger $.

In addition, there is at least one eigenmode with $\lambda_i=0$ corresponding to the time-independent steady state (TISS) $\text{\cite{Breuer, ARivas}}$. Moreover, for finite-dimensional systems, the steady state (that is, the eigenmode with Re$[\lambda_i]=0$) is unique under generic assumptions~\cite{Spohn1977_AlgebraicCondition, Evans1977_IrreducibleQDS, AlbertJiang2014_SymmetriesLindblad, Nigro, Yoshida2024_UniquenessGKSLE}. Exceptions to this generic uniqueness arise in the presence of additional structures such as strong symmetries~\cite{buvca2012note}, decoherence-free subspaces~\cite{lidar1998decoherence}, or strong dynamical symmetries~\cite{buvca2019non}, as will be discussed below.

Suppose that the steady state is unique, the normalized eigenmode with zero eigenvalue $\rho_0$ is the steady-state $\rho_{ss}=\rho_0/{\rm Tr}[\rho_0]$. The absolute value of the real part of the second maximal eigenvalue is termed the Lindbladian gap
\begin{align}
\label{Lindbladiangap}
\Delta_{\textrm{L}}:= -\max_{i\neq0}\textrm{Re}[\lambda_{i}],
\end{align}
which typically gives the timescale of the relaxation
$\text{\cite{Kessler, Minganti}}$ (except for the case of Ref.\cite{MoriShiraiPRL2020}). 
If $\hat{\mathcal L}$ is diagonalizable, the time evolution is simply written as
\begin{align}
\label{timedepedensity}
\rho(t)=\rho_{ss}+\sum_{i>0}c_{i}\exp(\lambda_{i}t)\rho_{i}.
\end{align}
Here $c_i$ can be expressed using the biorthogonal left eigenmodes $\hat{\mathcal L}^\dagger \ell_i=\lambda_i^*\ell_i$
normalized by $\mathrm{Tr}(\ell_i^\dagger\rho_j)=\delta_{ij}$, as $c_i=\mathrm{Tr}(\ell_i^\dagger\rho(0))$.

On the other hand, if the spectrum contains non-diagonalizable sectors, most notably EPs, several eigenvalues coalesce, and the Lindbladian acquires Jordan blocks. Consider an $n$‑dimensional Jordan block associated with the degenerate
eigenvalue $\lambda_{\mathrm{J}}$. Then the time evolution develops polynomial prefactors,
\begin{equation}
  \rho(t)=\rho_{ss}+
  \sum_{k=0}^{n-1}
  \Bigl(c_{\mathrm{J}}^{(k)}\,t^{\,k}\Bigr)\,
  e^{\lambda_{\mathrm{J}} t}\,
  \rho_{\mathrm{J}}^{(k)}
  \;+\;
  \sum_{i\notin\mathrm{J},\ i>0} c_i\,e^{\lambda_i t}\,\rho_i,
\end{equation}
where $k=0,1,\dots ,n-1$ labels the Jordan chain ($k=0$ is the ordinary eigenmode, whereas $k\ge 1$ the generalized ones). The $t^{\,k}$ arise directly from the Jordan structure and lead to non-trivial relaxation.


For a \emph{finite size} Lindbladian, a Jordan block cannot be attached to the steady-state eigenvalue \(\lambda = 0\); such a scenario would imply unphysical divergences of observables in time~\cite{Minganti}. Therefore, a special Lindbladian exceptional point (LEP), where multi-eigenmodes with zero eigenvalue \(\lambda = 0\) coalesce, exists only in the \emph{thermodynamic limit}. We call it \textit{zero-mode LEP}. \YNedit{(The finite-size approach toward this same forbidden steady-state EP is termed an \textit{asymptotic exceptional steady state}~\cite{HuBudich2025Asymptotic}.)}
\YNedit{The relevant finite-\(N\) LEP instead occurs at a nonzero eigenvalue, and its parameter value is slightly shifted from the thermodynamic-limit zero-mode LEP. At this LEP, the Jordan-chain contribution has the finite-\(N\) time dependence \(t^{\,k}e^{\lambda_J(N)t}\), where \(\lambda_J(N)\neq0\) is the coalesced eigenvalue. Its parameter value and \(\lambda_J(N)\) converge algebraically toward the thermodynamic-limit zero-mode LEP with increasing \(N\). Section~\ref{sec552} illustrates this scaling in a representative collective-spin model: the finite-\(N\) LEP approaches the thermodynamic-limit zero-mode LEP algebraically as the total spin \(S\) increases.}




\begin{figure}[t]
   \vspace*{-0.3cm}
     \hspace*{0cm}
\includegraphics[bb=0mm 0mm 90mm 150mm,width=0.37\linewidth]{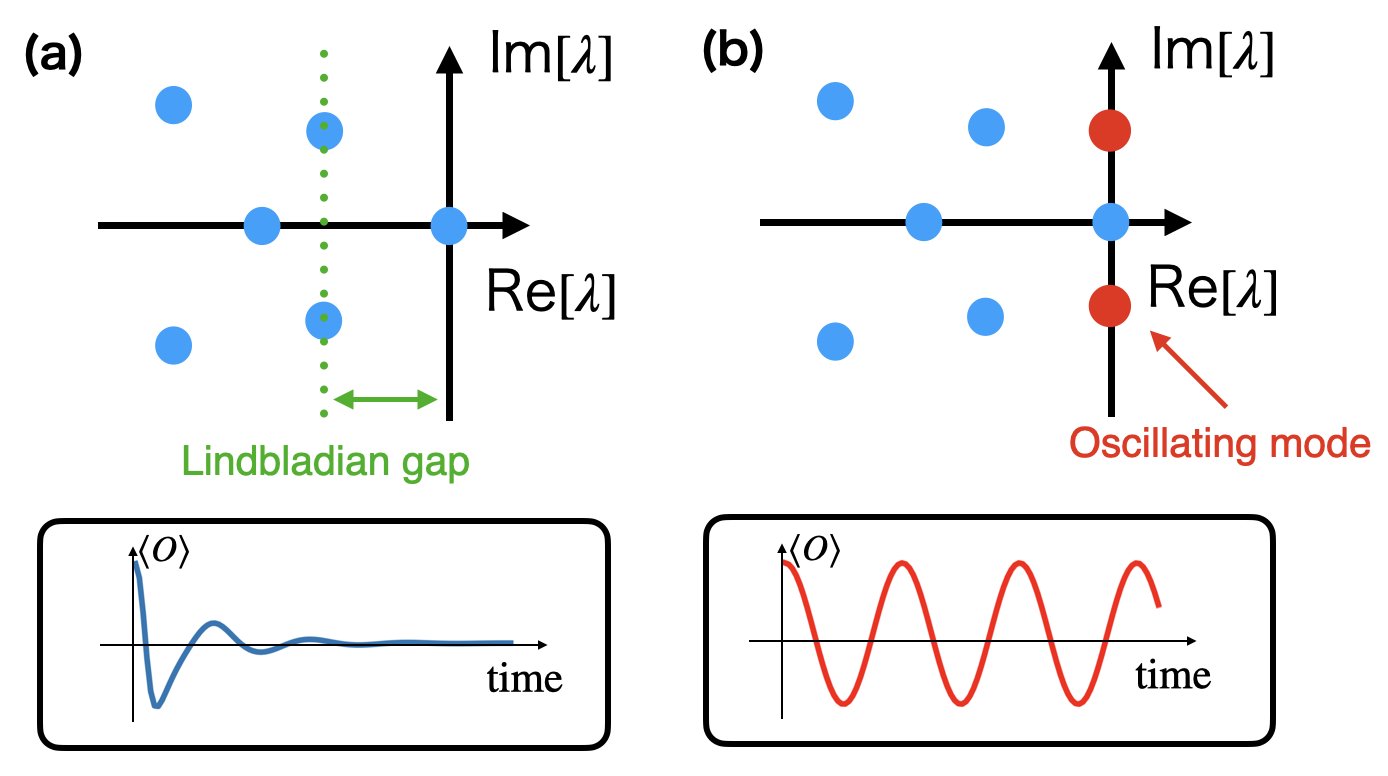}
\caption{\footnotesize\justifying(Top) Lindbladian eigenvalues and (Bottom) their corresponding dynamics of an observable $\braket{O}$ for (a) typical case and (b) persistent oscillations including CTCs. }
    \label{figeig}
\end{figure}

\subsubsection{\texorpdfstring{\YNblueedit{Natural representation}}{Natural representation}}
\YNblueedit{We now switch from Liouville space to the \emph{doubled Hilbert space} representation by using the \textit{natural representation}, or vectorization, which maps the density operator to a state vector and the Lindbladian to an operator acting on the doubled space~\cite{Hamazaki2026Monitored}.}
A density operator $\rho=\sum_{i,j}\rho_{i,j}\ket{i}\bra{j}$ is mapped to a pure state $\ket{\rho}=\sum_{i,j}\rho_{i,j}\ket{i}\ket{j}\in \mathcal H \otimes {\mathcal H}^*$ in the doubled Hilbert space. \YNblueedit{Under this representation}, the Lindbladian superoperator $\hat{\mathcal{L}}$ becomes a non-Hermitian matrix (operator) $\bar{\mathcal{L}}$ over a doubled Hilbert space,
\begin{align}
\label{lindmat}
\bar{\mathcal{L}}&=-i(H\otimes \1-\1\otimes H^{*})\nonumber\\
&+\sum_{\mu}\left(2L_{\mu}\otimes L_{\mu}^{*}-L_{\mu}^{\dagger}L_{\mu}\otimes\1-\1\otimes (L_{\mu}^{\dagger}L_{\mu})^{*}\right),
\end{align}
where $\1$ denotes the identity operator in the ket or bra spaces.
As a result, the GKSL equation becomes the Schr$\ddot{\rm{o}}$dinger-type equation with ``non-Hermitian Hamiltonian'' $i\bar{\mathcal{L}}$
\begin{align}
\label{matLin}
i\frac{d}{dt}\ket{\rho}=i\bar{\mathcal{L}}\ \ket{\rho}.
\end{align}
This representation is useful for exact diagonalization~\cite{am2015three}, symmetry classification~\cite{Sa, kawabata2023symmetry}, and constructing exact solutions~\cite{haga2023quasiparticles, shibata2019dissipative, shibata2019dissipative2}.

\subsection{Persistent oscillations and Purely imaginary eigenvalues}
\label{sec32}
In a closed system, the Lindbladian reduces to the commutator $\hat{\mathcal{L}}_0[H](\cdot)=-i[H,\cdot]$; all the eigenvalues are purely imaginary.
Dissipators generically shift these eigenvalues in the left half-plane, turning sustained oscillations into a relaxation process that converges to the steady state (Figure~\ref{figeig} (a)). 
Under certain symmetries or conditions, \textit{purely imaginary eigenvalues (PIEs) persist} despite dissipation (Figure~\ref{figeig} (b)). The presence of PIEs is necessary for the emergence of CTCs. See Section~\ref{sec6} for a detailed discussion of CTCs. Here, we present two representative conditions that yield PIEs in finite systems.

First, PIEs arise in the presence of a decoherence-free subspace~\cite{lidar1998decoherence}. By a decoherence-free subspace, we mean a subspace spanned by \textit{dark states} that are invariant under the dissipative part of the dynamics and preserved by the Hamiltonian. Specifically, we assume that there exist two dark states \(\ket{E_n}\) and \(\ket{E_m}\) such that
\begin{align}
\label{dfs}
H\ket{E_\alpha}=E_\alpha\ket{E_\alpha},
\qquad
L_\mu\ket{E_\alpha}=0,
\end{align}
for all Lindblad operators \(L_\mu\) and \(\alpha\in\{n,m\}\). 

Let us consider the coherence
\(\rho_{\rm dark}:=\ket{E_n}\!\bra{E_m}\).
The dissipator vanishes on \(\rho_{\rm dark}\) because all \(L_\mu\ket{E_\alpha}=0\). The action of the generator therefore reduces to the coherent part,
\(\hat{\mathcal{L}}[\rho_{\rm dark}] = -i(E_n-E_m)\,\rho_{\rm dark}\),
so that
\begin{align}
\rho_{\rm dark}(t)
= e^{-i(E_n-E_m)t}\,\rho_{\rm dark}(0),
\end{align}
which oscillates indefinitely at frequency \(\Omega=E_n-E_m\), independently of the dissipative rates. In other words, \(\rho_{\rm dark}\) is an eigenmode of the Lindbladian with a PIE \(\lambda=-i(E_n-E_m)\), protected by the decoherence-free subspace.

Second, PIEs can also arise from a strong dynamical symmetry~\cite{buvca2019non,Booker}. We say that the Lindbladian possesses a strong dynamical symmetry generated by an operator \(A\) if
\begin{align}
\label{sds}
[H, A] = \omega A, \qquad [L_\mu, A] = [L_\mu^\dagger, A] = 0,
\end{align}
with \(\omega\in\mathbb{R}\). The first relation means that \(A\) acts as a ladder operator for the Hamiltonian, shifting energies by \(\omega\), while the second ensures that the dissipative part of the dynamics respects the same symmetry.

Let \(\rho_{ss}\) be a steady state of the Lindbladian, \(\hat{\mathcal{L}}[\rho_{ss}]=0\). Using Eq.~\eqref{sds}, one can verify that the operators
\begin{align}
\rho_{n,m} := (A^\dagger)^n \rho_{ss} A^m
\end{align}
form a ladder of oscillatory eigenmodes of the Lindbladian with eigenvalues
\begin{align}
\lambda_{n,m} = i(n-m)\,\omega,
\end{align}
so that the coherences oscillate at frequencies fixed by the Hamiltonian, \(\Omega_{n,m}=(n-m)\omega\). Importantly, these modes are \emph{not} decoherence-free in general, in the sense that typically \(L_\mu \rho_{n,m} L_\mu^\dagger \neq 0\).
More general conditions for the appearance of PIEs in finite systems, based on dynamical symmetries, are discussed in Ref.~\cite{Buca2022SciPost}.




\subsection{Strong and Weak unitary symmetry in Lindbladians}
\label{sec33}
In open quantum systems, there are two distinct notions of unitary
symmetry, commonly termed \emph{strong} and \emph{weak} symmetries because the Lindbladian \(\hat{\mathcal L}\) acts on density operators rather than on state vectors.
A Lindbladian is said to have a strong symmetry~\cite{buvca2012note} when a
unitary operator $U$ commutes with the Hamiltonian and all Lindblad operators
\begin{align}
\label{strong}
[H, U]=0, \ \ \ [L_{\mu},U]=0 \quad (\forall\,\mu).
\end{align}
Under this condition the Lindbladian commutes with the left‐ and
right‐multiplication superoperators
\(
\hat{\mathcal U_k}(\rho)=U\rho
\)
and
\(
\hat{\mathcal U_b}(\rho)=\rho U^\dagger
\),
\begin{equation}
\label{strong2}
[\hat{\mathcal L},{\hat{\mathcal U}}_k]=
[\hat{\mathcal L},{\hat{\mathcal U}}_b]=0.
\end{equation}
Equivalently, in vectorized notation,
\(
{\bar{\mathcal U}}_k=U\otimes \1,\;
{\bar{\mathcal U}}_b=\1\otimes U^\dagger
\): $[\bar{\mathcal L},{\bar{\mathcal U}}_k]=
[\bar{\mathcal L},\bar{{\mathcal U}}_b]=0.$
Since $U$ commutes with $H$ and all $L_\mu$, the expectation
$\langle U\rangle=\mathrm{Tr}(U\rho)$ is conserved, $\frac{d}{dt}\braket{U}=0$.
In particular, for a continuous one-parameter unitary symmetry $U(\theta)=e^{i\theta O}$, the conservation of $\langle U(\theta)\rangle$ for all $\theta$ implies, by expanding to linear order in $\theta$, that $d\langle O\rangle/dt=0$, so that $O$ plays the role of a conserved Noether charge.
A basic example is the global $U(1)$ phase symmetry of a bosonic mode,  generated by $U(\theta)=\exp(i\theta\,a^\dagger a)$; this symmetry implies conservation of the particle number
$N=\langle a^\dagger a\rangle$.

In contrast, the Lindbladian is said to have a weak symmetry~\cite{buvca2012note} when it commutes with a unitary superoperator $\hat{\mathcal{U}}_{w}=\hat{\mathcal{U}}_{k}\hat{\mathcal{U}}_{b}$ such that
\begin{align}
\label{weak}
[\hat{\mathcal{L}},\hat{\mathcal{U}}_{w}]=0,
\end{align}
which ensures that eigenmodes can be classified into symmetry sectors, but does not imply the existence of conserved quantities. In summary, a strong unitary symmetry yields both a conserved charge and a decomposition of the Lindbladian into symmetry sectors, whereas a weak symmetry only ensures the latter and generally does not imply a conserved observable.




\begin{figure}[t]
   \vspace*{-3.0cm}
     \hspace*{-0.2cm}
\includegraphics[bb=0mm 0mm 90mm 150mm,width=0.52\linewidth]{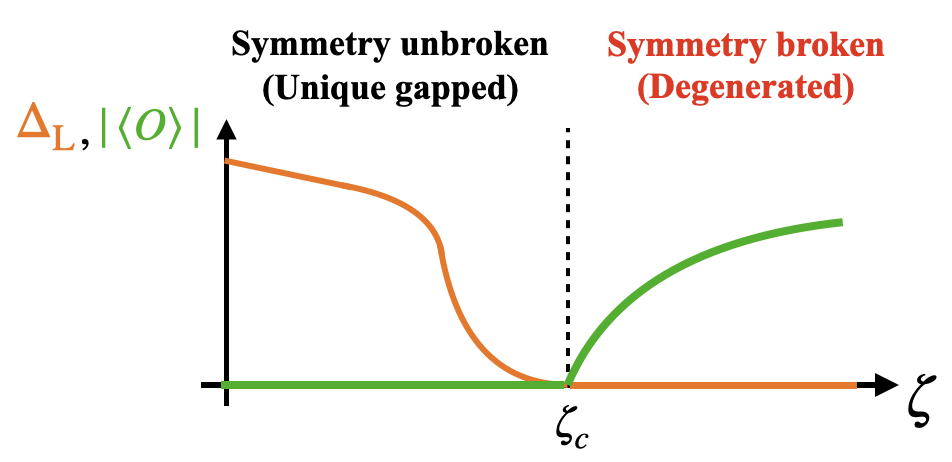}
\caption{\footnotesize\justifying Dissipative continuous phase transition with spontaneous breaking of a weak unitary symmetry: Beyond the critical point, multiple symmetry-broken steady states emerge. This indicates that the Lindbladian gap $\Delta_{\rm L}$ vanishes and an order parameter $|\braket{O}|$ takes a finite value. [Inspired by Refs.\cite{Minganti, fazio2024many}]}
\label{dpt}
\end{figure}

\subsection{Dissipative phase transitions}
\label{sec34}
In nonequilibrium systems, detailed balance is not always satisfied, and no universal thermodynamic potential (such as a free energy) can be defined in general. Consequently, phase transitions cannot be formulated via
the usual non-analyticities of a scalar potential. Instead, following Ref.~\cite{Minganti}, we define DPTs directly from the steady-state expectation values. 

Let $N$ denote the parameter controlling the thermodynamic limit $N\to\infty$
(e.g., the number of lattice sites in a lattice spin system), $O^{(N)}$ an intensive observable satisfying
$\mathrm{Tr}\!\left[ O^{(N)} \rho\right] = \mathcal{O}(1)$ as $N\to\infty$.
Let $\rho_{\mathrm{ss}}(\zeta,N)$ denote the time-independent steady state at finite $N$ controlled by a parameter $\zeta$ (assumed unique).


A DPT of order~$M$ is defined as occurring at $\zeta=\zeta_c$ if
\begin{align}
\label{phasetransition}
\lim_{\zeta\to\zeta_{c}}\left|\frac{\partial^{M}}{\partial\zeta^{M}}\lim_{N\to\infty}\textrm{Tr}[\rho_{ss}(\zeta,N)O_N]\right|=+\infty.
\end{align}
In particular, $M=1$ corresponds to a discontinuous (first-order) DPT, $M=2$ to a continuous (second-order) DPT, mirroring the Ehrenfest classification of equilibrium transitions but without invoking a thermodynamic function.
The singular behavior in Eq.~($\ref{phasetransition}$) is typically
linked to \emph{the closing of Lindbladian gap} ($\ref{Lindbladiangap}$)~\cite{Kessler, Minganti}, that is, the divergence of the relaxation time.


We now focus on a dissipative \textit{continuous} (second-order) phase transition that involves spontaneous breaking of a weak unitary symmetry~\cite{Minganti}. Similarly to conventional quantum phase transitions, this transition is characterized by the emergence of multiple eigenmatrices with $\lambda_{i}=0$, belonging to a different symmetry sector. 

For example, consider a model with discrete $\hat{\mathcal{Z}}_{2}$ symmetry, implemented by the superoperator $\hat{\mathcal{Z}}_{2}[\cdot]=Z_{2}\cdot Z_{2}^{\dagger}$. The superoperator $\hat{\mathcal{Z}}_{2}$ admits two eigenvalues, $\hat{\mathcal{Z}}_{2}\rho_{j}=\pm\rho_{j}$. Given that the steady state is unique in the unbroken phase ($\zeta<\zeta_c$), the following relation holds: $\hat{\mathcal{Z}}_{2}\rho_{ss}=\rho_{ss}$. In the symmetry-broken phase ($\zeta>\zeta_c$), in contrast, two eigenmodes with zero eigenvalue $\rho_{0}$ and $\rho_{1}$ degenerate, each belonging to different symmetry sectors. Here, we choose $\rho_{1}$ to be Hermitian and traceless. Since Tr$[\rho]=1$ for a steady state, symmetry-broken steady states can be expressed as: $\rho^{\pm}=(\rho_{0}\pm\rho_{1})/{\text{Tr}}[\rho_{0}]$ in the thermodynamic limit, with $\hat{\mathcal{Z}}_{2}\rho_{\pm}=\rho_{\mp}$. This indicates that the Lindbladian gap is open (closed) in the symmetry unbroken (broken) phase (Figure $\ref{dpt}$).

Since it provides a clear diagnostic of DPTs (and SSB when present) from the Lindbladian spectrum, without relying on a notion of LRO, this spectral approach is particularly useful in effectively zero-dimensional settings where the LRO in Eq.~\eqref{lro} is not well defined. It is now standard in theory~\cite{cabot2024nonequilibrium,Minganti,minganti2023dissipative,minganti2021liouvillian,Pausch2024QuditsDPT,Kessler,Casteels2017LiouvillianGapKerr,Vicentini2018CriticalSlowingBoseHubbard,Hwang2018OpenQRM_DPT,Hoening2012FermionicLatticeCriticalExponents} and has been used in experiments~\cite{Fink2018DPTPhotonCorrelations,Chen2023NatCommun,Beaulieu2025NatCommun,Benary2022NJP,Brookes2021SciAdv}, with recent extensions beyond the Markovian setting~\cite{Debecker2024NonMarkovianSpectral}. Appendix~\ref{phasebifur} presents an illustrative example.

\begin{table*}[htbp]
   \vspace*{0cm}
     \hspace*{0cm}
\scalebox{0.74}{
\begin{tabular}{|c||c|c|c|}  \hline
 Generator & Non-Hermitian Hamiltonian & \multicolumn{2}{|c|}{Lindbladian} \\ \hline \hline
  Symmetry & PT symmetry ($\ref{PT}$)&  L-$\mathcal{PT}$ symmetry ($\ref{HuberPT}$)  & $\mathcal{PT}$ symmetry for shifted Lindbladians ($\ref{ProzenPT2}$) \\ \hline
  Eigenvalues & Real$\to$ Complex conjugate pair  & \begin{tabular}{c}Existence of purely imaginary \\  eigenvalues (PIEs) $\to$ No PIEs \end{tabular} &\begin{tabular}{c} All the eigenvalues are on crosshairs \\ $\to$Some eigenvalues are not \end{tabular}   \\ \hline
 Eigenmodes & \begin{tabular}{c} PT symmetry breaking \\ for eigenstates \end{tabular}& \begin{tabular}{c} PT symmetry breaking  \\of  steady-state solutions\end{tabular} &  \begin{tabular}{c} PT symmetry breaking for\\  eigenmodes with decay\end{tabular} \\ \hline
 Dynamics & \begin{tabular}{c} Oscillating behavior$\to$ \\ Exponential decay/diverge  \end{tabular} &\begin{tabular}{c} Oscillating behavior$\to$ \\ Exponential decay \end{tabular} & No significant physical changes \\ \hline
   System size & Finite &Thermodynamic limit &  Finite \\ \hline
   \begin{tabular}{c} Transition \\ point  \end{tabular} & EP & CEP & EP \\ \hline
  \end{tabular}}
   \caption{\footnotesize\justifying Comparison of spectral PT transition in non-Hermitian Hamiltonians and shifted Lindbladians, and L-$\mathcal{PT}$ phase transitions. \YNblueedit{Here, ``System size'' indicates whether the thermodynamic limit is required for the transition itself to occur. L-$\mathcal{PT}$ phase transitions are dissipative phase transitions of steady states and therefore require the thermodynamic limit, whereas the PT transitions of non-Hermitian Hamiltonians and shifted Lindbladians are spectral transitions that can occur already in finite-dimensional systems.}}
   \label{HLcomp}
\end{table*}





\section{$\mathcal{PT}$ Symmetry in Lindbladians}
\renewcommand{\theequation}{4.\arabic{equation} }
\setcounter{equation}{0}
\label{sec4}
The previous section provided an explanation of unitary symmetries in Lindbladians. We now turn to the main topic of this review: \textit{L‑${\mathcal{PT}}$ symmetry}. 
Although unitary symmetries are well established, studies on ``$\mathcal{PT}$ symmetry'' in Lindbladians and transitions accompanied by its breaking are still in their infancy. Consequently, several criteria are employed, and the term is often used in multiple inequivalent senses.
In Section~\ref{sec41}, we explain the physical interpretation of the L‑${\mathcal{PT}}$ symmetry ($\ref{HuberPT2}$) adopted in this review. In Section~\ref{sec42}, we introduce other criteria and compare them with the L‑$\mathcal{PT}$ symmetry ($\ref{HuberPT2}$).

\subsection{Lindbladian $\mathcal{PT}$ symmetry}
\label{sec41}
In Section~\ref{sec13}, we introduce the L-$\mathcal{PT}$ symmetry in Eq.~($\ref{HuberPT2}$); for convenience, we rewrite it here as
\begin{align}
\label{HuberPT}
\hat{\mathcal{L}}[\mathbb{PT}(H); \mathbb{PT}(L_\mu),\mu=1,2,...]=\hat{\mathcal{L}}[H;L_\mu,\mu=1,2,...],
\end{align}
with $\mathbb{PT}(O)=PTO^{\dagger}(PT)^{-1}$~\cite{Huber1, Nakanishi3}. 
\footnote{Here we briefly comment on ``strong'' and ``weak'' forms of $\mathcal{PT}$ symmetries. In analogy with the unitary case, we refer to Eq.~\eqref{HuberPT} as a ``weak'' \(\mathcal{PT}\) symmetry. By contrast, we call the symmetry a ``strong'' \(\mathcal{PT}\) symmetry when \(\mathbb{PT}(H)=H\) and \(\mathbb{PT}(L_\mu)=L_\mu\).
However, unlike the unitary case, this distinction does not seem to lead to essential differences (e.g., the presence or absence of conserved quantities). Therefore, as far as is currently known, there is little reason to emphasize the distinction between ``strong'' and ``weak'' here.
}
Here, we assume that the parity operator $P$ and time-reversal operator $T$ satisfy Eq.~($\ref{opPT}$). 
If Lindbladian satisfies the L-$\mathcal{PT}$ symmetry condition~\eqref{HuberPT}, 
Lindbladian can be written in the symmetric form as
\begin{align}
\label{PTsymm}
\hat{\mathcal{L}}\rho=\hat{\mathcal{L}}_0[H_{\mathrm{PT}}]\rho
+\frac{1}{2}\sum_{\mu}(\mathcal{\hat{D}}[L_{\mu}]+\mathcal{\hat{D}}[\mathbb{PT}(L_\mu)])\rho,
\end{align}
where we define $H_{\mathrm{PT}}:=\tfrac12(H+PTH(PT)^{-1})$. This form will be used in Sec.~\ref{sec71}.
It is worth noting why the time–reversal operation in the L-$\mathcal{PT}$ symmetry is implemented by combining the \emph{Hermitian adjoint} and the complex conjugation \(L_\mu\!\to KL_\mu^{\dagger}K\) rather than simple complex conjugation. In non-Hermitian systems, the action \(i\to -i\) which indicates time-reversal flips the sign of every anti-Hermitian term and thus exchanges amplification (gain) and decay (loss).  
By contrast, in the GKSL formalism, dissipation is encoded in dissipative terms $\hat{\mathcal{D}}[L_\mu]$; the roles of gain and loss are interchanged only when \(L_\mu\) itself is replaced by its adjoint \(L_\mu^{\dagger}\), not by its complex conjugate \(L_\mu^*\).
\YNedit{For example, for a bosonic mode, Hermitian conjugation maps a loss jump \(L_\mu\propto a\) (e.g., spontaneous emission or particle loss) to the corresponding gain jump \(L_\mu^\dagger\propto a^\dagger\) (e.g., incoherent pumping).}
Therefore, we define the time-reversal action on jump operators as the combined operation of the Hermitian adjoint and complex conjugation.

\YNedit{We emphasize that this gain--loss interpretation should not be understood as requiring every L-$\mathcal{PT}$-symmetric Lindbladian to contain two separate dissipative channels with fine-tuned dissipation rates. The actual requirement is that the jump-operator set be invariant under}
\begin{equation}
\label{PTjumpmap}
\YNedit{L_\mu \mapsto PTL_\mu^\dagger(PT)^{-1},}
\end{equation}
\YNedit{up to relabeling and irrelevant phase factors.
Thus, there are two common possibilities: a jump channel may be mapped to a distinct \(\mathcal{PT}\)-partner channel, or the same jump channel may be returned to itself, up to an irrelevant phase of the corresponding Lindblad operator. This latter situation applies to the single-collective-spin class central to this review. (For example, under a suitable \(PT\) operation, the collective spin-lowering channel \(L\propto S_-\) can itself satisfy Eq.~\eqref{HuberPT}; in the Schwinger-boson representation discussed in Sec.~\ref{sec57}, \(S_-\propto ab^\dagger\) removes a boson from one mode and adds it to the other, so this single collective channel already represents balanced gain-and-loss dissipation.)}

This microscopic jump-operator condition~\eqref{HuberPT} should be distinguished from a symmetry condition imposed directly on the Lindbladian superoperator. In general, the L-\(\mathcal{PT}\) condition is not equivalent to a simple commutation or anticommutation relation between the Lindbladian and a superoperator, and therefore does not by itself guarantee that the Lindbladian eigenmodes are PT-symmetric.
Nevertheless, the constraints imposed by the L-$\mathcal{PT}$ symmetry ($\ref{HuberPT}$) can give rise to non-trivial DPTs. (We call such DPTs L-$\mathcal{PT}$ phase transitions.)
The similarities and differences with non-Hermitian PT transitions are summarized in Table $\ref{HLcomp}$. We explore L-$\mathcal{PT}$ phase transitions in detail in Sections~\ref{sec5}, ~\ref{sec6}, and~\ref{sec7}.

\subsection{Other criteria of $\mathcal{PT}$ symmetries in Lindbladians }
\label{sec42}
The term ``$\mathcal{PT}$ symmetry'' has been used in several inequivalent ways for Lindbladians. We briefly summarize those criteria (without adopting them in this review) to contrast them with the L-$\mathcal{PT}$ symmetry ($\ref{HuberPT}$).
\subsubsection{$\mathcal{PT}$ symmetry for a shifted Lindbladian }
The shifted Lindbladian is defined by subtracting a scalar multiple of the identity superoperator:
\begin{align}
\hat{\mathcal{L}}^{\prime}:=\hat{\mathcal{L}}-\gamma \hat{1},
\end{align}
where $\hat{1}$ is the identity superoperator and the scalar shift $\gamma:=\textrm{Tr}\hat{\mathcal{L}}/\textrm{Tr}\hat{1}$ (with Tr taken in Liouville space) ensures that $\hat{\mathcal{L}}^{\prime}$ is traceless.
The shifted Lindbladian is said to be PT-symmetric if it satisfies the condition
$\hat{\mathcal{P}}(\hat{\mathcal{L}}^{\prime})^{\dagger}\hat{\mathcal{P}}^{-1}=-\hat{\mathcal{L}}^{\prime}$, that is,
\begin{align}
\label{ProzenPT2}
\hat{\mathcal{P}}(i\hat{\mathcal{L}}^{\prime})^{\dagger}\hat{\mathcal{P}}^{-1}=i\hat{\mathcal{L}}^{\prime},
\end{align}
where $\hat{\mathcal{P}}$ is a parity superoperator and the superscript $\dagger$ denotes the adjoint generator defined in Eq.~\eqref{eq:adjoint}~\cite{Prosen1}. \footnote{Because matrix $i\bar{\mathcal{L}}$ in Eq.~\eqref{matLin} plays the role of a non-Hermitian Hamiltonian, it is natural to define PT symmetry of the shifted Lindbladian $\hat{\mathcal{L}}^{\prime}$ through the commutation relation
\begin{align}
\label{ProzenPT}
[i\bar{\mathcal{L}}^{\prime},\bar{\mathcal{P}}\bar{\mathcal{T}}]=0,
\end{align}
where $\bar{\mathcal{P}}\bar{\mathcal{T}}$ is a PT operator acting on the doubled Hilbert space.
In a finite Hilbert space, every PT-symmetric Hamiltonian is necessarily pseudo-Hermitian~\cite{MostafazadehA1, MostafazadehA2, MostafazadehA3, zhang2020pt}. Consequently, once ($\ref{ProzenPT}$) holds, the shifted Lindbladian $i\bar{\mathcal{L}}^{\prime}$ satisfies the pseudo-Hermiticity relation ($\ref{ProzenPT2}$) in the doubled Hilbert space. \YNblueedit{Via the natural (vectorized) representation introduced in Eq.~\eqref{lindmat}, this is equivalent to the shifted Lindbladian superoperator $\hat{\mathcal{L}}^{\prime}$ satisfying Eq.~\eqref{ProzenPT2}.} Here, the parity superoperator $\hat{\mathcal{P}}$ in Eq.~\eqref{ProzenPT2} is not necessarily identical to the parity operator $\bar{\mathcal{P}}$ in Eq.~\eqref{ProzenPT}.
}

When the shifted Lindbladian with a non-degenerate Hamiltonian satisfies condition ($\ref{ProzenPT2}$), a spectral transition occurs~\cite{Prosen1}: below threshold all the eigenvalues are on the crosshairs with lines of symmetry $l_{V}=-\gamma +i\mathbb{R} \ $ and $l_{h}=\mathbb{R}$, while above threshold some move off-axis.
However, this transition is not a DPT, as the steady-state property does not change. An illustrative example is provided in Appendix~\ref{appprosenpt}.

\subsubsection{Modular conjugation and anti-$\mathcal{PT}$ symmetry }
Lindbladian \(\bar{\mathcal{L}}\), as represented by the doubled Hilbert space in Eq.~($\ref{lindmat}$), necessarily possesses an anti-unitary symmetry known as \textit{modular conjugation symmetry}~\cite{Sa,kawabata2023symmetry} in this representation. This symmetry is implemented by an anti-unitary operator \(J\) acting on the double Hilbert space \(\mathcal{H} \otimes \mathcal{H}^*\), defined as
\begin{align}
J(A \otimes B)J^{-1} = B \otimes A, \qquad J z J^{-1} = z^*,
\end{align}
for any operators $A, B$ acting on the ket and bra spaces, and any complex scalar $z$.
Here, \(J\) exchanges the ket and bra spaces of the vectorized density operator and performs complex conjugation.
One finds that the Lindbladian anti-commutes with the anti-unitary operator $J$,
\begin{align}
\label{ms}
\{i\bar{\mathcal{L}},J\}= 0.
\end{align}
The modular conjugation symmetry can be interpreted as an \textit{anti-$\mathcal{PT}$ symmetry} in Lindbladians since the matrix $i\bar{\mathcal{L}}$ corresponds to the non-Hermitian Hamiltonian. This guarantees that the Lindbladian spectrum is symmetric under complex conjugation, $\{\lambda,\lambda^*\}$.

In addition to the intrinsic modular conjugation symmetry $J$, one can also impose an anti-$\mathcal{PT}$ symmetry defined by
\begin{align}
\label{antilpt}
\{i\hat{\mathcal{L}},\hat{\mathcal{P}}\hat{\mathcal{T}}\}=0.
\end{align}
(This condition ($\ref{antilpt}$) is termed ``$\mathcal{PT}$ symmetry'' in the studies of non-reciprocal phase transitions, in contrast to the criterion adopted in our framework.
\footnote{
If the Lindbladian satisfies the symmetry ($\ref{antilpt}$), its corresponding mean-field equations typically inherit the n-anti-PT symmetry~\eqref{nonlinearantiPT}.
In the literature on non-reciprocal phase transitions,
the same conditions ($\ref{antilpt}$) and ($\ref{nonlinearantiPT}$) are instead referred to as ``$\mathcal{PT}$ (PT) symmetry''~\cite{Fruchart, nadolny2025nonreciprocal}. This difference arises due to the property that the order parameter dynamics of our open quantum systems are described by a form similar to a nonlinear Schr$\ddot{\rm{o}}$dinger equation, $i\partial_t{\bf{q}}={\bf{f}}({\bf{q}})$, while those considered in active matter systems are overdamped, $\partial_t{\bf{q}}={\bf{g}}({\bf{q}})$, where the factor ``$i$'' in front of the time-derivative on the left-hand side is missing compared to the open quantum system analog.})
Such anti-$\mathcal{PT}$ symmetries ($\ref{antilpt}$) indicate that the mean-field equation has an n-anti-PT symmetry~\eqref{nonlinearantiPT} and its symmetry characterizes a non-reciprocal phase transition; however, the resulting physical consequences are qualitatively different from those of L-$\mathcal{PT}$ phase transitions~\cite{nadolny2025nonreciprocal}. The difference between non-reciprocal phase transitions and L-$\mathcal{PT}$ phase transitions is discussed in detail in Sect.~\ref{sec54}.

\section{Mean-field theory of Lindbladian $\mathcal{PT}$ phase transitions}
\renewcommand{\theequation}{5.\arabic{equation} }
\setcounter{equation}{0}
\label{sec5}
We begin with a mean-field analysis to explore the characteristics of L-$\mathcal{PT}$ phase transitions, initially focusing on collective spin systems where the mean-field approximation is valid in the thermodynamic limit. We then extend this framework to systems with long-range dissipation or spatially extended bipartite bosonic systems that exhibit strong $U(1)$ symmetry. Along the way, we uncover how n-PT symmetry and its breaking can trigger persistent oscillations, CEPs, and links to non-reciprocal phase transitions.

In Section~\ref{sec51}, we introduce collective spin models and present their paradigmatic example, the driven Dicke model (DDM), which serves as a toy model for exploring the role of L-$\mathcal{PT}$ symmetry and L-$\mathcal{PT}$ phase transitions. In Section~\ref{sec52}, we derive the mean-field equation for single-collective spin models and demonstrate that they exhibit n-PT symmetry. In Section~\ref{sec53}, we perform a linear stability analysis and reveal that the L-$\mathcal{PT}$ symmetry can generically lead to persistent oscillations in single-collective spin systems. Furthermore, we show that a DPT occurs with spontaneous n-PT symmetry breaking, and the transition point is CEP. In addition, we estimate the critical exponents characterized by the order parameter and the relaxation time. In Section~\ref{sec54}, we explore the connection between L-$\mathcal{PT}$ phase transitions and non-reciprocal phase transitions. 
In Sections~\ref{sec55} and~\ref{sec56}, we analyze specific examples, the DDM and the dissipative Lipkin-Meshkov-Glick (LMG) model, respectively. Finally, in Section~\ref{sec57}, we generalize our framework to systems with long-range dissipation or spatially extended bipartite bosonic systems with a strong $U(1)$ symmetry. 

\subsection{Collective spin system and Driven Dicke model}
\label{sec51}
Our exploration starts with collective spin models (fully-connected two-level systems), which offer a clean window to the L-$\mathcal{PT}$ phase transitions. Accordingly, we begin with an overview of collective spin systems and their prototypical example, the DDM. 

\subsubsection{Collective spin system}
\label{sec511}
Collective spin systems, where a large ensemble of two-level systems behaves as a single macroscopic spin, serve as a versatile and powerful platform in various branches of physics. These systems offer not only an analytical advantage but also access to rich many-body phenomena that are otherwise difficult to study in conventional lattice-based models.

In quantum optics, collective-spin models provide a compact yet faithful description of light–matter interactions.
The archetype is the Dicke model~\cite{dicke1954coherence, hepp1973equilibrium, kirton2019introduction}, which describes $N$ identical two-level atoms that are uniformly coupled to a single electromagnetic mode.
In the small-sample limit and within the fully symmetric subspace, it predicts Dicke superradiance, where the peak emission rate scales as $N^2$ instead of $N$.



In atomic, molecular, and optical physics 
ultracold atomic gases such as spinor Bose–Einstein condensates offer a highly controllable platform for collective spin dynamics, exhibiting phenomena from long-lived spin-mixing oscillations~\cite{chang2005coherent} to strong spin and spin-nematic squeezing beyond the standard quantum limit~\cite{mao2023quantum}. Their tunability and low disorder make them a clean testbed for the nonequilibrium many-body dynamics of collective spins.

In condensed-matter and nuclear physics, collective-spin models are key tools for studying magnetism and quantum criticality. A canonical example is the Lipkin–Meshkov–Glick (LMG) model~\cite{lipkin1965validity}, an all-to-all–coupled spin ensemble in a transverse field that undergoes a quantum phase transition in the thermodynamic limit~\cite{dusuel2005continuous, castanos2006classical, leyvraz2005large, dusuel2004finite, ma2009fisher} and now serves as a standard prototype for collective-spin physics. 



A key advantage of collective spin models is the drastic reduction of Hilbert-space dimension due to permutational symmetry. 
While a generic system of $N$ two-level particles lives in a Hilbert space of dimension $2^N$, restricting to the fully symmetric subspace yields an effective spin-$S$ description with $S = N/2$ and dimension $2S+1$. 
This reduced description still captures essential many-body effects such as quantum phase transitions and nonequilibrium steady states, and is widely used in theoretical and numerical studies of open quantum systems~\cite{ribeiro2008exact, ribeiro2019integrable, huber2021phase}.
For convenience, the basic properties of the collective spin operators are summarized in Appendix~\ref{colmath}.

\subsubsection{Driven Dicke model: master equation and experimental realizations}
Collective spin systems also provide fertile ground for exploring nonequilibrium quantum many-body physics. A representative example is the driven Dicke model (DDM), which describes a coherently driven ensemble of two-level atoms with collective dissipation~\cite{Carmichael,Walls,Puri}.

\begin{figure*}[t]
  \centering
\includegraphics[width=\textwidth]{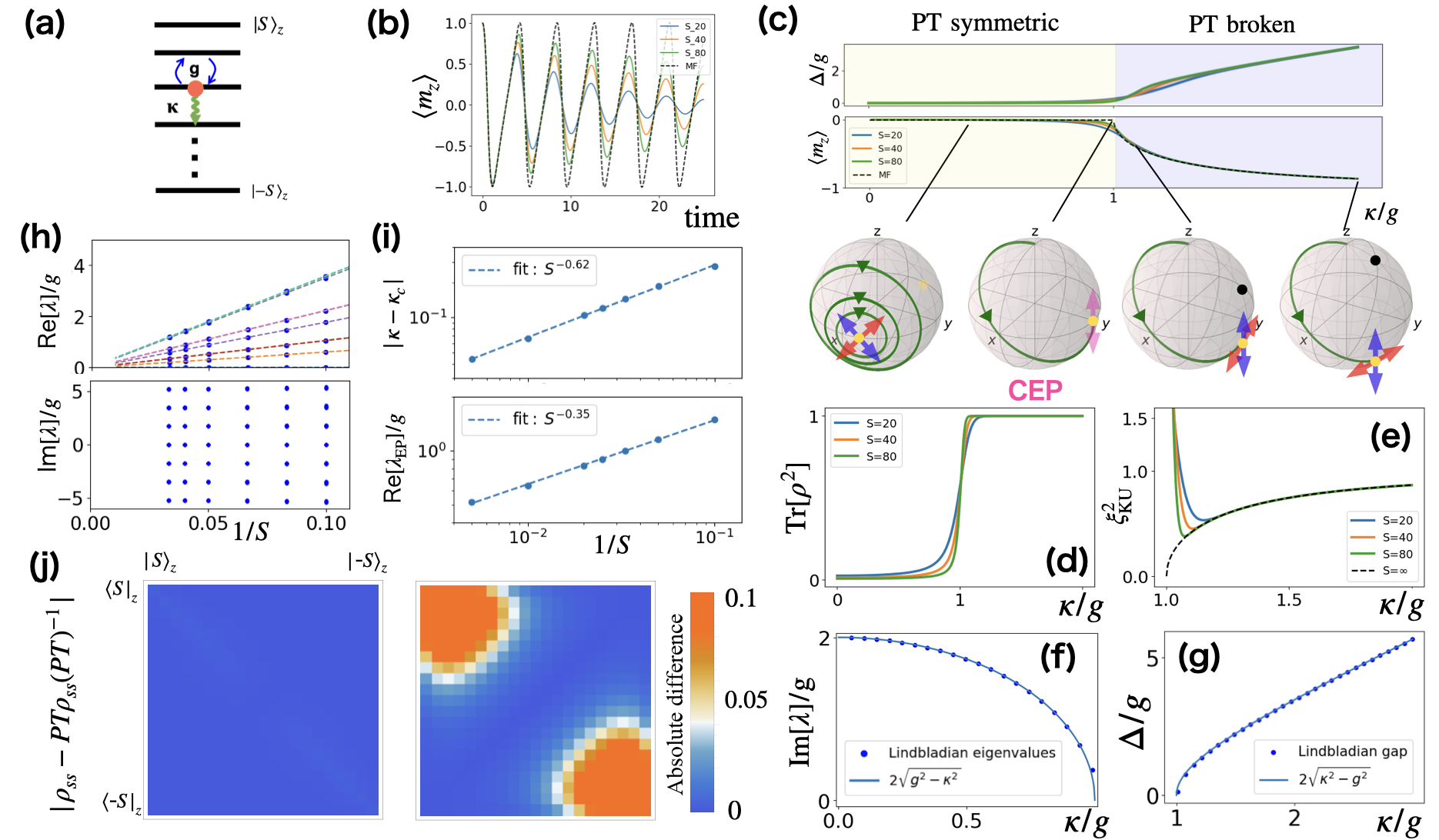}
\caption{\footnotesize\justifying Numerical calculation for the DDM. (a) Illustration of the DDM. (b) The magnetization dynamics for $S=20,40,80$ and mean-field analysis ($S=\infty$). (c) Top: The Lindbladian gap in the TISS. Middle: The normalized magnetization $\braket{m_{z}}$ in the TISS. Viewing $\braket{m_{z}}$ as the order parameter, the Lindbladian gap closes in the disordered phase ($\braket{m_{z}}=0$), while it remains open in the ordered phase ($\braket{m_{z}}\neq0$). This behavior is different from the conventional spectral theory for DPTs with unitary symmetry breaking. Bottom: Mean-field dynamics (green arrow), stable (orange), and unstable (black) fixed points, and the representative collective excitation modes (blue and red or pink arrows). The transition point is a CEP. (d) Purity and (e) Kitagawa-Ueda spin squeezing parameter $\xi_{\rm KU}$ in the TISS. (f) The non-zero imaginary part with the maximum real part of the Lindbladian eigenvalues and (g) the Lindbladian gap for $S=200$ (blue dot) and the mean-field solution (light blue line) with $g=1$. (h) Top: the $S$-dependence of the maximum real parts. Bottom: imaginary parts of eigenvalues with maximum real parts. (i) The $S$-dependence of (top)  $|\kappa-\kappa_c|$ and (bottom) real part for the LEP with maximum real part. The dashed line shows the possible $S^{-0.62}$, $S^{-0.35}$ scaling, respectively. (j)  $|\rho_{ss}-PT\rho_{ss} (PT)^{-1}|$ for (left) $\kappa/g=0.5$ (right) $\kappa/g=1.5$ and $S=10$ where $|\rho|$ implies the matrix taking the absolute value for each element of the density matrix $\rho$. \YNblueedit{The color bar represents the corresponding absolute difference.} Here, elements are computed on the $S_{z}$ basis. 
 }
    \label{DDMfig}
\end{figure*}

The DDM is described by the following GKSL master equation $\text{\cite{drummond1978volterra, Walls, Puri, Carmichael, Lawande, gross1982superradiance}}$:
\begin{align}
\label{DDM}
\frac{d}{dt}\rho=-2ig[S_{x},\rho]+\frac{\kappa}{S}\hat{\mathcal{D}}[S_{-}]\rho,
\end{align}
where $S$ represents the total spin, which can be an integer or half-integer. Here, $g$ and $\kappa$ are the strengths of interaction and dissipation, respectively (Fig.~$\ref{DDMfig}$ (a)). Here,
\begin{align}
S_{\alpha}=\frac{1}{2}\sum_{i=1}^N\sigma_{i}^{\alpha},\quad (\alpha=x,y,z)
\end{align} 
is a collective spin operator along $\alpha$ with the Pauli matrices $\sigma^{\alpha}$, and $S_{\pm}:=S_{x}\pm iS_{y}$ is a ladder operator. 
An atom-only master equation of the form in Eq.~\eqref{DDM} arises in several microscopic settings, for example, for a collectively coupled atomic ensemble in a single-mode cavity in the bad-cavity limit, or for atoms radiating collectively into free space.
As a concrete example, Appendix~\ref{deriddm} outlines a free-space derivation starting from $N$ identical two-level atoms confined to a subwavelength volume and coupled to the quantized electromagnetic field.

Experimentally, a closely related scenario has been realized in a pencil-shaped cold-atom ensemble in free space: Ferioli \textit{et al.} implemented an effective DDM using a cloud of laser-cooled atoms optically driven along its main axis, and showed that, by introducing an effective atom number $N_{\mathrm{eff}}$ that accounts for finite-size, the DDM predictions for the magnetization and the scaling of the superradiant photon emission along the main axis of the cloud quantitatively reproduce the observed behavior~\cite{Ferioli}. 
However, subsequent work has argued that the apparent transition is more appropriately understood as driven-dissipative spatial phase separation rather than as a genuine phase transition of a spatially uniform collective spin~\cite{goncalves2024driven}.


Related driven–dissipative critical behavior has also been observed in a cavity-QED implementation of the cooperative-resonance-fluorescence (CRF) model~\cite{Song2025SciAdv}. In an atom-only description with purely collective decay, the CRF model is reduced to the DDM.
In the experimentally relevant cavity-QED regime, however, the cavity mode remains a dynamical degree of freedom and cannot be neglected. By probing short evolution times, Ref.~\cite{Song2025SciAdv} reports an observation of a continuous superradiant phase transition; at longer times, single-particle spontaneous emission slowly steers the system toward a qualitatively different steady state, thereby transforming the transition into a discontinuous one.

\subsubsection{Driven Dicke model: phase transition and physical properties}
The DDM is known to exhibit a continuous time crystal (CTC) in the weakly dissipative regime.
Fig.~$\ref{DDMfig}$ (b) shows that the normalized magnetization oscillates, with the relaxation time increasing as the total spin $S$ increases, while the oscillation frequency remains almost invariant. In the mean-field analysis, which corresponds to the large total spin limit $S\to\infty$, it periodically oscillates without decay. This behavior indicates that continuous time-translation symmetry is broken into a discrete one, that is, the realization of a CTC. 
\footnote{\footnotesize The CTC in the DDM was originally introduced in a boundary--bulk setting, where the bulk is treated as an effective environment and continuous time-translation symmetry breaking manifests in a macroscopic boundary subsystem~\cite{Iemini}. Accordingly, they were termed \emph{boundary time crystals} (BTCs). However, in some parts of the literature, the term ``BTC'' has also been used more broadly, as a shorthand for time crystals in open quantum systems whose oscillations become strictly persistent only in the thermodynamic limit, even without an explicit boundary--bulk construction. 
}

The important feature of the DDM is the non-commutativity between the long time limit and the large spin limit, $\lim_{S\to\infty}\lim_{t\to\infty}\neq\lim_{t\to\infty}\lim_{S\to\infty}$. The former case concerns \textit{a time-independent steady state (TISS)}, $\rho_{ss}(S)=\lim_{t\to\infty}\rho(t, S)$. By contrast, the latter case shifts attention to time-dependent behavior, with particular interest in CTCs.

Moreover, the TISS undergoes a continuous phase transition that does not break any unitary symmetry (middle of Fig.~$\ref{DDMfig}$ (c)) $\text{\cite{Walls}}$. For this transition, in the thermodynamic limit, the normalized magnetization $\braket{m_z}:=\braket{S_z}/S$ vanishes for $\kappa<g$ while it remains finite for $\kappa>g$, so it can be regarded as the order parameter. For conventional continuous phase transitions, multiple steady states become degenerate in a symmetry broken phase (see Fig.~$\ref{dpt}$). In contrast, for the DDM, the Lindbladian gap is open in the ordered phase (top of Fig.~$\ref{DDMfig}$ (c)). These results show that this DPT is outside the scope of conventional spectral theory discussed in Section~\ref{sec33}. 

Furthermore, both the statistical and quantum properties change abruptly at the transition point in the TISS~\cite{Hannukainen, Cabot, Montenegro, AC, cabot2025quantum}.
Fig.~\ref{DDMfig}~(d) plots the purity, ${\rm Pur}[\rho]:=\operatorname{Tr}[\rho^{2}]$, which measures how mixed the density matrix is. In the disordered phase with $\braket{m_z}=0$, the purity is nearly zero, signifying a highly mixed state, whereas in the ordered phase with $\braket{m_z}\neq0$, it approaches unity, indicating a pure state.
Fig.~\ref{DDMfig}~(e) shows the Kitagawa–Ueda spin-squeezing parameter $\xi^{2}_{\rm KU}$, which represents the reduction of fluctuations of a collective-spin component orthogonal to the mean-spin direction~\cite{ma2011quantum}. This shows that it stays below unity near the transition within the ordered phase, indicating spin squeezing. For \(|\braket{{\bf S}}|=\sqrt{\braket{S_x}^2+\braket{S_y}^2+\braket{S_z}^2}\simeq N/2\), it can also be regarded as an entanglement witness parameter~\cite{ma2011quantum}, so this result reveals the build-up of quantum entanglement. 
Section~\ref{sec7} will discuss the statistical and quantum properties in detail.

Furthermore, the DDM has been studied from multiple perspectives: quantum thermodynamics~\cite{Carollo2024}; quantum trajectory analysis~\cite{Cabot2023}; measurement-induced phase transition~\cite{KrishnaPRL2023_MICTC}; resource-theory via non-stabilizerness~\cite{Passarelli2025MagicBTC}; and AC sensors~\cite{Gribben2025BoundaryTimeCrystalsAC}.

In the following, we use mean-field theory and linear stability analysis to reveal that persistent periodic dynamics and continuous phase transitions without symmetry breaking of unitary symmetries can be understood from the viewpoint of L-$\mathcal{PT}$ symmetry. 



\subsection{Nonlinear PT symmetry for mean-field equation}
\label{sec52}
We now turn to the center of our analysis, the mean-field theory of collective-spin systems with L-$\mathcal{PT}$ symmetry.
The commutator of the normalized collective spin operators $m_{\alpha} = S_{\alpha}/S$ ($\alpha = x, y, z$) vanishes in the thermodynamic limit, i.e., 
\begin{align}
\braket{[m_{\alpha}, m_{\beta}] }= i \epsilon^{\alpha\beta\gamma} \braket{m_{\gamma}}/S \to 0
\end{align}
as $S \to \infty$, where $\epsilon^{\alpha\beta\gamma}$ is the totally antisymmetric Levi–Civita symbol with $\epsilon^{xyz}=1$. Therefore, one can expect that the mean-field approximation is valid at large $S$,
\begin{align}
\braket{m_{\alpha} m_{\beta}} \approx \braket{m_{\alpha}} \braket{m_{\beta}},
\end{align}
and the dynamics can be effectively described by a closed set of equations for the expectation values of collective observables. Indeed, in the large $S$ limit the mean-field approximation becomes exact for initial states whose cumulants of order $n\geq2$ vanish~\cite{da2023sufficient, carollo2021exactness}. 

The mean-field equation of single-collective spin is written in the form
\begin{align}
\label{csmean}
\partial_t{\bf{M}}={\bf{g}}({\bf{M}}),
\end{align}
where a function ${\bf{g}}:=(g_{x},g_{y},g_{z})^{T}\in\mathbb{R}^{3}$ and the magnetization vector ${\bf{M}}:=(X, Y, Z)\in\mathbb{R}^{3}$. Here, we define $X:=\lim_{S\to\infty}\braket{m_{x}}$, $Y:=\lim_{S\to\infty}\braket{m_{y}}$, and $Z:=\lim_{S\to\infty}\braket{m_{z}}$. In our notation, the nonlinear symmetry is defined through the nonlinear function
\begin{align}
\label{f=ig}
{\bf{f}}({\bf{M}}):=i{\bf{g}}({\bf{M}}).
\end{align}

To take the well-defined thermodynamic limit $S\to\infty$, we assume that the Hamiltonian and Lindblad operators are written as 
\begin{align}
\label{scaling}
    H=:Sh,\quad L_\mu=:\sqrt{S}l_\mu,
\end{align}
where $h$ and $l_\mu$ are finite sums of operator monomials built from an $S$-independent number of normalized magnetization operators $\{m_\alpha\}$. 
We begin by establishing the following lemma, which will serve as a key ingredient of our framework. \\ 

\noindent \textit{{\bf Lemma}\ \ Consider a dissipative single–collective spin model satisfying the scaling~\eqref{scaling} and having an
L-$\mathcal{PT}$ symmetry [Eq.~\eqref{HuberPT}]. Then the following relation holds:
\begin{align}
\label{lemma}
\lim_{S\to\infty}\braket{[i\hat{\mathcal L}^\dagger,\hat{\mathcal P}\hat{\mathcal T}]\,m_\alpha}=0,
\end{align}
where $\hat{\mathcal{P}}[\cdot]:=P\cdot P^{-1}$ and $\hat{\mathcal{T}}[\cdot]:=T\cdot T^{-1}$ are parity and time-reversal superoperators satisfying $[\hat{\mathcal{P}},\hat{\mathcal{T}}]=0$, and $\hat{\mathcal{P}}^2=\hat{\mathcal{T}}^2=1$.}

\begin{proof}
Using the explicit form of the adjoint generator [Eq.~\eqref{eq:adjoint}], we find 
\begin{align}
\label{ilpt2}
i\hat{\mathcal{L}}^\dagger \hat{\mathcal{P}}\hat{\mathcal{T}}m_\alpha=& -[H,\hat{\mathcal P}\hat{\mathcal T}m_\alpha]\nonumber\\
&+i\sum_{\mu}([L^\dagger_\mu,\hat{\mathcal{P}}\hat{\mathcal{T}}m_\alpha]L_\mu-L^\dagger_\mu[L_\mu,\hat{\mathcal{P}}\hat{\mathcal{T}}m_\alpha]\bigr).
\end{align}
On the other hand, we find 
\begin{align}
\label{ptil2}
&\hat{\mathcal{P}}\hat{\mathcal{T}}(i\hat{\mathcal{L}}^\dagger) m_\alpha
= -[\mathbb{PT}(H),\hat{\mathcal P}\hat{\mathcal T}m_\alpha]\nonumber\\
&\quad
-i\sum_{\mu}\Bigl(
[\mathbb{PT}(L_\mu),\hat{\mathcal{P}}\hat{\mathcal{T}}m_\alpha]\mathbb{PT}(L_\mu^\dagger)\nonumber\\
&\qquad\qquad
-\mathbb{PT}(L_\mu)[\mathbb{PT}(L_\mu^\dagger),\hat{\mathcal{P}}\hat{\mathcal{T}}m_\alpha]\Bigr),
\end{align}
where we use $\hat{\mathcal{P}}\hat{\mathcal{T}}(l_\mu)=\mathbb{PT}(l_\mu^\dagger)$.

The L-$\mathcal{PT}$ symmetry [Eq.~\eqref{HuberPT}] implies that replacing the Hamiltonian and Lindblad operators to their $\mathbb{PT}$-transformed ones
\begin{align}
H\to\mathbb{PT}(H),\quad \{L_\mu\}\to\{\mathbb{PT}(L_\mu)\},\quad (\mu=1,2,...)
\end{align}
leaves the system invariant. Therefore, for L-$\mathcal{PT}$ symmetric systems, Eq.~\eqref{ptil2} can be simplified to
\begin{align}
\label{ptil3}
\hat{\mathcal{P}}\hat{\mathcal{T}}(i\hat{\mathcal{L}}^\dagger) m_\alpha&= -[H,\hat{\mathcal P}\hat{\mathcal T}m_\alpha]\nonumber\\
&-i\sum_{\mu}([L_\mu,\hat{\mathcal{P}}\hat{\mathcal{T}}m_\alpha]L_\mu^\dagger-L_\mu[L_\mu^\dagger,\hat{\mathcal{P}}\hat{\mathcal{T}}m_\alpha]\bigr).
\end{align}
Subtracting Eq.~\eqref{ilpt2} from Eq.~\eqref{ptil3} gives,
\begin{align}
\label{comm2}
&[i\hat{\mathcal{L}}^\dagger,\hat{\mathcal{P}}\hat{\mathcal{T}}] m_\alpha\nonumber\\
&=i \sum_\mu\Bigl(\bigl[[L_\mu^\dagger,\hat{\mathcal{P}}\hat{\mathcal{T}}m_\alpha],L_\mu\bigr]-\bigl[L^\dagger_\mu,[L_\mu,\hat{\mathcal{P}}\hat{\mathcal{T}}m_\alpha]\bigr]\Bigr).
\end{align}

Given the scaling of Lindblad operators in Eq.~\eqref{scaling}, Eq.~\eqref{comm2} becomes
\begin{align}
\label{comm22}
&[i\hat{\mathcal{L}}^\dagger,\hat{\mathcal{P}}\hat{\mathcal{T}}] m_\alpha\nonumber\\
&=i S\ \sum_\mu\Bigl(\bigl[[l_\mu^\dagger,\hat{\mathcal{P}}\hat{\mathcal{T}}m_\alpha],l_\mu\bigr]-\bigl[l^\dagger_\mu,[l_\mu,\hat{\mathcal{P}}\hat{\mathcal{T}}m_\alpha]\bigr]\Bigr).
\end{align}
Each term on the right-hand side contains two commutators of normalized collective spin operators, whose expectation values scale as $\mathcal O(1/S^2)$. Taking into account the overall factor $S$, we finally arrive at
\begin{align}
\label{comm}
\braket{[i\hat{\mathcal L}^\dagger,\hat{\mathcal P}\hat{\mathcal T}]\,m_\alpha}=\mathcal O\ (1/S)\xrightarrow[S\to\infty]{}0 .
\end{align}
\end{proof}


We now present Theorem 1, the main result of this section. It follows as an immediate corollary of the general theorem in Ref.~\cite{Nakanishi3}, here specialized to dissipative single--collective spin models. 
{In the following, we assume the scaling~\eqref{scaling} and the L-$\mathcal{PT}$ symmetry~\eqref{HuberPT} with $P=\prod_i \sigma_i^x$ and $T=K$. }\\

\noindent\textbf{Theorem 1 (Central result)}
\textit{{The mean-field equation~\eqref{csmean} possesses an n-PT symmetry~\eqref{nonlinearPT} with $\tilde{P}=\mathrm{diag}(1,1,-1)$.}}
\\[0.3em]



\noindent 

\begin{proof}

We express the action of the adjoint generator on the spin components as
\begin{equation}
i\hat{\mathcal{L}}^\dagger\hat{m}_{\alpha}=: iG_{\alpha}(m_x,m_y,m_z),
  \quad (\alpha=x,y,z),
  \label{eq:defF}
\end{equation}
where $G_{\alpha}$ is an operator-valued functional of $(m_x,m_y,m_z)$. By construction, $G_\alpha$ is Hermitian and all of its expansion coefficients are real. Using \(\hat{\mathcal P}\hat{\mathcal T}\,m_\alpha = m_\alpha\) for \(\alpha=x,y\) and \(\hat{\mathcal P}\hat{\mathcal T}\,m_z=-m_z\), we obtain
\begin{equation}
\label{eqptil}
\hat{\mathcal{P}}\hat{\mathcal{T}}(i\hat{\mathcal{L}}^\dagger\hat{m}_{\alpha})=-iG_{\alpha}(m_x,m_y,-m_z),
\end{equation}
and
\begin{align}
i\hat{\mathcal{L}}^\dagger (\hat{\mathcal{P}}\hat{\mathcal{T}}\hat{m}_{\alpha})
&=\begin{cases}
iG_{\alpha}(m_x,m_y,m_z), & (\alpha=x,y)\\[3pt]
-iG_{\alpha}(m_x,m_y,m_z), & (\alpha=z).
\end{cases}
\label{eqilpt}
\end{align}
By the lemma, the quantum expectation values of \eqref{eqptil} and \eqref{eqilpt} agree up to corrections of order $\mathcal O(1/S)$ at large $S$. Hence 
\begin{align}
&\braket{G_{\alpha}(m_x,m_y,-m_z)}\nonumber\\
&=\begin{cases}
-\braket{G_{\alpha}(m_x,m_y,m_z)}+\mathcal O(1/S), & (\alpha=x,y)\\[3pt]
\braket{G_{\alpha}(m_x,m_y,m_z)}+\mathcal O(1/S), & (\alpha=z).
\end{cases}
\end{align}
Invoking the mean-field approximation \(\braket{\hat{\mathcal{L}}^\dagger m_\alpha}=\langle G_\alpha(m_x,m_y,m_z)\rangle\simeq g_\alpha(X,Y,Z)\) and taking the thermodynamic limit \(S\to\infty\), we obtain
\begin{align}
\label{evenodd}
g_{\alpha}(X,Y,-Z)
&=\begin{cases}
-g_{\alpha}(X,Y,Z), & (\alpha=x,y)\\[3pt]
g_{\alpha}(X,Y,Z), & (\alpha=z),
\end{cases}
\end{align}
which is equivalent to
\begin{align}
[{\bf f}(\tilde{P}\tilde{T}{\bf M})]_\alpha&=ig_\alpha(X,Y,-Z)\nonumber\\
&=\begin{cases}
-ig_{\alpha}(X,Y,Z), & (\alpha=x,y)\\[3pt]
ig_{\alpha}(X,Y,Z), & (\alpha=z).
\end{cases}\nonumber\\
&=[\tilde{P}\tilde{T}\ {\bf f}({\bf M})]_\alpha,
\end{align}
with ${\bf f}=i{\bf g}$ [Eq.~\eqref{f=ig}].
Therefore, the mean-field equation has the n-PT symmetry with \(\tilde P=\mathrm{diag}(1,1,-1)\) and $(\tilde{T}{\bf f})={\bf f}^*$, as claimed.

\end{proof}

The key point is that, with the appropriate scaling of Lindblad operators in Eq.~\eqref{scaling}, the commutator between the (adjoint superoperator of) Lindbladian multiplied by the imaginary unit and the $\mathcal{PT}$ superoperator \textit{seems to be effectively} restored in the thermodynamic limit [Eq.~\eqref{lemma}]. As a consequence, the mean-field equation possesses the n-PT symmetry.
Note, however, that we do not claim 
\begin{align}
\lim_{S\to\infty}[\,i\hat{\mathcal L}^\dagger,\hat{\mathcal P}\hat{\mathcal T}\,]=0,
\end{align} 
which is equivalent to Eq.~\eqref{firstilpt} in the thermodynamic limit. Our statement applies only after the commutator acts on
$m_\alpha$ (or, more generally, on any operator built from an
$S$-independent finite number of $\{m_\alpha\}$) and an expectation value
is taken.

\subsection{Linear stability analysis}
\label{sec53}
Since the nonlinear mean-field equation inherits the n-PT symmetry $(\ref{nonlinearPT}$) for single-collective spin systems with the L-$\mathcal{PT}$ symmetry $(\ref{HuberPT})$, each fixed point must belong to either the PT-symmetric or the PT-broken one (see Section~\ref{sec22}). 
We clarify how its dynamical character depends on the symmetry breaking of fixed points by a linear stability analysis. In particular, this analysis reveals (i) the emergence of persistent oscillations protected by PT symmetry, (ii) a pair of stable and unstable symmetry broken solutions, (iii) CEPs, and (iv) the evaluation of critical exponents. Sections~\ref{sec531} and~\ref{sec532} are based on Ref.\cite{Nakanishi3}.


\subsubsection{Linear stability analysis for PT-symmetric fixed points}
\label{sec531}
We now perform a linear stability analysis around a fixed point ${\bf{M}}_{0}:=(X_{0}, Y_{0}, Z_0)^{T}$. 
This can be investigated by using the Jacobian matrix $J$ with $\partial_t\delta{\bf{M}}=J\delta{\bf{M}}$, where $\delta{\bf{M}}:=(\delta Y,\delta Z)^{T}$ represents the fluctuation vector around the fixed point in the $Y-Z$ plane, with $\delta Y(t):= Y(t)-Y_{0}$ and $\delta Z(t):= Z(t)-Z_{0}$. 
Here, we eliminated the $X$ component by using the constraint $X^2+Y^2+Z^{2}=1$ (unit-sphere normalization).

In the PT-symmetric fixed point (i.e., $Z_0=0$), the Jacobian can be simplified in the form
\begin{align}
\label{LL}
J=\left.\begin{pmatrix}
0&\alpha\\
\beta&0\\
\end{pmatrix}\right|_{ss},
\end{align}
where $\alpha:=\partial_{Z}g_{y},\ \beta:=(\partial_{Y}g_{z}-\frac{Y}{X}\partial_{X}g_{z})\in\mathbb{R}$ depends on $X$ and $Y$. Here, the quantities with ``${}|_{ss}$'' describe the value at the fixed point. Importantly, diagonal terms vanish due to the oddness of the functions $g_x, g_y$ in Eq.~\eqref{csmean} and the evenness of $g_z$ for $Z$.

The collective excitation spectra and excitation modes are given by 
\begin{align}
\lambda=\pm\sqrt{\alpha\beta}, \ \ \ \text{and}\ \ \ \ \delta{\bf{M}}&=(\pm\sqrt{\alpha/\beta},1)^T.
\end{align}
Physically, the real and imaginary part of the collective excitation spectrum $\lambda$ represents the growth rate of the fluctuations and their oscillation frequency. If the collective excitation spectrum contains a positive real part, the fixed point is unstable. Conversely, if all the real parts are non-positive, the fixed point is stable. If all collective excitation spectrum is purely imaginary, the fixed point is referred to as \textit{a center}~\cite{strogatz2018nonlinear}, indicating persistent oscillations with amplitude dependent on the initial state. 
\YNedit{We remark that these stability notions refer to deterministic mean-field dynamics, corresponding to the thermodynamic limit \(N\to\infty\). At finite \(N\), quantum fluctuations in the GKSL dynamics generally destroy strictly persistent oscillations; the oscillatory modes acquire small negative real parts, and the system ultimately relaxes to a TISS. Nevertheless, the corresponding relaxation time grows with system size, so finite-\(N\) dynamics can exhibit long-lived oscillations over an increasingly broad time window before the eventual decay. This finite-size behavior is confirmed numerically for the DDM in Sec.~\ref{sec552}.}

The PT-symmetric fixed point is unstable when $\alpha\beta>0$, but behaves as a center when $\alpha\beta<0$.
As $\alpha\beta\to0_{-}$, the fixed point loses stability and the oscillation period 
\begin{align}
\label{period}
T=2\pi/\sqrt{-\alpha\beta}
\end{align}
diverges, signaling the occurrence of a DPT. At the transition point, two collective modes typically coalesce unless $\alpha$ and $\beta$ simultaneously vanish; the critical point is characterized by the exceptional point (i.e., CEP), irrespective of whether the phase transition is continuous or discontinuous.

\subsubsection{Linear stability analysis for PT-broken fixed points}
\label{sec532}
We next perform the linear stability analysis around the PT-broken fixed points (i.e, $Z_0\neq0$).
The Jacobian can be written in the form
\begin{align}
\label{LLL}
J=\left. \begin{pmatrix}
\gamma_1 Z&\alpha\\
\beta&\gamma_2 Z\\
\end{pmatrix}\right|_{ss},
\end{align}
where $\alpha,\ \beta, \gamma_1$, $\gamma_2\in\mathbb{R}$ depend on $X,\ Y$ and $Z^2$, which are the same for both fixed points ${\bf M}_0$, $\tilde{P}\tilde{T}{\bf M}_0$.

Diagonalizing Eq.~($\ref{LLL}$), excitation spectra and excitation modes are obtained as
\begin{align}
\label{lambda}
\lambda_{\pm}=\left.\frac{1}{2}\left(R\pm\sqrt{R^2-Q}\right)\right|_{ss},
\end{align}
and
\begin{align}
\delta {\bf{M}}=\left(\left.\frac{(\gamma_1-\gamma_2)Z\mp\sqrt{R^2-Q}}{2\beta}\right|_{ss},1\right)^T,
\end{align}
with $R:=(\gamma_1+\gamma_2)Z$, and $Q:=4(\gamma_{1}\gamma_{2}Z^2-\alpha\beta)$, 
respectively. Hence, the excitation spectrum of a pair of PT-broken fixed points ${\bf M}_0$, $\tilde{P}\tilde{T}{\bf M}_0$ is the same except for the sign of \(R\).

To examine the stability of fixed points, we analyze different cases for $R$ and $Q$ under the assumption that $\gamma_1 + \gamma_2 \neq 0$. 
When $Q < 0$, the real parts of the excitation spectrum $\lambda_{\pm}$ are always of opposite sign, one positive and one negative, regardless of the value of $R$. As a result, both PT-broken fixed points ${\bf M}_0$, $\tilde{P}\tilde{T}{\bf M}_0$ are unstable.
In contrast, when $Q > 0$, the real parts of $\lambda_{\pm}$ are both negative (positive) for $R < 0$ ($R > 0$). This shows that a pair of stable and unstable PT-broken fixed points emerges, indicating that the steady-state does not degenerate.
Furthermore, in the case of a continuous phase transition where $\left.Z\right|_{\mathrm{ss}} = \left.\alpha\beta\right|_{\mathrm{ss}} = 0$, the transition point becomes a CEP.

\begin{figure}[t]
   \vspace*{-0.5cm}
     \hspace*{-0.1cm}
\includegraphics[bb=0mm 0mm 90mm 150mm,width=0.51\linewidth]{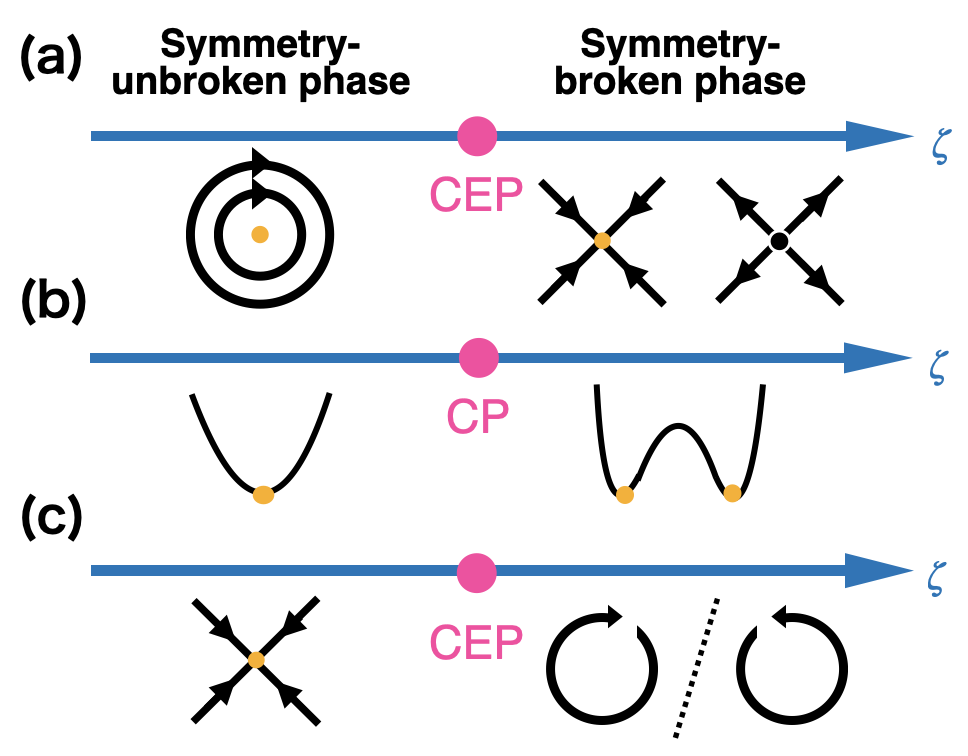}
\caption{\footnotesize\justifying Illustrative explanation of (a) L-$\mathcal{PT}$ phase transition, (b) nonequilibrium phase transition with $Z_{2}$ symmetry breaking, and (c) (standard continuous) non-reciprocal phase transition. For case (b), the transition from a unique stable fixed point to two degenerated stable fixed points occurs with $Z_{2}$ symmetry breaking. Landau description can often capture nonequilibrium phase transitions with unitary symmetry breaking in systems described by Eq.~\eqref{modelA}. Meanwhile, the nonlinear dynamical system for a L-$\mathcal{PT}$ phase transition (\textit{resp.} non-reciprocal phase transition) has an n-PT (\textit{resp.} n-anti-PT) symmetry (see Sec.~\ref{sec22}) and phase transition occurs as a dissipation (\textit{resp.} non-reciprocity) strength $\zeta$ increases from the unbroken phase with closed orbits (\textit{resp.} stable fixed points) to the symmetry-broken phase with pairs of stable and unstable fixed points (\textit{resp.} two symmetry-broken stable limit cycles). Furthermore, the transition point is associated with a CEP for both cases. These behaviors are out of scope of the potential-based Landau description. }
    \label{recipro}
\end{figure}

In summary, as shown in Fig.~\ref{recipro}~(a), the stable fixed point is a center as long as it remains PT-symmetric. When the n-PT symmetry is spontaneously broken, this center bifurcates into a pair of stable and unstable fixed points. This transition point is a CEP in the limit from the PT-broken phase.


\subsubsection{Critical exponents for continuous L-$\mathcal{PT}$ phase transitions}
\label{sec533}
Critical exponents encode the universal scaling behavior near a transition and usually identify its universality class. We now determine the critical exponents associated with magnetization $x_Z$ and relaxation time $\nu_t$ in continuous L-$\mathcal{PT}$ phase transitions, where the order parameter and the characteristic timescale scale as $Z|_{\mathrm{ss}} \sim (\kappa - \kappa_c)^{x_Z}$ and $\tau \sim (\kappa - \kappa_c)^{-\nu_t}$, respectively. Here, $\kappa_c$ denotes the critical dissipation rate at the transition point.

\medskip
\noindent\emph{Order parameter.}  
We begin by expanding the even function $f_z$ with $Z$ in Eq.~($\ref{csmean}$) (see Eq.~\eqref{evenodd}) as a power series in $Z$:
\begin{align}
g_z(X, Y, Z) = \sum_{n=0}^\infty c_{2n}(X, Y) Z^{2n},
\end{align}
where the coefficients $c_{2n}$ depend on $X$ and $Y$. At the transition point, both $Z$ and $g_z$ vanish in the steady state, implying that $c_0 = 0$.
Near the transition point, assuming
\begin{align}
\label{cq}
c_0 \sim (\kappa - \kappa_c)^{q},
\end{align}
\footnote{Generically, $c_0(\kappa)$ is an analytic function of the control
parameter with a simple zero at $\kappa_c$. Tuning a single parameter to
reach the transition only enforces $c_0(\kappa_c)=0$, while the derivative
$\partial_\kappa c_0|_{\kappa_c}$ remains finite and nonzero unless an extra
symmetry or further fine tuning is imposed. In this generic situation, one
can write $c_0(\kappa)\propto(\kappa-\kappa_c)$ and hence set $q=1$ in
Eq.~\eqref{cq}.}
and $c_{2} \neq 0$ with $q\in \mathbb{R}$, the steady-state condition $g_z = 0$ yields
\begin{align}
c_0 + c_{2} Z^{2}|_{\mathrm{ss}} = 0,
\end{align}
leading to the scaling behavior:
\begin{align}
Z|_{\mathrm{ss}} = \pm \sqrt{-c_0 / c_{2}} \sim (\kappa - \kappa_c)^{q/2},
\end{align}
which gives the critical exponent $x_Z= q/2$ in the PT-broken phase.

\medskip
\noindent\emph{Relaxation time.}  
Subsequently, we show that the relaxation time scales as $\tau \sim (\kappa - \kappa_c)^{-q/2}$. The relaxation time is defined as the inverse of the real part of the minimum excitation eigenvalue in Eq.~($\ref{lambda}$). Near the transition $\alpha\beta=0$, so set
$\alpha\beta\sim(\kappa-\kappa_c)^{r}$ with $r\in\mathbb{R}$.
\begin{enumerate}
\item $r>q$:  $\alpha\beta$ vanishes faster than $c_0$ and does not alter the scaling of $\tau$.
\item $r<q$ and $\alpha\beta>0$: $Q<0$ near criticality, destabilizing the fixed point; this conflicts with a continuous phase transition and is therefore discarded.
\item $r<q$ and $\alpha\beta<0$:  $\sqrt{R^2-Q}$ becomes purely imaginary, again leaving the scaling of $\tau$ unchanged.
\end{enumerate}

In all admissible cases, $\lambda_{\pm}\;\propto\;Z|_{\mathrm{ss}}
  \;\propto\;(\kappa-\kappa_c)^{q/2},
  \tau=\lambda_{\pm}^{-1}
  \;\propto\;(\kappa-\kappa_c)^{-q/2},$
therefore, both the magnetization and the relaxation time share the same
critical exponents,
\begin{align}
\label{nux}
 \nu_t=x_Z=q/2.   
\end{align}

\subsection{L-$\mathcal{PT}$ phase transitions, nonequilibrium phase transitions with breaking of unitary symmetry, and non-reciprocal phase transitions}
\label{sec54}
Equilibrium phase transitions are commonly characterized by a thermodynamic free-energy functional, as reviewed in Sec.~\ref{sec11}. Phases correspond to (global) minimizers of $\mathcal{F}$, and a transition occurs when the stability of a minimizer changes. Although this ``landscape'' picture is static, near equilibrium, the relaxational dynamics of the order parameter is a gradient flow,
\begin{equation}
\partial_t \phi(\mathbf r,t) = -\,\frac{\delta \mathcal{F}[\phi]}{\delta \phi(\mathbf r,t)},
\label{modelA}
\end{equation}
where $\mathcal F$ includes the standard gradient terms, and stochastic noise may be added to represent thermal fluctuations if required~\cite{hohenberg1977theory}. We note that whenever the dynamics admits the gradient form~\eqref{modelA}, the Landau description remains applicable even for nonequilibrium systems, i.e., in the absence of an underlying thermodynamic free-energy function~\cite{AransonKramer2002RMP, CrossHohenberg1993RMP, HohenbergKrekhov2015PhysRep} (Appendix~\ref{phasebifur} gives an example in open quantum systems). In the following, we assume that the order parameter is spatially uniform. 

We take as a basic example a system with a single $Z_2$-symmetric order parameter with $\phi\to-\phi$, whose Landau free energy is given by~\eqref{free}. As the control parameter is tuned through a critical value, the single minimum at $\phi=0$ splits into a pair at $\phi=\pm \phi_0\neq0$, producing the $Z_2$ symmetry breaking (supercritical pitchfork bifurcation) shown in Fig.~\ref{recipro}(b). 

Now, we consider nonequilibrium systems with $N$-component order parameters $\boldsymbol\phi=(\phi_1,\ldots,\phi_N)\in\mathbb{R}^N$, and the time evolution of order parameters can be described by a nonlinear dynamical equation,
\begin{align}
\partial_t\boldsymbol\phi={\bf g}(\boldsymbol\phi),
\end{align}
with ${\bf g}\in\mathbb{R}^N$.
Following Ref.~\cite{Fruchart}, we call a nonlinear dynamical system
\emph{conservative} if it is of gradient-flow (potential-derived) type, namely if it derives
from some potential \(F\) (e.g., a free energy) such that \(g_a(\phi) = -\partial_{\phi_a} F\). When the dynamical system is conservative,
\begin{align}
J_{ab}
\;=\; -\,\partial_{\phi_b}\partial_{\phi_a} F
\;=\; J_{ba} \, ,
\end{align}
so \(J\) is symmetric and hence a normal operator. 
By contrast, when no such potential exists, it is possible that
\begin{align}
J_{ab} \neq J_{ba} \, ,
\end{align}
i.e., \(J\) is not symmetric, and phase transitions or dynamical bifurcations beyond the potential-based Landau framework can occur.
A representative example is a non-reciprocal phase transition, in which the system's effective couplings violate detailed balance.
~\cite{Fruchart}.

Moreover, we extend the order parameters ${\boldsymbol{\phi}}$ and the dynamical map ${\bf g}$ to complex variables, ${\bar{\boldsymbol{\phi}}}\in\mathbb{C}^N$ and ${\bar{\bf g}}:\mathbb{C}^N\to\mathbb{C}^N$. The standard continuous non-reciprocal phase transition can be described by an effective theory with $O(2)$ symmetry~\cite{Fruchart}. The group $O(2)$ consists of a continuous $U(1)$ subgroup and a discrete $Z_2$ symmetry generated by complex conjugation. Therefore, equation $\partial_t \bar{\boldsymbol{\phi}}=\bar{\bf g}(\bar{\boldsymbol{\phi}})$ is invariant under complex conjugation in the sense that
\begin{equation}
\label{ccsymm}
\bar{\bf g}(\bar{\boldsymbol{\phi}}^*)=\bar{\bf g}(\bar{\boldsymbol{\phi}})^*.
\end{equation}
At a non-reciprocal phase transition, this $Z_2$ symmetry is spontaneously broken: the system undergoes a transition from a static ordered phase, in which the continuous $U(1)$ symmetry is broken, to a time-periodic phase in which 
the complex-conjugation symmetry is broken, and two stable limit cycles with opposite chiralities emerge [Fig.~\ref{recipro}(c)]. The transition point is characterized by a CEP. In terms of the nonlinear map $\mathbf{f}=i\bar{\mathbf{g}}$, the complex-conjugation symmetry~\eqref{ccsymm} is equivalent to an anti-n-PT symmetry~\eqref{nonlinearantiPT} with $\tilde{P}=\mathrm{Id}$.

Thus, at the mean-field level, the mathematical structures of non-reciprocal phase transitions and L-\(\mathcal{PT}\) phase transitions are very similar. \YNedit{However, it should be noted that L-\(\mathcal{PT}\) phase transitions and non-reciprocal phase transitions lead to qualitatively different physical consequences.
In L-\(\mathcal{PT}\) phase transitions, the time dependence originates from a PT-symmetric center in the unbroken phase, where mean-field trajectories form marginal closed orbits around the fixed point.
The mean-field dynamics then undergoes a transition, accompanied by n-PT symmetry breaking, from this center to a stable--unstable pair of fixed points.
By contrast, in non-reciprocal phase transitions, the anti-n-PT-symmetric phase has a stable fixed point, which can correspond to a stationary polar or aligned state, whereas the broken phase supports time-dependent ordered states such as attracting limit cycles or spatially traveling waves.}

Thus, both exhibit oscillations somewhere in parameter space, but the origins are distinct: the former relies on symmetry protection and is therefore fragile to perturbations that break the n-PT symmetry, whereas the latter do not rely on an exact symmetry and persist under small symmetry-breaking perturbations, as in the explicit complex-conjugation (i.e., the n-anti-PT symmetry in our notation) breaking cases discussed in Ref.~\cite{Fruchart}.


We also note that (second-order) non-reciprocal phase transitions can occur in open quantum systems whose Lindbladian is anti-\(\mathcal{PT}\)-symmetric [Eq.~\eqref{antilpt}]~\cite{nadolny2025nonreciprocal}.

\subsection{Continuous L-$\mathcal{PT}$ phase transition  --- DDM}
\label{sec55}
\subsubsection{Mean-field analysis}
To confirm our mean-field framework for L-$\mathcal{PT}$ phase transitions, we begin by analyzing the DDM ($\ref{DDM}$). This model exhibits the L-$\mathcal{PT}$ symmetry ($\ref{HuberPT}$) when the parity operator is chosen as a $\pi$-rotation along the $x$-axis defined as $P=\prod_i \sigma_i^x$. 
The parity operator acts on each spin operator as follows:
\begin{align}
\label{PSP}
PS_{z}P^{-1}=-S_{z},\ \ PS_{\pm}P^{-1}=S_{\mp}.
\end{align}
Then, $\mathbb{PT}(S_{x})=S_{x}$ and \(\YNedit{\mathbb{PT}(S_{\pm})=S_{\pm}}\).
\YNedit{We note that, under the same \(PT\) transformation, adding jump operators of the form \(L_\mu=S_\alpha\) \((\alpha=x,y,z)\), \(L_\mu=S_+\), or \(L_\mu=S_-\), with arbitrary dissipation rates, also preserves the L-\(\mathcal{PT}\) symmetry, since these operators are mapped to themselves up to irrelevant phase factors. Consequently, no fine tuning of the corresponding dissipation rates is required.}


The time evolution in the large total spin limit $S\to\infty$ is given by
\begin{align}
\label{timem22}
\partial_{t}{\bf{M}}=
\begin{pmatrix}
2\kappa ZX\\
-2gZ+2\kappa ZY \\
2gY-2\kappa (1-Z^{2}) \\
\end{pmatrix}.
\end{align}
As guaranteed by Theorem 1, the nonlinear dynamical system ($\rm\ref{timem22}$) has the n-PT symmetry ($\ref{nonlinearPT}$) with $\tilde{P}=diag(1,1,-1)$. For $\kappa<g$, there are two PT-symmetric fixed points,
\begin{align}
(X, Y, Z)=\left(\pm\sqrt{1-\left(\kappa/g\right)^{2}},\kappa/g,0\right).
\end{align}
Meanwhile, for $\kappa>g$, there are two PT-broken fixed points,
\begin{align}
\label{brokenfp}
(X, Y, Z)=\left(0,g/\kappa,\pm\sqrt{1-\left(g/\kappa\right)^{2}}\right).
\end{align}
Eqs.~\eqref{timem22} and \eqref{brokenfp} give the exponent $q=1$ in Eq.~\eqref{cq}, and consequently the critical exponents are $ \nu_t=x_Z=1/2$ [see Eq.~\eqref{nux}].


The Jacobian around the PT-symmetric fixed points is given by
\begin{align}
J=2\begin{pmatrix}
0&-(g^{2}-\kappa^{2})/g\\
g&0 \\
\end{pmatrix}.
\end{align}
The collective excitation spectrum and excitation modes are given by $\lambda=\pm2i\sqrt{g^{2}-\kappa^{2}}$, 
$\delta {\bf{M}}=(\mp i(\sqrt{g^{2}-\kappa^{2}})/g,1)^{T}$,
indicating the presence of persistent closed orbits. 
Two collective excitation modes coalesce at the transition point, i.e. it becomes a CEP. 
\footnote{Although the enhancement of fluctuation has been reported around CEPs~\cite{Hanai2}, its effect for the DDM cannot be observed at the mean-field level. This is because the fluctuation vector orthogonal to the zero excitation mode coincides with other excitation modes. One can avoid this problem by adding other terms to the Hamiltonian, such as the two-body interaction along the $z$-axis $S_{z}^{2}$~\cite{Nakanishi3}, generalized DDM.}

Similarly, the Jacobian around the PT-broken fixed points is obtained as
\begin{align}
J=2\left.\begin{pmatrix}
\kappa Z&0\\
g&2\kappa Z \\
\end{pmatrix}\right|_{ss}.
\end{align}
Then, the excitation spectrum and excitation modes are given by $\lambda=\{2\kappa Z,\ 4\kappa Z\}$, $\delta {\bf{M}}=\{(-\kappa Z/g,1)^{T},\ (0,1)^{T}\}$.
Therefore, as discussed in Section~\ref{sec532}, one is stable while the other is unstable.
A CEP also emerges at $\kappa=g$ from the ordered phase.

So far, we have classified the PT symmetry for fixed points. We now explore the PT symmetry of time-dependent solutions (e.g. an oscillating solution) as well.
As discussed in Sec.~\ref{sec22}, for a system governed by a time-independent generator, we say that ${\bf{q}}(t)$ is PT-symmetric if ${\bf{q}}(t)=\tilde{P}\tilde{T}{\bf{q}}(t-t_{0})$ with $t_{0}\in \mathbb{R}$.
Ref.~\cite{Carmichael} gives a periodic solution with $X(t)=0,\ Y(0)=0,\ Z(0)=1$ as
\begin{align}
\label{zt}
Z(t)=-\sqrt{\Bigl(\frac{g}{\kappa}\Bigr)^2-1}\frac{\sin(\Omega t+\phi)}{\cos(\Omega t+\phi)-g/\kappa},
\end{align}
with frequency $\Omega=2\sqrt{g^{2}-\kappa^2}$ and $\phi=\cos^{-1}(\kappa/g)$.
This solution is PT-symmetric since $Z(t)=-Z(-t+t_{s})$ with $t_s=-2\phi/\Omega$.

\subsubsection{Lindbladian eigenvalues and Lindbladian exceptional point}
\label{sec552}
As discussed in Section~\ref{sec31}, the Lindbladian eigenvalues and eigenmodes encode, in principle, the essential features of the dynamics of the system. Therefore, we expect that centers and the non-degenerate stable fixed point imply the existence of PIEs and a unique steady-state with a finite Lindbladian gap, respectively.

Figure \ref{DDMfig}(f) plots the eigenvalue whose real part is largest among all non-zero modes; its non-zero imaginary part sets the oscillation frequency of the CTC. Figure \ref{DDMfig}(g) shows the Lindbladian gap, whose inverse dictates the relaxation time. Figure \ref{DDMfig}(h) displays the $S$–dependence of the real and imaginary parts of that leading eigenvalue: the magnitude of the real part scales approximately as \(1/S\), while the imaginary part remains essentially constant. In the large-spin limit the eigenvalue thus becomes purely imaginary, in line with the center picture and fully consistent with the mean-field predictions.


Furthermore, a CEP implies the presence of a zero-mode LEP, where multiple eigenmodes with zero eigenvalues coalesce [see Sec.~\ref{sec313}]. Figure~\ref{DDMfig}(i) quantifies the finite-size behavior: the distance to the critical dissipation and the real part of the LEP with the largest real part obey power-law scalings with the total spin, $|\kappa-\kappa_c|\sim S^{-0.62}$ and $\mathrm{Re}[\lambda_{\mathrm{EP}}]/g\sim S^{-0.35}$. Hence, as $S\to\infty$, $\kappa$ approaches $\kappa_c$ and $\mathrm{Re}[\lambda_{\mathrm{EP}}]\to 0$ algebraically, so that at $\kappa=\kappa_c$ the Lindbladian exhibits a zero-mode LEP and the gap closes in the thermodynamic limit.

\begin{figure}[t]
   \vspace*{0.8cm}
     \hspace*{0cm}
\includegraphics[bb=0mm 0mm 90mm 150mm,width=0.52\linewidth]{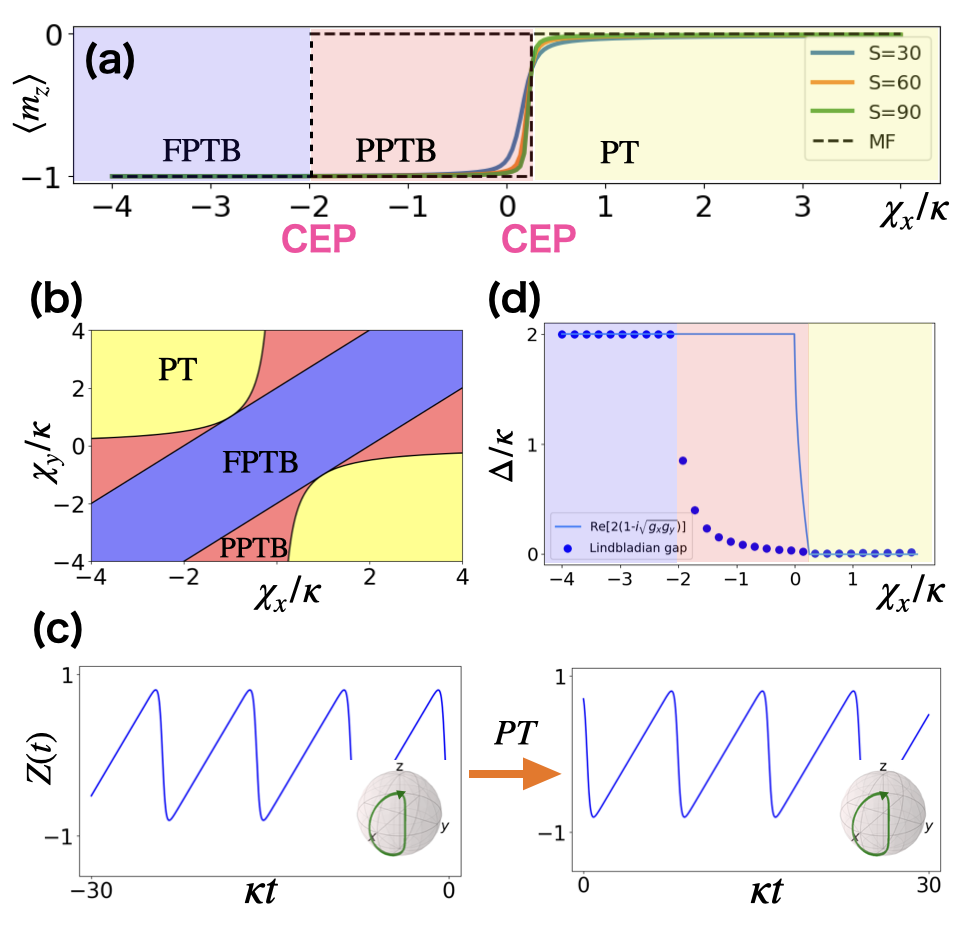}
\caption{\footnotesize\justifying The phase diagram and numerical calculation for the dissipative LMG model ($\ref{LMG}$). (a) The normalized magnetization in the TISS for finite $S$ and mean field solution ($S = \infty$) with $\chi_{y}=-4$. (b) Phase diagram: In the yellow (blue) region, there are only PT-symmetric (PT-broken) fixed points. In the red region, both PT-symmetric and PT-broken fixed points coexist. (c) Dynamics and the PT transformed one ($Z\ \to -Z$,  $t\to\ -t$) with $\chi_{x}/\kappa=1,\ \chi_{y}/\kappa=-2$. Inset: Their trajectories. They describe the same attractor. (d) The Lindbladian gap for $S = 100$ (blue dot) and the real part of the mean-field excitation spectra (light blue line) with $\chi_{y}=-4$.}
    \label{lpt}
\end{figure}

\subsection{Discontinuous L-$\mathcal{PT}$ phase transition --- Dissipative LMG model}
\label{sec56}
Let us next consider the dissipative LMG model described by 
\begin{align}
\label{LMG}
\partial_{t}\rho=-\frac{i}{S}[(\chi_{x}S_{x}^{2}+\chi_{y}S_{y}^{2}),\rho]+\frac{\kappa}{S}\mathcal{D}[S_{-}]\rho,
\end{align}
where the coefficients $\chi_{x},\ \chi_{y}$ are determined by the surface anisotropy and the coefficient $\kappa$ is the dissipation strength~\cite{lee2014dissipative, ferreira2019lipkin}. In this model, a discontinuous (first-order) phase transition occurs (Fig.~\ref{lpt} (a)).

The time evolution in the large total spin limit $S\to\infty$ is given by
\begin{align}
\label{timem222}
\partial_{t}{\bf{M}}=
\begin{pmatrix}
2\chi_{y}YZ+2\kappa ZX\\
-2\chi_{x}XZ+2\kappa ZY \\
2(\chi_{x}-\chi_{y})XY-2\kappa (1-Z^{2}) \\
\end{pmatrix}.
\end{align}
As guaranteed by Theorem 1, the nonlinear dynamical system ($\rm\ref{timem222}$) has the n-PT symmetry ($\ref{nonlinearPT}$) with $\tilde{P}=diag(1,1,-1)$. 


For $\kappa<\kappa_{c1}:=(\chi_{x}-\chi_{y})/4$, there are four PT-symmetric fixed points,
${\bf{M_{1}}}=(\pm M_{+},\pm M_{-},0)$,
with 
\begin{align}
M_{\pm}:=\sqrt{(1\pm\sqrt{1-(\kappa/\kappa_{c1})^{2}})/2}
\end{align}
and
${\bf{M_{2}}}=(\pm M_{-},\pm M_{+},0).$
In addition, there are two PT-broken fixed points,
${\bf{M_{3}}}=(0,0,\pm1).$ Figure~\ref{lpt} (b) shows the phase diagram~\cite{ferreira2019lipkin}. In the yellow (blue) region, there are only PT-symmetric (PT-broken) fixed points. In the red region, both PT-symmetric and PT-broken fixed points coexist. Based on the PT symmetry of fixed points, we refer to these phases as PT phase, partially PT broken (PPTB) phase, and fully PT broken (FPTB) phase. 

The Jacobian around the PT-symmetric fixed points is given by
\begin{align}
J=\left.\begin{pmatrix}
0&-2\chi_{x}X+2\kappa Y\\
\frac{2}{X}(\chi_{x}-\chi_{y})(X^{2}-Y^{2})&0 \\
\end{pmatrix}\right|_{ss}.
\end{align}
The fixed points ${\bf{M_{1}}}$ (\textit{resp}. ${\bf{M_{2}}}$) are centers for $\kappa<\kappa_{c1}\ (\textit{resp}.\ \kappa_{c2}:=\sqrt{-\chi_{x}\chi_{y}})$ and are destabilized at $\kappa=\kappa_{c1}$ (\textit{resp}. $\kappa_{c2}$). These results indicate that persistent closed orbits exist for $\kappa<\kappa_{c1}$. In fact, Fig.~\ref{lpt} (c) shows the presence of a PT-symmetric oscillation.
Moreover, two collective excitation modes coalesce at the transition points $\kappa=\kappa_{c1},\kappa_{c2}$, that is, these are CEPs. Note that the line with $\chi_{x}=-\chi_{y}$ is special in that the CEP does not emerge since the non-diagonalized elements $\alpha$ and $\beta$ simultaneously vanish~\cite{Nakanishi3}.

The Jacobian around the PT-broken fixed points is given by
\begin{align}
\label{Jlmg}
J=2Z|_{ss}\begin{pmatrix}
\kappa &\chi_{y}\\
-\chi_{x}&\kappa  \\
\end{pmatrix},
\end{align}
where we choose the fluctuation vector in the $X-Y$ plane $\delta{\bf{M}}^{\prime}:=(\delta X,\delta Y)^{T}$ because a matrix element diverges for the fixed point with $X=Y=0$.
The collective excitation spectra and excitation modes are given by
$\lambda=2Z(\pm\sqrt{-\chi_{x}\chi_{y}}+\kappa),$ 
$\delta {\bf{M}}=(-i\chi_{y}/\chi_{x},1)^{T}.$
Therefore, the fixed point with $Z<0$ is stable and the other fixed point with $Z>0$ is unstable. Note that the Jacobian $(\ref{Jlmg})$ does not take the form of Eq.~$(\ref{LLL})$ because the choice of planes is different. We adopted the $Y-Z$ (\textit{resp}. $X-Y$) plane in Eq.~($\ref{LLL}$) (\textit{resp}. Eq.~($\ref{Jlmg}$)).

We summarize the differences from the DDM for the LMG model in the PT-broken phase:
(i) Absence of a CEP from the ordered phase, (ii) Steady-state is a dark state, which is defined as $\rho_{\rm dark}:=\ket{\psi}\bra{\psi}$ with $L_\mu\ket{\psi}=0$ for all $\mu$, (iii) Linear relaxation-time scaling $\tau\sim |\kappa-\kappa_{c2}|^{-1}$.


Lastly, we compare the mean-field results and Lindbladian eigenvalues.
Fig.~$\ref{lpt}$ (d) compares the Lindbladian gap with the collective excitation spectrum. One can expect that the Lindbladian gap vanishes in the PT and PPTB phases because of the presence of persistent oscillations without decay, while in the FPTB phase it corresponds to the real part of the mean-field excitation spectra. 
Fig.~$\ref{lpt}$ (d) indicates that the mean-field prediction coincides well with the Lindbladian gap for a large $S$.

\subsection{Generalization to broader settings} 
\label{sec57}
So far, we have focused on collective-spin models, in which the total spin is conserved and there is no spatial structure. Here, we extend our mean-field argument to settings where these conditions are relaxed.
\subsubsection{Long-range dissipation}
\label{sec571}
We consider a family of dissipative couplings interpolating between local and long-range dissipation that preserves the L-$\mathcal{PT}$ symmetry but breaks conservation of the total spin~\cite{Passarelli}. It can be written in Kac-normalized form as
\begin{align}
\label{kac}
L_i(\eta)=\sum_{j=1}^N g_{ij}(\eta)\,\sigma^-_j,\quad
g_{ij}(\eta)=\frac{K^{(N)}(\eta)}{D(|i-j|)^{\eta}},
\end{align}
with 
\begin{align}
\sum_j g_{ij}(\eta)=1,\ \  D(r)=\min(r,N-r)+1,
\end{align}
and $K^{(N)}(\eta)=1/\sum_{r=0}^{N-1}D(r)^{-\eta}$. In the $\eta\to\infty$ (\textit{resp.} $\eta\to0$) limit, one has $g_{ij}\to\delta_{ij}$ (resp. $g_{ij}\to1/N$), so that the dissipator reduces to local dissipation $L_i\to\sigma_-^{\,i}$ (resp. collective dissipation $L_i\to \sum_{j=1}^N\sigma^-_{\,j}/N$). 

The Lindbladian with a PT-symmetric Hamiltonian then takes the form
\begin{align}
    \hat{\mathcal{L}}{\rho}=-i[H,\rho]+\sum_i\hat{\mathcal{D}}[L_i(\eta)]\rho,
\end{align}
which is invariant under the L-$\mathcal{PT}$ transformation with $PT=\prod_i \sigma_i^x\  K$. Throughout this discussion, we impose periodic boundary conditions.

To quantify how the long-range dissipation restores PT symmetry at the level of macroscopic observables, we evaluate the commutator of the adjoint generator with the $\mathcal{PT}$ superoperator, $\hat{{\mathcal P}}\cdot=P\cdot P^{-1}$, ${\hat{\mathcal T}}\cdot=T\cdot T^{-1}$, acting on the normalized magnetization,
\begin{equation}
M_\alpha := \frac{1}{N}\sum_{k=1}^N \sigma_k^\alpha.
\end{equation}
In analogy with Eq.~\eqref{comm2}, this commutator can be expressed as
\begin{align}
\label{comm4}
[i\hat{\mathcal{L}}^\dagger,\hat{\mathcal{P}}\hat{\mathcal{T}}] M_\alpha
&=i\sum_i\Bigl(\bigl[[L^\dagger_i(\eta),\hat{\mathcal{P}}\hat{\mathcal{T}}M_\alpha],L_i(\eta)\bigr]\nonumber\\
&-\bigl[L_i^\dagger(\eta),[L_i(\eta),\hat{\mathcal{P}}\hat{\mathcal{T}}M_\alpha]\bigr].
\end{align}
Among the terms on the right-hand side of Eq.~\eqref{comm4}, the double commutator $\sum_{i=1}^{N}[[L_i^\dagger(\eta),\hat{\mathcal{P}}\hat{\mathcal{T}}M_\alpha],L_i(\eta)]$ can be evaluated as
\begin{align}
&\sum_{i=1}^{N}\bigl[[L^\dagger_i(\eta),\hat{\mathcal{P}}\hat{\mathcal{T}}M_\alpha],L_i(\eta)\bigr]\nonumber\\
&=\frac{1}{N}\sum_{j=1}^{N}\Bigl(\sum_{i=1}^{N} g_{ij}^2(\eta)\Bigr)\,
\bigl[[\sigma_j^{+},\hat{\mathcal{P}}\hat{\mathcal{T}}\sigma_j^\alpha],\sigma_j^{-}\bigr]
=:G_\eta^{(N)}\,\mathcal{S}_\alpha,
\label{eq:double_comm}
\end{align}
where we defined $\mathcal{S}_\alpha:=\frac{1}{N}\sum_j\bigl[[\sigma_j^{+},\hat{\mathcal{P}}\hat{\mathcal{T}}\sigma_j^\alpha],\sigma_j^{-}\bigr]$, which is an operator of $\mathcal{O}(1)$ norm, and, by translational invariance
\begin{equation}
G_\eta^{(N)}:=\sum_{i=1}^{N} g_{ij}^2(\eta)=[K^{(N)}(\eta)]^2\sum_{r=0}^{N-1}D(r)^{-2\eta}
\label{eq:F_def}
\end{equation}
is independent of $j$.
From the large-$N$ behavior of the sums defining $G_\eta^{(N)}$, one finds
\begin{align}
&\lim_{N\to\infty}G_\eta^{(N)}=
\begin{cases}
0 & 0\le \eta\le1,\\
\text{const.}>0, & \eta>1,
\end{cases}
\label{eq:F_scaling}   
\end{align}
Another term on the right-hand side in Eq.~\eqref{comm4} can be bounded in the same way, so the lemma and Theorem~1 apply for $\eta\leq1$. 

By contrast, for $\eta>1$ the same double commutator remains of $\mathcal{O}(1)$ and, although L-$\mathcal{PT}$ symmetry is preserved, the n-PT symmetry of the mean-field dynamics is \textit{explicitly broken}; therefore, the linear stability analysis of Sec.~\ref{sec53} fails, and the oscillations protected by the n-PT symmetry do not emerge. In particular, this includes the local dissipator
\begin{align}
\hat{\mathcal{L}}_{\mathrm{loc}}[\rho]=\sum_{i=1}^{N}\hat{\mathcal{D}}[\sigma^-_{\,i}]\,\rho.
\end{align}
This explicit n-PT symmetry breaking for dissipator~\eqref{kac} is consistent with the phase transition at $\eta=1$~\cite{Passarelli}.

\subsubsection{Spatially extended bipartite bosonic systems with strong $U(1)$ symmetry}
Next, we consider spatially extended bosonic systems with strong $U(1)$ symmetry. 
We take a lattice with $l$ sites and bosonic annihilation operators $a_i$ at each site, and assume that the total particle number $N := \sum_i a_i^\dagger a_i$ is conserved and large ($N \gg 1$). This conservation is necessary to prevent unphysical divergences in observables; for example, in a linear balanced gain–loss bosonic system, expectation values can grow without bound in time~\cite{Nakanishi}. We introduce normalized bosonic operators $b_i := a_i/\sqrt{N}$ and impose the large-$N$ scaling
\begin{equation}
\label{scaling2}
  H = N h,\quad L_\mu = \sqrt{N}\, l_\mu,
\end{equation}
where $h$ and $l_\mu$ are intensive $O(1)$ operators that can be written as finite polynomials, with $N$-independent degree, of the normalized bosonic operators $\{b_i,b_i^\dagger\}$. In the large total particle number limit, the mean-field dynamics of the local observables
\begin{equation}
  q_i := \lim_{N\to\infty}\langle b_i\rangle 
\end{equation}
is governed by a closed set of nonlinear equations of the form
\begin{equation}
  i\,\partial_t \mathbf{q} = \mathbf{f}(\mathbf{q}),
\end{equation}
with $\mathbf{q}:=(q_{1},q_{2},\dots,q_{l})^{\mathsf T}\in\mathbb{C}^{l}$ and 
$\mathbf{f}:=(f_{1},f_{2},\dots,f_{l})^{\mathsf T}:\mathbb{C}^{l}\to\mathbb{C}^{l}$).
Furthermore, we assume that it satisfies the L-$\mathcal{PT}$ symmetry [Eq.~\eqref{HuberPT}], where the parity operator $P$ acts as a permutation on the lattice sites and satisfies $P^{2}=1$, and $T=K$.
Above the consideration, the following theorem holds~\cite{Nakanishi3}.\\

\noindent\textbf{Theorem 2} \textit{ The mean-field equation possesses an n-PT symmetry [Eq.~\eqref{nonlinearPT}], where the parity operator $\tilde P$ is a corresponding permutation in a nonlinear dynamical system. 
}\\

\begin{figure}[t]
   \vspace*{-2.4cm}
     \hspace*{0.5cm}
\includegraphics[bb=0mm 0mm 90mm 150mm,width=0.4\linewidth]{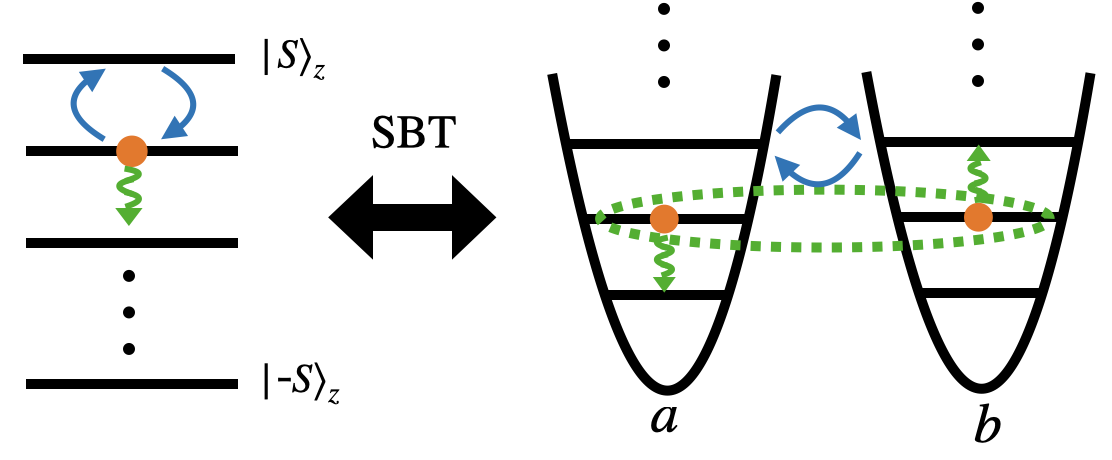}
\caption{\footnotesize\justifying Illustration of the Schwinger boson transformation (SBT). For example, the collective lowering operator $S_-$ maps onto a bilinear product of two bosonic modes, yielding a nonlinear gain–loss dissipator $ab^\dagger$. In this SBT representation, the resulting model is physically analogous to the basic PT-symmetric system introduced in Fig.~\ref{figPT4}.}
    \label{figsbsb}
\end{figure}

\noindent A single collective spin can be mapped to a two-mode bosonic particle-number-conserving system via the Schwinger-boson representation~\cite{Pires}, 
\begin{align}
\label{sb}
    m_-=a b^\dagger/(2S),\ \ m_z=(a^\dagger a-b^\dagger b)/(2S),
\end{align}
where $a,b$ are bosonic annihilation operators. In this representation, Theorem~2 provides a natural generalization of Theorem~1. The parity operator $P=\prod_i \sigma_i^x$ can be interpreted as exchanging the two bosonic modes, as illustrated in Fig.~\ref{figsbsb}. A detailed proof of Theorem~2 is given in Ref.~\cite{Nakanishi3}.

Subsequently, we specialize to a bipartite lattice with a spatially uniform solution, which provides the minimal setting for spontaneous n-PT symmetry breaking in dissipative bosonic systems.
In a bipartite system with sublattices $A$ and $B$, such a solution is characterized by two order parameters,
\begin{equation}
q_A:=\frac{2}{l}\sum_{i\in A}\langle q_i\rangle,\qquad
q_B:=\frac{2}{l}\sum_{i\in B}\langle q_i\rangle.
\end{equation}
Under strong $U(1)$ symmetry, the total boson number is conserved, and the mean-field equations become independent of the individual phases $\theta_{A}$ and $\theta_{B}$ in the polar representation $q_{A}=r_{A}e^{i\theta_{A}}$ and $q_{B}=r_{B}e^{i\theta_{B}}$. Together with the assumption of conserved total density, $r_{\rm A}^2+r_{\rm B}^2=\mathrm{const.}$, the mean-field dynamics reduces to
\begin{align}
\label{521}
&\partial_{t}r_{A}=\frac{1}{r_{A}}\,G(r_{A},r_{B},\Delta\theta_{AB}),\nonumber\\
&\partial_{t}r_{B}=-\frac{1}{r_{B}}\,G(r_{A},r_{B},\Delta\theta_{AB}),\nonumber\\
&\partial_{t}\Delta\theta_{AB}=H(r_{A},r_{B},\Delta\theta_{AB}),
\end{align}
with $\Delta\theta_{AB}:=\theta_{A}-\theta_{B}$, where the function $G$ ($H$) is symmetric (antisymmetric) under the exchange of amplitudes $r_{A}$ and $r_{B}$, i.e., $G(r_{A},r_{B})=G(r_{B},r_{A})$ and $H(r_{A},r_{B})=-H(r_{B},r_{A})$.

The set of mean-field equations in Eq.~\eqref{521} is equivalent to that of the single-collective-spin case under identification
\begin{align}
m_x&=r_A r_B\cos\Delta\theta_{AB},\nonumber\\
m_y&=-r_A r_B\sin\Delta\theta_{AB},\quad
m_z=r_A^2-r_B^2.
\end{align}
Therefore, by the same reasoning as for a single collective spin, the transition to a dynamical phase is again characterized by a CEP. An example of such a spatially extended bosonic model is presented in the Supplemental Material of Ref.~\cite{Nakanishi3}.

We note that the above discussion effectively discards the individual phase degrees of freedom $\theta_A$ and $\theta_B$ and, with them, the possibility of limit-cycle behavior. If these degrees of freedom are retained, persistent oscillations can in principle survive even in the PT-broken phase; indeed, Ref.~\cite{solanki2025generation} reports the existence of limit cycles in the PT-broken regime.

\section{Continuous-time crystals and Lindbladian $\mathcal{PT}$ symmetry}
\renewcommand{\theequation}{6.\arabic{equation} }
\setcounter{equation}{0}
\label{sec6}
In the previous section, we have shown that the L-$\mathcal{PT}$ symmetry can produce persistent oscillations, i.e., break a continuous-time translation symmetry for collective spin systems. The prototypical example is the DDM~\eqref{DDM}, which is known to exhibit CTC-like persistent oscillations.
Here, we discuss the relation between CTCs and persistent oscillations arising from L-$\mathcal{PT}$ symmetry. To avoid ambiguity, we first define CTCs precisely and specify the criterion we adopt throughout.


In Sections~\ref{sec61},~\ref{sec62}, we provide a brief review of time crystals in closed and periodically driven systems, respectively. In Section~\ref{sec63}, we organize the concept and criteria of CTCs in Markovian open quantum systems.
In Section~\ref{sec64}, we discuss the relation between CTCs in Markovian open quantum systems and L-$\mathcal{PT}$ symmetry in dissipative single-collective spin systems. 
Finally, in Section~\ref{sec65}, we examine a two-collective spin model that incorporates balanced gain and loss.



\subsection{Time crystal in closed systems}
\label{sec61}
An ordinary spatial crystal is a phase of matter characterized by the spontaneous breaking of continuous spatial-translation symmetry down to a discrete lattice subgroup, resulting in periodic structures in space.
By analogy, a continuous-time crystal is a phase of matter in which continuous time-translation symmetry is spontaneously broken down to a discrete subgroup, leading to persistent time-periodic behavior.


The notion of a time crystal was originally introduced in the quantum many-body setting by Wilczek~\cite{Wilczek} and in classical mechanics by Shapere and Wilczek~\cite{shapere2012classical}.
However, these original proposals were criticized in subsequent work~\cite{bruno2012comment,bruno2013impossibility}.
Most notably, Watanabe and Oshikawa~\cite{Watanabe} have established a general no-go theorem: for closed quantum systems governed by time-independent short-range Hamiltonians, neither the ground state nor any Gibbs state can exhibit a time crystal.

This no-go theorem was proved by properly extending the notion of SSB to temporal degrees of freedom. As explained in Sec.~\ref{secSSB}, SSB can be formulated via a small symmetry-breaking field [Eq.~\eqref{eq:quasiavg}] or via LRO [Eq.~\eqref{lro}]. The former is inapplicable to time-translation symmetry, since breaking it requires an explicitly time-dependent field, and then the ground state or a Gibbs state is not well defined. Watanabe and Oshikawa therefore formulated time crystals in terms of LRO extended to the temporal direction.

Concretely, a state qualifies as a CTC if the space-time correlator of a local observable approaches a non-constant periodic function $f(t)$ in time $t$ in the thermodynamic limit:
\begin{align}
\label{timecry}
\lim_{V\to\infty}\braket{o_{\bf x}(t)o_{\bf y}(0)}= f(t),
\end{align}
for sufficiently large distances $ |{\bf{x-y}}|$.
Here, $o_{\bf x}(t):=e^{iHt}o_{\bf x}e^{-iHt}$ denotes a local observable in the Heisenberg picture, and $V$ denotes the system volume. In terms of the integrated order parameter, $O(t)= \sum_{{\bf x}\in\Lambda}\,o_{\bf x}(t)$, the condition~\eqref{timecry} reads
\begin{align}
\label{TCmacro}
\lim_{V\to\infty}\frac{1}{V^{2}}\, \langle O(t)\,O(0)\rangle = f(t).
\end{align}
In a closed system with a time-independent short-range Hamiltonian, the no-go theorem implies that neither the ground state nor any Gibbs state can satisfy the condition~\eqref{timecry} or~\eqref{TCmacro}.

One might think that the constraints on large system sizes and large distances are unnecessary for the detection of CTCs. However, without these constraints, even essentially trivial cases, such as chains of unconnected finite systems, exhibit a CTC. 

We finally comment on \textit{symmetry-protected time crystals}~\cite{Khemani2019BriefHistory}. These can be viewed as equilibrium spatiotemporal order enabled by an extra conserved charge. When the corresponding symmetry is spontaneously broken, the associated order parameter can undergo persistent phase rotation at a frequency fixed by the conjugate chemical potential $\mu$ (or field). For a $U(1)$ particle-number symmetry, while the absolute phase of a single isolated condensate is not directly observable, the relative phase $\Delta\phi$ between two weakly coupled condensates is, with
$d\Delta\phi/dt=(\mu_1-\mu_2)/\hbar$, leading to AC Josephson oscillations.
Closely related ideas have been explored in proposals for time-crystalline behavior in superfluid systems~\cite{Wilczek2013SuperfluiditySpaceTime}.
However, it is not considered as ``genuine time crystals'' with the strict viewpoint, where one typically requires stability against generic perturbations that break any symmetry other than time translation~\cite{zaletel2023colloquium,else2020discrete}.
In Sec.~\ref{sec63}, we will also treat symmetry-protected time crystals separately.

\subsection{Time crystal in periodically driven systems}
\label{sec62}
\begin{figure}[t]
   \vspace*{0.7cm}
     \hspace*{0.2cm}
\includegraphics[bb=0mm 0mm 90mm 150mm,width=0.28\linewidth]{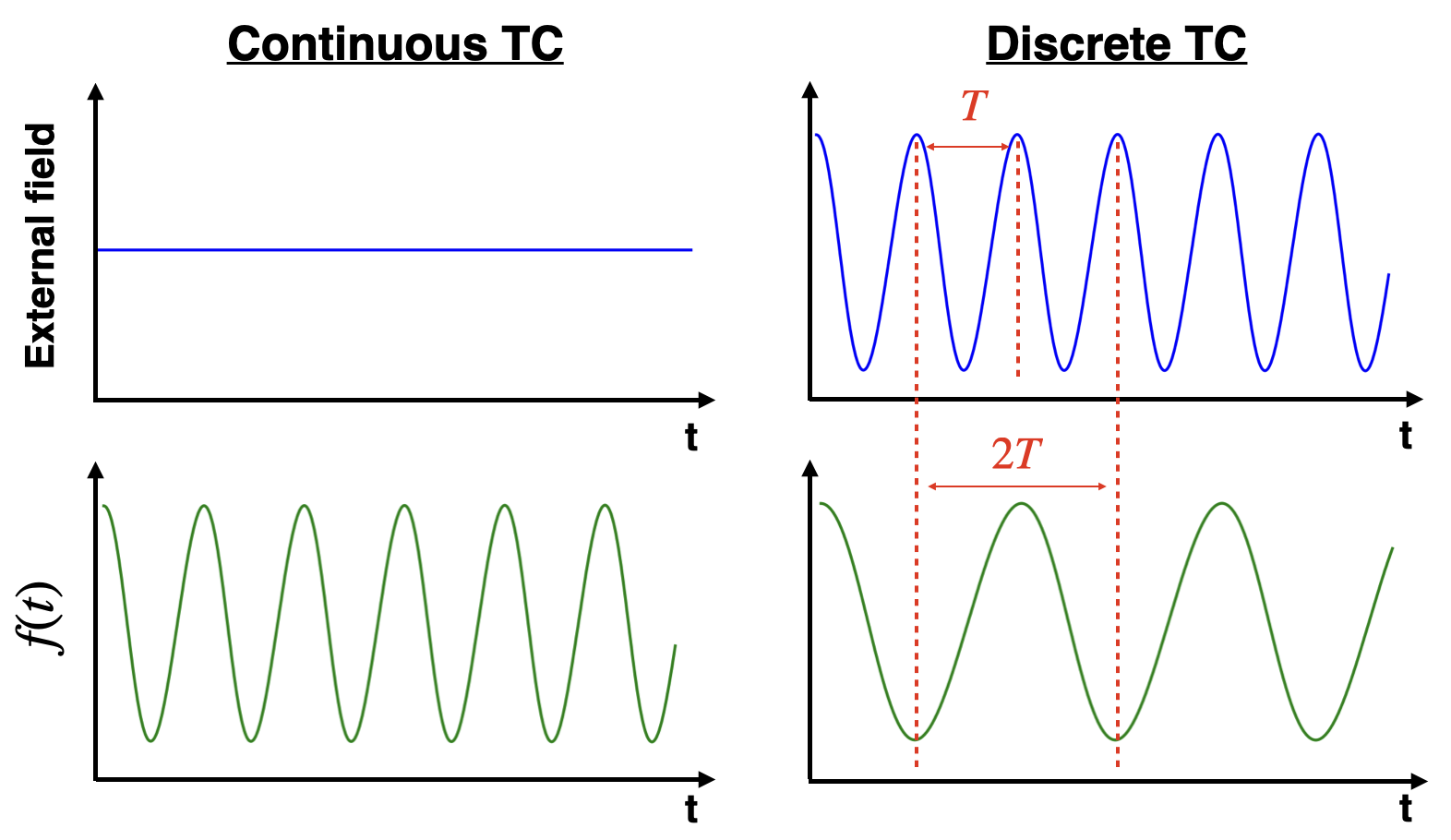}
\caption{\footnotesize\justifying Difference between continuous- and discrete-time crystals. The left column sketches a continuous-time crystal: the generator is time-independent (flat blue line), yet a correlation function or macroscopic observable (green) exhibits self-sustained oscillations with an intrinsic period, evidencing the breaking of continuous time-translation symmetry. The right column depicts a discrete-time crystal: the system is driven periodically with period $T$ (blue), but the observable oscillates with the doubled period $mT$ ($m=2$) (green); this subharmonic response signals spontaneous breaking of the discrete time-translation symmetry imposed by the drive.}
    \label{timecrystalref}
\end{figure}

While the no-go theorem rules out equilibrium CTCs in a broad class of closed systems, it does not apply to nonequilibrium steady states. This motivated the study of $T$-periodically driven (Floquet) systems
with $H(t{+}T)=H(t)$, leading to the notion of \emph{discrete-time crystals}~\cite{else2020discrete, zaletel2023colloquium, Khemani2019BriefHistory, sacha2015modeling, sacha2017time, huang2018clean, yao2017discrete, else2016floquet, Khemani2016PhaseStructure, Russomanno2017LMG_DTC, MunozArias2022_pSpin_AllToAll_DTC, Zhu2019DickeTC_NJP, Gong2018DTC_DickeCircuitQED, Khemani2017DefiningTimeCrystals, vonKeyserlingk2016AbsoluteStability, Campeny}. 
These are phases of matter in which the discrete time-translation symmetry \(t\!\mapsto\! t+T\) is spontaneously broken to \(t\!\mapsto\! t+mT\) with \(m>1\) and $m\in\mathbb{Z}$ (see Figure $\ref{timecrystalref}$).

A standard formal definition of a discrete-time crystal is based on the existence of spatiotemporal long-range correlations~\cite{vonKeyserlingk2016AbsoluteStability}. 
While this correlation function-based criterion provides a sharp theoretical characterization, it is rarely directly accessible in experiments; accordingly, it is customary in practice to adopt an alternative operational definition~\cite{Khemani2017DefiningTimeCrystals, huang2018clean, Campeny}.
For an integer $n$, let the stroboscopic expectation be defined as $O_n := \mathrm{Tr}[O(nT)\,\rho]$. The following three properties characterize a discrete-time crystal:
\begin{enumerate}
\renewcommand{\labelenumi}{(A-\arabic{enumi})}
\item \textbf{Time-translation symmetry breaking:} There exists an integer \(m>1\) such that \(O_{n+m}=O_n\) for all integers \(n\), while \(O_{n+1}\neq O_n\).
\item \textbf{Persistence:} The subharmonic oscillation persists for arbitrarily long times in the thermodynamic limit.
\item \textbf{Rigidity:} The subharmonic period $mT$ is fixed without fine-tuning of initial conditions and parameters, and is stable against small and \textit{local} perturbations; that is, the response frequency is locked to $2\pi/(mT)$.
\end{enumerate}

According to these criteria, signatures of discrete-time crystals have been experimentally observed on various platforms, including trapped ions~\cite{Zhang,kyprianidis2021observation}, nuclear spins~\cite{choi2017observation, rovny2018observation}, and other systems~\cite{frey2022realization, smits2018observation, zhang2022digital}. More detailed reviews of discrete-time crystals are provided in Refs.~\cite{else2020discrete,zaletel2023colloquium, Khemani2019BriefHistory}.


We finally comment on discrete-time crystals in collective-spin models~\cite{Russomanno2017LMG_DTC, MunozArias2022_pSpin_AllToAll_DTC, Zhu2019DickeTC_NJP, Gong2018DTC_DickeCircuitQED}. In such fully connected systems, which lack a notion of spatial distance, the definition in terms of spatiotemporal correlation function, is not directly applicable; instead we use the temporal correlation function of a macroscopic observable $M$ and the thermodynamic limit taken as $N\to\infty$ (number of spins) rather than $V\to\infty$. 
In such models, the collective observable $M$ is typically governed by mean-field dynamics, so that
\begin{align}
\braket{M(nT)M(0)}\approx M_nM_0,
\end{align}
and the nontrivial temporal structure is encoded in the one-point function $M_n:=\braket{M(nT)}$.
This type of time-crystalline behavior is characterized by criteria (A-1)–(A-3), with the stability requirement (A-3) taken in a weak form.
Specifically, it does not require robustness against generic short-range interactions that would invalidate the mean-field description. They are often called \textit{mean-field time crystals}~\cite{else2020discrete} and discussed separately from other time crystals. In the next section, we will also treat mean-field time crystals separately.

\subsection{Time crystals in open quantum systems}
\label{sec63}


Besides periodically driven systems, another extensively studied avenue is open-system realizations, where coupling to an environment can stabilize nonequilibrium long-time periodic dynamics. Such a phase breaks the continuous time-translation symmetry and is closer to the original concept of a time crystal.
We refer to CTCs in open systems as \textit{dissipative continuous-time crystals (DCTCs)}.

Based on this notion, a number of theoretical works have predicted the emergence of DCTCs in driven-dissipative collective spin systems~\cite{Iemini, Piccitto, dos}, collective spin systems outside the symmetric subspace~\cite{Solanki2024CTCOutsideSym}, bosonic~\cite{Minganti4, Lled2, cabot2024nonequilibrium, Seibold2020PRA, Lledo, Bakker, Li2024_TimeCrystal_SingleMode}, spin-boson~\cite{NieZheng2023PRA, Mattes2023PRA_EntangledTC}, atom-cavity~\cite{Kessler2019EmergentLimitCycles}, fermionic~\cite{Booker, Alaeian2022ExactMultistability}, spin-1/2 systems~\cite{Yang2025CommunPhys}, spin-1/2 system with long-range dissipation~\cite{Passarelli}, four-level systems~\cite{xiang2024self}, and spin-1 systems~\cite{wang2025boundary,RussoPohl2025PRL_QDCTC}. 
In addition, experimental realizations of DCTCs have been reported across atom-cavity systems~\cite{Kongkhambut, Kessler2020FromCTCtoDTC}, Rydberg-atom platforms~\cite{Wu}, semiconductor spin ensembles with electron–nuclear coupling~\cite{greilich2024robust}, microcavity exciton–polariton condensates~\cite{CarraroHaddad2024Science}, room-temperature Rydberg gases~\cite{Jiao2025NatCommun}, and spin gases~\cite{Huang2024SpinGases}. 




\subsubsection{Our criterion for DCTCs}

As mentioned above, a variety of models with persistent periodic oscillations have been termed time crystals. However, to our knowledge, there is no consensus definition of DCTCs, and any classification depends on the choice of robustness criteria, that is, on the precise notion of ``phase of matter'' one adopts. This review prioritizes organizing the robustness and fragility of oscillatory phenomena to some class of perturbations rather than attempting to settle the terminology. Building on this viewpoint, we first define DCTCs in a narrow sense (hereafter referred to as \emph{strict DCTCs}), and then introduce two classes with relaxed robustness requirements.

As a starting point, to accommodate extensions to effectively zero-dimensional settings that lack a notion of spatial distance (such as collective-spin and bosonic models), we adopt an operational criterion by analogy with the discrete-time crystal conditions (A-1)–(A-3). We then require strict DCTCs to satisfy the following three conditions (B-1)-(B-3). 

\begin{enumerate}
\renewcommand{\labelenumi}{(B-\arabic{enumi})}
\item \textbf{Time-translation symmetry breaking:} Under a time-independent generator, there exists at least one observable whose expectation value exhibits periodic oscillations.
\item \textbf{Persistence:} The periodic oscillation persists for arbitrarily long times in
the thermodynamic limit.
\item \textbf{Robustness:} The oscillations emerge without fine-tuning of initial conditions and parameters and are stable against small and \textit{local} perturbations. \footnote{In discrete-time crystals, such robustness (often referred to as rigidity) manifests itself as subharmonic locking to the external drive, whereas in continuous-time crystals without a drive it instead manifests itself as persistence of the oscillation rather than frequency locking.} 
\end{enumerate}
Throughout, the perturbations in (B-3) are understood as generic time-independent perturbations of the Lindbladian; no symmetry is assumed to be preserved, except for time-translation invariance.

Beyond this analogy to discrete-time crystals, we further wish to focus on oscillations that are genuinely induced by system–environment couplings, and therefore impose the additional condition (B-4).
\begin{enumerate}
\renewcommand{\labelenumi}{(B-\arabic{enumi})}
\setcounter{enumi}{3} 
\item \textbf{Dissipation-induced origin:}
Persistent oscillations must not be reducible to purely Hamiltonian dynamics on a dissipation-decoupled sector.
\end{enumerate}
In summary, we define a strict DCTC as any phase of matter that satisfies conditions (B-1)–(B-4).

However, in many proposals of open quantum systems, the full robustness condition (B-3) is not satisfied.
We therefore also define two subclasses that retain (B-1), (B-2), and (B-4) but impose a weaker form of robustness in (B-3):
\textit{mean-field DCTCs} and \textit{symmetry-protected DCTCs}.

\begin{enumerate}
\item Mean-field DCTCs

By relaxing condition (B-3) so that robustness is required only within a mean-field description
(i.e., we do not demand stability against perturbations that invalidate the mean-field approximation),
we define a \emph{mean-field DCTC}.

This class includes many proposed collective-spin or bosonic models without spatial structure. In the thermodynamic limit, these are captured by nonlinear mean-field dynamics with only a few degrees of freedom, and time-crystalline behavior often manifests itself either as stable limit cycles or as families of periodic orbits around center-type fixed points.

\item Symmetry-protected DCTCs

By relaxing (B-3) and demanding robustness only against perturbations that respect certain symmetries besides time-translation symmetry, we define a \emph{symmetry-protected DCTC}. 
This class includes periodic oscillations induced by strong dynamical symmetries~\eqref{sds}.
\end{enumerate}

In the following, we provide a more detailed explanation of conditions (B-1), (B-2), and (B-4), and then comment on condition (B-3) in relation to noise in experiments.

\subsubsection{Time-translation symmetry breaking and persistence}
We now specify what we mean by persistent periodic oscillations in open quantum systems.
Let us consider a system for which the thermodynamic limit is obtained as a parameter $N\to\infty$, in the same way as DPTs~\eqref{phasetransition}.
We diagnose persistent periodic oscillations through an
intensive order parameter (order $\mathcal{O}(1)$ in $N$) $O^{(N)}$:
\begin{align}
\label{timecrystal}
\lim_{N\to\infty}\mathrm{Tr}\ \!\big[O^{(N)}\rho(t,N)\,\big]= f(t),
\end{align}
where the function $f(t)$ is non-constant and satisfies a condition for a sufficiently late time $t_0$:
\begin{align}
\label{timecrystal2}
f(t) &= f(t + T),\ \ \text{for } \ t > t_0,
\end{align}
with the oscillation period $T$.

In the GKSL formalism, persistent periodic oscillations are reflected in characteristic features of the Lindbladian spectrum. 
Recalling the mode decomposition in Eq.~\eqref{timedepedensity}, nondecaying oscillatory contributions can persist at long times only if the relevant decay rates vanish; this corresponds to eigenvalues on the imaginary axis,
\begin{align}
\label{pie}
\lambda_{j}=i\omega_{j}\ \ \textrm{with}\ \ \omega_{j}\in\mathbb{R}.
\end{align}
Moreover, if one demands \emph{strictly periodic} (rather than quasiperiodic) long-time dynamics, the set of oscillation frequencies should be commensurate~\cite{Booker},
\begin{align}
\label{yuri}
\omega_{j}/\omega_{k}\in\mathbb{Q}\quad \forall\, j,k .
\end{align}
Accordingly, conditions~\eqref{pie} and~\eqref{yuri} have been widely used as spectral diagnostics for persistent periodic oscillations~\cite{Iemini, Piccitto, Lled2, Booker, Minganti4, NieZheng2023PRA, cabot2024nonequilibrium, KrishnaPRL2023_MICTC, Seibold2020PRA, Nakanishi2}.

\subsubsection{Dissipation-induced origin}
Since we regard DCTCs as intrinsically generated by system--environment coupling, we exclude situations in which the long-time oscillations are entirely reducible to unitary dynamics on a dissipation-decoupled sector. More precisely, we rule out cases where the dissipative part of the GKSL generator vanishes on the sector that supports the oscillations, so that the dynamics there reduces to $\hat{\mathcal{L}}_0[\cdot]=-i[H_d,\cdot]$, where $H_d$ is the Hamiltonian projected onto the dissipation-decoupled sector. 
A paradigmatic example is dynamics confined to a decoherence-free subspace, where the Lindblad dissipator vanishes on the relevant sector (see Sec.~\ref{sec32}); any purely imaginary Lindbladian eigenvalues then yield persistent oscillations determined solely by $H$. 

We also exclude oscillations for which the influence of dissipation disappears in a limiting procedure, for example, in the thermodynamic limit or in the vanishing-dissipation limit, so that the oscillatory sector becomes effectively decoupled from the environment.

\subsubsection{Robustness against noise}

While condition (B-3) formulates robustness against a specified class of perturbations of the generator, experimental stability is often also assessed operationally via robustness to noise.
Typical noise sources include white noise~\cite{Yang2025CommunPhys,Jiao2025NatCommun}, quantum noise~\cite{Kessler2019EmergentLimitCycles, Kongkhambut}, and temporal noise~\cite{Kongkhambut,Huang2024SpinGases,Wu}.
A convenient quantitative metric is \emph{relative crystalline fraction}, defined as the relative spectral weight at the frequency of CTCs without noise in the long-time Fourier spectrum; it remains near one for sufficiently small perturbation for DCTCs~\cite{Kongkhambut, Wu, Yang2025CommunPhys, Jiao2025NatCommun}.

\end{comment}

\subsection{L-$\mathcal{PT}$ symmetry and DCTCs}
\label{sec64}

As discussed in Sec.~\ref{sec5}, the mean-field theory of L-$\mathcal{PT}$ phase transitions predicts that the PT-symmetric phase supports persistent periodic dynamics of macroscopic observables, as characterized by Eqs.~($\ref{timecrystal}$), ($\ref{timecrystal2}$). 
In this section, we argue that these L-$\mathcal{PT}$-induced oscillations satisfy the persistence requirement and have a dissipation-induced origin in the sense of Sec.~6.3, while their robustness is generally weaker and depends on the class of perturbations considered (i.e., mean-field and symmetry-protected robustness).

\subsubsection{Time-translation symmetry breaking, persistence, and dissipation-induced origin}
First, we verify conditions (B-1) and (B-2). 
At the mean-field level, the PT-symmetric fixed point is a center, so the trajectories in its vicinity form closed orbits and yield persistent oscillations. In a quantum-trajectory picture, the GKSL evolution corresponds to an ensemble average over stochastic trajectories, which may explore multiple centers in the semiclassical phase space; strict periodicity therefore requires that their frequencies be mutually commensurate [Eqs.~\eqref{pie},~\eqref{yuri}], whereas incommensurate frequencies lead to quasi-periodic dynamics.


Furthermore, condition (B-4) is typically satisfied since their oscillation period generically depends on the dissipative rates Eq.~\eqref{period}, further distinguishing these dynamics from those that arise in decoherence-free subspaces~\eqref{dfs}. 

\subsubsection{Robustness}
Next, we discuss condition (B-3), that is, robustness against perturbations. For a class of perturbations that preserve the L-\(\mathcal{PT}\) symmetry and the total spin $S$, the mean-field dynamics exhibits the n-PT symmetry and persistent oscillations appear.
This class includes standard collective dissipators, incoherent pumping via \(m_{+}\), decay via \(m_{-}\), and dephasing via \(m_{\alpha}\) (\(\alpha=x,y,z\)), respect the L-\(\mathcal{PT}\) symmetry.

However, we remark that these oscillations are extremely fragile for a class of perturbations that explicitly violate the n-PT symmetry~\eqref{nonlinearPT} (and L-$\mathcal{PT}$ symmetry~\eqref{HuberPT}). 
For example, an infinitesimal longitudinal field proportional to $S_z$ suffices to destroy the behavior of the time crystal~\cite{Piccitto, dos}. (Similarly, noise that breaks the n-PT symmetry is expected to destabilize these oscillations.) Therefore, these are symmetry-protected oscillations. 

We further consider robustness against perturbations that break total-spin conservation but preserve the L-$\mathcal{PT}$ symmetry. As a minimal example, we add a nearest-neighbor Ising interaction
\begin{align}
H_{\rm ising}=J_{zz}\sum_{\langle ij\rangle}\sigma_i^z\sigma_j^z,
\end{align}
which satisfies $[H_{\rm ising},PT]=0$ with $PT=\prod_i\sigma_i^xK$. 
Hence, the L-$\mathcal{PT}$ symmetry is preserved microscopically.

Within mean-field approximation, assuming a spatially uniform magnetization $Z=\langle\sigma_i^z\rangle$, we decouple
\begin{align}
H_{zz}\approx 2J_{zz}z_c Z S_z - S z_c J_{zz}Z^2,
\end{align}
with the coordination number $z_c$, so the mean-field equations retain the n-PT symmetry and the PT-symmetric oscillations remain stable at this level. Note that it remains an open question whether they are stable against generic local perturbations that invalidate the mean-field description.

In contrast, for dissipation that breaks conservation of the total spin, local dissipation generically destroys the n-PT symmetry of the mean-field dynamics, even if the Lindbladian retains L-$\mathcal{PT}$ symmetry, as discussed in Sec.~\ref{sec571}. Consequently, these oscillations are fragile against such perturbations even in regimes where the mean-field approximation is otherwise valid and are stable only with respect to long-range dissipation that preserves the n-PT symmetry.

Hence, we conclude that oscillations arising from L-$\mathcal{PT}$ symmetry, when the frequencies of a stable PT-symmetric center are mutually commensurate, fall into the intersection of our mean-field and symmetry-protected DCTC classes. 
In other words, they require both weakenings of (B-3): they qualify as DCTCs only in the simultaneous sense of being mean-field \emph{and} symmetry-protected, rather than under either weakening alone. 



We note that, unlike symmetry-protected time crystals in isolated systems, where the oscillation can be viewed as motion along a spontaneously broken symmetry orbit generated by a conserved charge, the time-translation symmetry breaking here is not induced by such symmetry rotations. 

\subsection{Two-collective spin model with balanced gain and loss}
\label{sec65}
The advantage of a symmetry-based analysis is that its qualitative conclusions are often insensitive to microscopic details.
So far, we have mainly focused on a single-collective spin system; here, we turn to a two-collective spin model with $\mathrm{L}$-$\mathcal{PT}$ symmetry implemented by a distinct PT operator, whose Lindbladian spectrum can be obtained exactly (around the TISS) in the thermodynamic limit. 

This is achieved by combining the Holstein–Primakoff (HP) approximation~\cite{Holstein}, which maps collective spin operators to bosonic ones, with the technique of third quantization~\cite{Prosen2, Prosen}, which provides an exact solution to the Lindbladian eigenproblem for systems with quadratic Hamiltonians and linear Lindblad operators. This combination allows us to treat, beyond mean-field analysis, quantities such as Lindbladian eigenvalues, purity, and entanglement negativity, which can be analytically obtained. (Purity and negativity are treated in Sec.~\ref{sec72}.)


We note that Theorem 2 in Sec.~\ref{sec57} ensures that the nonlinear dynamical equation associated with this model respects the n-PT symmetry. However, the linear stability framework developed in Sec.~\ref{sec53} can not be applied since this model does not map bipartite bosonic systems via the Schwinger boson transformation [Eq.~\eqref{sb}]. 

\subsubsection{Model and Holstein-Primakoff mapping}
\begin{figure*}[t]
   \vspace*{0.4cm}
     \hspace*{0.1cm}
\includegraphics[bb=0mm 0mm 90mm 150mm,width=0.26\linewidth]{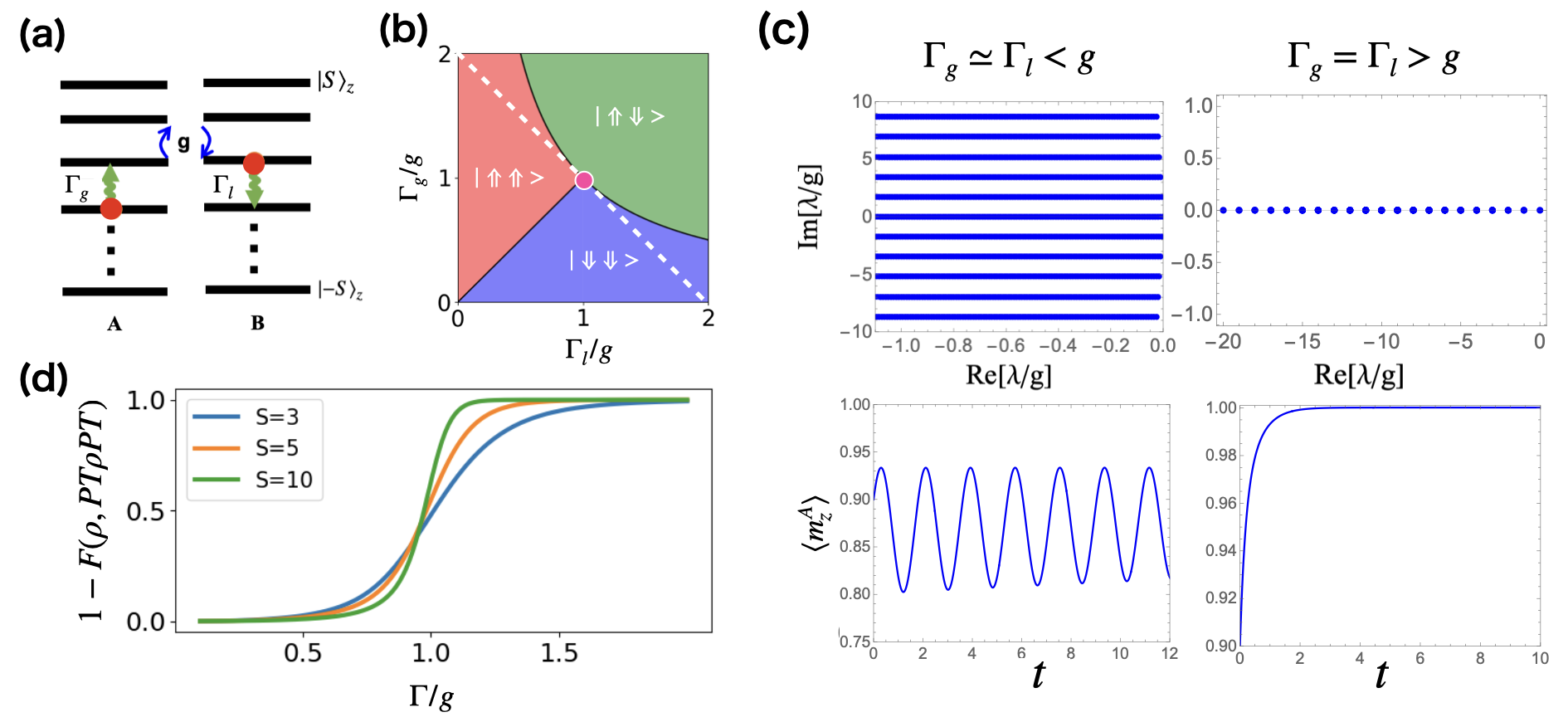}
\caption{\footnotesize\justifying (a) Schematic picture of the two-collective spin model with gain and loss ($\ref{2spindefinition}$). (b) Phase diagram: Red, blue, and green regions indicate the ferromagnetic (FM) phase with $Z_{A}=Z_{B}=1$, the FM phase with $Z_{A}=Z_{B}=-1$, the anti-ferromagnetic (AM) phase with $Z_{A}=1,\  Z_{B}=-1$, respectively. The dashed white line is a LEP, in particular, the tricritical point $\Gamma_{g}=\Gamma_{l}=g$ (pink circle) is a zero-mode LEP. (c) Lindbladian eigenvalues (around the TISS) and the time evolution of the normalized magnetization $\braket{m_{z}^{A}}(t)$. We set parameters as $\Gamma_{g}=\Gamma_{l}=2$, and $g=1$ in the AM (PT broken) phase and $\Gamma_{g}=0.5, \Gamma_{l}=0.495$, and $g=1$ in FM phase  (in the vicinity of the PT-symmetric line $\Gamma_{g}=\Gamma_{l}=\Gamma$)~\cite{Nakanishi}. (d) Infidelity between the TISS and PT-transformed one. 
}
    \label{figfig}
\end{figure*}

We consider a model comprising two collective spins, $A$ and $B$, coupled by an $XX$-type interaction and subject to asymmetric gain and loss~\cite{Huber1, Huber2, Nakanishi},
\begin{align}
\label{2spindefinition}
&H=\frac{g}{2S}(S^{A}_{+}S^{B}_{-}+ \textrm{H.c.}),\nonumber\\
L_{1}=&\sqrt{\frac{\Gamma_{g}}{2S}}S^A_+,\ \ \ L_{2}=\sqrt{\frac{\Gamma_{l}}{2S}}S^B_-.
\end{align}
Here, $g$ is the interaction strength, while $\Gamma_{g}$ and $\Gamma_{l}$ are the gain and loss rates, respectively. A schematic of the setup is shown in Fig.~$\ref{figfig}$ (a).
This model has the L-$\mathcal{PT}$ symmetry ($\ref{HuberPT}$) when the gain and loss rates are balanced, that is, $\Gamma_{g}=\Gamma_{l}$. Here, the parity operator $P$ is the exchange of the $A$ and $B$ subspaces. 
\YNedit{In contrast to the single-collective-spin case discussed above, this condition is a genuine tuning between two independent dissipators; any imbalance between the full Lindblad channels explicitly breaks the L-\(\mathcal{PT}\) symmetry.}

To solve the eigenproblem of this model, we use the HP transformation~\cite{Holstein}, which provides an exact mapping of the collective spin operators $S_{z},\ S_{\pm}$ into a bosonic mode with an annihilation operator $a$,
\begin{align}
\label{Hol}
S_{-}=\sqrt{2S-a^{\dagger}a}\ a,\ \ \ \ \ \ S_{z}=-S+a^{\dagger}a.
\end{align}
In the case where the state is the spin-down state $\braket{S_{z}}\simeq-S$ with $\braket{a^{\dagger}a}/(2S)\ll 1$, one can approximate the spin operators by
\begin{align}
\label{Hol2}
S_{-}\simeq\sqrt{2S}a,\ \ \ \ \ \ S_{z}=-S+a^{\dagger}a.
\end{align}
Applying the HP approximation to Eq.~($\ref{2spindefinition}$) yields a quadratic bosonic Hamiltonian with linear jump operators. In this case, second-order observables (e.g.~$\braket{a^{\dagger}a}$) can be evaluated in a closed form. These analytic expressions directly yield the steady-state phase boundaries; the resulting phase diagram is presented in Fig.~$\ref{figfig}$ (b)~\cite{Huber1,Nakanishi}. The red (blue) region denotes the ferromagnetic (FM) phase with spin-up $\ket{\Uparrow\Uparrow}$ (spin-down $\ket{\Downarrow\Downarrow}$) state, and the green region denotes the antiferromagnetic (AM) phase with spin-up and spin-down $\ket{\Uparrow\Downarrow}$ state. 



\subsubsection{Lindbladian spectrum from third quantization}
We next apply the third quantization to obtain the Lindbladian eigenvalues near the TISS. A detailed description of the third quantization formalism is provided in Appendix~\ref{Thirdquantization}.
Applying this method to the FM $\ket{\Uparrow\Uparrow}$ phase (the red region of Fig.~$\ref{figfig}$ (b)), we obtain the entire Lindbladian spectrum (around the TISS) as
$\lambda^{\mathrm{FM}}=-2[(m_{1}+m_{2})\beta_{+}+(m_{3}+m_{4})\beta_{-}]$,
with $m_i\in\mathbb{N}_0$ and 
\begin{align}
\label{betaFM}
\beta^{\mathrm{FM}}_{\pm}=\frac{1}{4}\left(\Gamma_g-\Gamma_l\pm\sqrt{(\Gamma_g+\Gamma_l)^2-4g^2}\right).
\end{align} 
Here, $\beta^{\mathrm{FM}}_{\pm}$ are the eigenvalues of the matrix 
\begin{align}
\label{FMX}
\textbf{X}=\frac{1}{2}\left(
    \begin{array}{rrrr}
     \Gamma_g & ig&0&0\\
      ig& -\Gamma_l &0&0\\
   0 & 0&\Gamma_g&-ig\\
      0&0&-ig& -\Gamma_l 
    \end{array}
  \right).
\end{align}
Similarly, in the FM $\ket{\Downarrow\Downarrow}$ phase, $\beta^{\mathrm{FM}}_{\pm}$ are obtained from Eq.~\eqref{betaFM} by interchanging $\Gamma_g$ and $\Gamma_l$.

In the limit to the AM phase $\Gamma_{g}\Gamma_{l}\to g^{2}$,
$\beta^{\mathrm{FM}}_{\pm}\to\frac{1}{2}|\Gamma_g-\Gamma_l|,\ 0$,
indicating that the Lindbladian gap is closed at the phase boundary. \footnote{We have analyzed the eigenvalues near the TISS (stable fixed point). According to Ref.~\cite{ferreira2019lipkin}, the full spectrum of the Lindbladian generally includes contributions not only from stable fixed points but also from unstable ones. However, when focusing on the behavior near the steady state, it is sufficient to consider only the influence of stable fixed points. 
At a phase boundary, the spectral structure may reflect the features of both phases.} 
Furthermore, in the limit $|\Gamma_{g}-\Gamma_{l}|\to 0$,
\begin{align}
\label{purely}
\beta^{\mathrm{FM}}_{\pm}\to\pm \frac{i}{2}\sqrt{g^{2}-\Gamma^{2}},
\end{align}
showing that the Lindbladian gap is closed and an infinite number of PIEs appear. As a result, one can obtain the eigenvalue structure for $\Gamma_g\simeq\Gamma_l$ as shown in Fig.~$\ref{figfig}$ (c). 

On the line $\Gamma_g/g+\Gamma_l/g=2$, the matrix $\bf X$ ($\ref{FMX}$) is non-diagonalizable, that is, an LEP emerges (the white dashed line in Fig.~$\ref{figfig}$ (b)). In particular, the tricritical point,  $\Gamma_g/g=\Gamma_l/g=1$, becomes a zero-mode LEP (pink circle in Fig.~$\ref{figfig}$ (b)).

Applying a similar analysis to the AM phase, we find that the matrix $\bf{X}$ is Hermitian as
\begin{align}
\label{XX}
\textbf{X}=\frac{1}{2}\left(
    \begin{array}{rrrr}
     \Gamma_g & 0&0&-ig\\
      0& \Gamma_l &-ig&0\\
   0 & ig&\Gamma_g&0\\
      i g &0&0& \Gamma_l 
    \end{array}
  \right),
\end{align}
with eigenvalues 
\begin{align}
\label{betaAM}
\beta^{\mathrm{AM}}_{\pm}=\frac{1}{4}\left(\Gamma_g+\Gamma_l\pm\sqrt{(\Gamma_g-\Gamma_l)^2+4g^2}\right).
\end{align}
This indicates that all Lindbladian eigenvalues are real, as shown in Fig.~$\ref{figfig}$ (c). This means that physical quantities exponentially decay to a unique steady state without oscillation.
In the limit $g^{2}\to\Gamma_{g}\Gamma_{l}$,
$\beta^{\mathrm{AM}}_{\pm}\to\frac{1}{2}(\Gamma_g+\Gamma_l),\ 0$, showing that the Lindbladian gap is closed. In this phase, (i) LEP does not appear even in the limit to a tricritical point from the AM ordered phase, (ii) the steady state is a dark state, (iii) the relaxation time scales linearly with the dissipation rate, and (iv) is unique and gapped. These behaviors coincide with the PT-broken phase for the dissipative LMG model discussed in Section~\ref{sec56}.

Thus, even for the two-collective spin model, the DPT features discussed in Sec.~\ref{sec5} appear.
\YNedit{In particular, persistent oscillations are protected by L-\(\mathcal{PT}\) symmetry and are not robust against symmetry-breaking perturbations.
We remark that even if the gain--loss balance condition is violated, as long as the imbalance is small, \(|\Gamma_g - \Gamma_l| = \epsilon\), Eq.~\eqref{betaFM} near the phase boundary gives a decay rate of order \(\epsilon\), so the relaxation time \(\tau \sim 1/\epsilon\) remains sufficiently long.
Therefore, the exact transition in this two-spin realization requires fine tuning of the gain--loss balance, while a small residual imbalance mainly cuts off the ideal persistent dynamics on the long time scale set by \(1/\epsilon\), so oscillations originating from L-\(\mathcal{PT}\) symmetry should still be experimentally observable.}

\section{Quantum and Statistical properties for Lindbladian $\mathcal{PT}$ phase transitions}
\renewcommand{\theequation}{7.\arabic{equation} }
\setcounter{equation}{0}
\label{sec7}
We here investigate the quantum and statistical properties beyond the mean field theory, most notably the PT symmetry breaking, purity, and quantum entanglement indicators in the TISS $\rho_{ss}(S) = \lim_{t \to \infty} \rho(t, S)$. 
This state in the time crystal phase is known not to be captured by the mean-field approximation, even in the large-spin limit, as evidenced by inequality $\braket{S_\alpha^2} \neq \braket{S_\alpha}^2$~\cite{huber2021phase, da2023sufficient}.

In Sections~\ref{sec71} and~\ref{sec72}, we examine two representative cases: the DDM ($\ref{DDM}$) and the two-collective-spin model with balanced gain and loss, defined in Eq.~($\ref{2spindefinition}$), respectively.

\subsection{Driven Dicke model}
\label{sec71}
\subsubsection{PT symmetry of the TISS}

\noindent To explore the role of the L-$\mathcal{PT}$ symmetry ($\ref{HuberPT}$) beyond the mean-field theory, we first analyze the PT symmetry of the TISS. For the DDM, the TISS is unique for any finite $S$ and has been solved exactly~\cite{Puri} as
\begin{align}
\label{ss}
\rho_{ss}(S)=\frac{1}{D}\sum_{n,n^{\prime}=0}^{2S}\left(i\frac{\kappa}{g}\frac{S_{-}}{S}\right)^{n^{\prime}}\left(-i\frac{\kappa}{g}\frac{S_{+}}{S}\right)^{n},
\end{align}
where $D$ is the normalization constant given by, 
\begin{align}
D=\sum_{m=0}^{2S}\frac{(2S+m+1)!(m!)^{2}}{(2S-m)!(2m+1)!}\left|\frac{ig S}{\kappa}\right|^{-2m}.
\end{align}
With this exact expression, the symmetry properties of the TISS can be evaluated explicitly. 
Remarkably, we find that the TISS $\rho_{\mathrm{ss}}(S)$ restores the PT symmetry
in the large-spin limit for $\kappa/g<1$~\cite{Nakanishi2}.
More precisely, the PT-transformed steady state approaches the original one as
\begin{equation}
  \lim_{S\to\infty}\ 
  \bigl\|
    \rho_{ss}(S)
    - {PT}\,\rho_{{ss}}(S)\,({PT})^{-1}
  \bigr\|
  = 0,
  \label{eq:PT_TISS}
\end{equation}
where $\|\cdot\|$ denotes the operator norm,
\begin{equation}
  \|X\|
  := \sup_{\|\psi\|=1} \|X\psi\|,
\end{equation}
and $PT= \prod_i\sigma_i^x K$.

The proof is based on the fact that as $S\to\infty$, the following commutators vanish:
\begin{align}
\label{comDDM}
\lim_{S\to\infty}\left\|\ \biggl[\Bigl(-i\frac{\kappa}{g}\frac{S_{+}}{S}\Bigr)^{n}, \Bigl(i\frac{\kappa}{g}\frac{S_{-}}{S}\Bigr)^{n^{\prime}}\biggr]\ \right\|=0,
\end{align}
for $0\leq n,\ n^{\prime}\leq 2S$. This follows directly from the standard SU(2) commutation relations of the collective spin operators. In contrast, when the dissipation rate crosses the critical value ($\kappa/g>1$), the TISS loses the PT symmetry even in the thermodynamic limit, $\lim_{S\to\infty}\|\rho_{ss}(S)- {PT}\,\rho_{{ss}}(S)\,({PT})^{-1}\|\neq0$. The detailed proof has been shown in Ref.\cite{Nakanishi2}.  

To confirm it numerically, we evaluate the quantity $|\rho_{ss}-PT\rho_{ss}(PT)^{-1}|$, where $|\rho|$ means that the matrix takes the absolute value of each element of the matrix $\rho$.
Fig.~\ref{DDMfig} (j) shows $|\rho_{ss}-PT\rho_{ss}(PT)^{-1}|$ in the (left) PT-symmetric phase and (right) PT-broken phase. In the PT phase, all the elements of the difference are nearly zero, indicating that the TISS is invariant under the PT transformation. In contrast, some elements have finite values in the PT-broken phase, consistent with expectations in the large-$S$ limit.

It is important to stress that the PT symmetry of the TISS \eqref{ss} is nontrivial.
For ordinary symmetries, one usually has a linear (anti-)commutation relation between the generator of the dynamics and a symmetry operator [e.g., Eqs.~\eqref{strong2}, \eqref{weak}, \eqref{antilpt}].
Such a relation block-diagonalizes the eigenproblem according to the symmetry and ensures that each eigenmode is an eigenstate of the symmetry operator, or else breaks the symmetry spontaneously.
By contrast, the L-$\mathcal{PT}$ symmetry \eqref{HuberPT} cannot, in general, be written as a commutator or anti-commutator with any linear superoperator.
As a result, there is no a priori reason for the eigenmodes to respect the PT symmetry.


In Appendix~\ref{imagptpt}, we present numerical evidence that the PT-symmetry of the right eigenmodes whose eigenvalues have a nonzero imaginary part is restored in the thermodynamic limit, that is, $\lim_{S\to\infty}PT\rho_i(PT)^{-1} {\propto}\ \rho_i$. This is consistent with the result that the periodic oscillation around the center [Eq.~\eqref{zt}] is n-PT symmetric.

\subsubsection{Purity}
Next, we investigate the purity, ${\rm Pur}[\rho]=\operatorname{Tr}[\rho^{2}]$, which quantifies how mixed the density matrix is.
Figure $\ref{DDMfig}$ (d) shows that, as $S$ increases, the purity tends to zero throughout the PT-unbroken phase, but rises toward unity once the PT symmetry is broken. This behavior indicates a transition from a highly mixed state, characterized by ${\rm Pur}[\rho]\sim 1/d$, where $d$ is the Hilbert space dimension of Hamiltonian, to the pure state in the thermodynamic limit. 


Although how the L-$\mathcal{PT}$ symmetry \eqref{HuberPT} constrains the purity in the TISS is not yet fully understood, 
it suffices to show that the maximally mixed (infinite-temperature) state $\rho\propto\1$ remains linearly stable in the weak dissipation limit~\cite{Huber2}.


Using the L-$\mathcal{PT}$-symmetric form of the Lindbladian~Eq.\eqref{PTsymm}, we start by expanding the density operator on the energy eigenbasis of~$H_{PT}$,
\begin{equation}
  \rho
  \;=\;
  \sum_{n,m}\rho_{n,m}\,
  \lvert E_n\rangle\langle E_m\rvert,
  \label{eq:rho_decomp}   
\end{equation}
where $H_{PT}\lvert E_n\rangle = E_n\lvert E_n\rangle$. Because $[H_{PT},PT]=0$, $\lvert E_n\rangle$ can be chosen as parity-time
eigenstates, 
\begin{equation}
  PT\lvert E_n\rangle = \zeta_n\lvert E_n\rangle,
  \qquad
  |\zeta_n|=1.
\end{equation}

For zero dissipation rates, any \emph{diagonal} $\rho$ is stationary under the dissipationless
Lindbladian $\mathcal L_0\rho= -i[H_{PT},\rho]$ in~Eq.\eqref{PTsymm}.  Let us therefore set
$\rho_{n,m} = \delta_{n,m}/d + \delta\rho_{n,m}$ and evaluate the evolution
of $\delta\rho_{n,m}$ to first order in the dissipation rates $\gamma_\mu$ (using
$L_\mu=\mathcal{O}(\sqrt\gamma_\mu)$)
\begin{align}
  \dot{\delta\rho}_{n,m}
  \;&=\;
  -i(E_n-E_m)\rho_{n,m}
  \nonumber\\
      &+\sum_\mu\frac{2}{d}\,
    \langle E_n\rvert\!\
      \bigl[
        L_\mu,L_\mu^{\dagger}
      \bigr]+
      \bigl[
        \mathbb{PT}(L_\mu),\mathbb{PT}(L_\mu^{\dagger})
      \bigr]\
    \lvert E_m\rangle.
  \label{eq:delta_rho_dot}   
\end{align}

Focusing first on coherences between non-degenerate levels $E_n\neq E_m$,
their evolution is given by
\begin{align}
  \delta\rho_{n,m}(t)
  \;&\simeq\;
  -\,
  \frac{2i}{d(E_n-E_m)}\Bigl[1-e^{-i(E_n-E_m)t}\Bigr]\nonumber\\
  &\times\sum_\mu
  \langle E_n\rvert\ 
    \bigl[
        L_\mu,L_\mu^{\dagger}
      \bigr]+
      \bigl[
        \mathbb{PT}(L_\mu),\mathbb{PT}(L_\mu^{\dagger})
      \bigr]\
  \lvert E_m\rangle,
  \label{eq:delta_rho_coh}      
\end{align}
hence $|\delta\rho_{n,m}|\to0$ as $\gamma_\mu\to0$.

Next we focus on the diagonal sector with $E_n = E_m$.
We assume that the Hamiltonian is nondegenerate. 
\footnote{For the degenerate case, coherences can build up unless every symmetry $A$ commuting with $H_{PT}$ also satisfies $[PT,A]=0$~\cite{Huber2}. Otherwise, states of equal energy but opposite parity ($\zeta_n\neq\zeta_m$) can destabilize the infinite-temperature state.}
In this case, the coherent part vanishes and Eq.~($\ref{eq:delta_rho_dot}$)
reduces to
\begin{equation}
  \dot{\delta\rho}_{n,n}
  \;=\;
  \frac{2}{d} \sum_\mu\,
    \langle E_n\rvert\!\
      \bigl[
        L_\mu,L_\mu^{\dagger}
      \bigr]+
      \bigl[
        \mathbb{PT}(L_\mu),\mathbb{PT}(L_\mu^{\dagger})
      \bigr]\ 
    \lvert E_n\rangle.
  \label{eq:delta_rho_diag1}   
\end{equation}
Since $\mathbb{PT}(L_\mu)=PTL_\mu^\dagger(PT)^{-1}$, we obtain
\begin{align}
  &\langle E_n\rvert\ \bigl[
        \mathbb{PT}(L_\mu),\mathbb{PT}(L_\mu^{\dagger})
      \bigr]\lvert E_n\rangle\notag\\
  &= \langle E_n\rvert PT
       \bigl[
        L_\mu^\dagger,L_\mu
      \bigr](PT)^{-1}
     \lvert E_n\rangle
  = \langle E_n\rvert 
       \bigl[
        L_\mu^\dagger,L_\mu
      \bigr]
     \lvert E_n\rangle
     \notag\\
  \label{eq:comm_rel}   
\end{align}
so that $\dot{\delta\rho}_{n,n}=0$.
Thus, the maximally mixed state $\rho\propto\1$ is stable in the presence of a small amount of dissipation.

\subsubsection{Spin squeezing}



\noindent Next, we investigate the spin squeezing, which represents the reduction of fluctuations of a collective-spin component orthogonal to the mean-spin direction. We use the Kitagawa-Ueda (KU) squeezing parameter~\cite{kitagawa1993squeezed,ma2011quantum} as a measure of spin squeezing:
\begin{align}
\label{wk}
  \xi^{2}_{\rm KU}
  =\min_{\mathbf n_{\!\perp}}
    \frac{2\,(\Delta S_{\mathbf n_{\!\perp}})^{2}}
         {S},  
\end{align}
with
\[
  (\Delta S_{\mathbf n_{\!\perp}})^{2}
    \equiv
    \bigl\langle S_{\mathbf n_{\!\perp}}^{2}\bigr\rangle
    -\bigl\langle S_{\mathbf n_{\!\perp}}\bigr\rangle^{2},
\qquad
  S_{\mathbf n_{\!\perp}}
    =\mathbf S\!\cdot\!\mathbf n_{\!\perp}.
\]
Here we denote the collective spin by
\(
\mathbf{S}=(S_x,S_y,S_z)
\),
The unit vector
\(
\langle \mathbf{S}\rangle/|\langle \mathbf{S}\rangle|
\)
sets the mean–spin direction, and the minimization below is taken over all unit vectors
\(
\mathbf{n}_{\perp}
\)
orthogonal to \(\langle \mathbf{S}\rangle/|\langle \mathbf{S}\rangle|\).
A value \(\xi^{2}_{\rm KU}<1\) certifies spin squeezing. 

For the present model, the squeezing parameter has a closed-form expression in the thermodynamic limit~\cite{lee2014dissipative,carollo2022exact, Buonaiuto, BarberenaRey2024PRA}:
\begin{align}
\label{exxi}
\lim_{S\to\infty}\xi^{2}_{\rm KU}
=\sqrt{\,1-\frac{g^{2}}{\kappa^{2}}\,},
\qquad
(\kappa>g).
\end{align}
Hence \(\xi^{2}_{\rm KU}<1\) throughout the ordered (PT-broken) phase and
\(
\xi^{2}_{\rm KU}\to 0
\)
as \(\kappa\to g\), indicating the enhancement of spin squeezing.

Figure~\ref{DDMfig}(e) plots the KU spin squeezing parameter for several finite total spins \(S\) in the PT-broken phase.
For every \(S\) considered, \(\xi^{2}<1\), confirming that the steady state is squeezed.
Moreover, \(\xi^{2}\) drops sharply and approaches zero near the transition point as \(S\) increases, consistent with the thermodynamic prediction of vanishing squeezing parameter.

The other criterion of Wineland is defined by
\begin{align}
\label{xiw}
\xi^{2}_{\rm W}
=\frac{2S\,(\Delta S_{\mathbf{n}_{\perp}})^{2}}{|\langle \mathbf{S}\rangle|^{2}}.
\end{align}
When the Bloch vector attains its maximal length
\(
|\langle \mathbf{S}\rangle|=S
\),
Eq.~\eqref{wk} coincides with \(\xi_{\rm W}^{2}\)~\cite{wineland1992spin}. 
The spin squeezing parameter $\xi^2_{\rm W}$ is known to be an entanglement witness for which all separable states satisfy \(\xi_{\rm W}^{2}\!\ge\!1\).
More generally, a state is \emph{\(k\)-producible} if it is a convex mixture of pure states that factorize into blocks of size at most \(k\); any \(k\)-producible state obeys \(\xi_{\rm W}^{2}\!\ge\!1/k\), so observing \(\xi_{\rm W}^{2}\!<\!1/k\) certifies entanglement depth \(>\!k\) (i.e., at least \((k\!+\!1)\)-partite entanglement)~\cite{SorensenMolmer2001,Hyllus2012,Toth2012, ma2011quantum}. 

In the PT-broken phase, the steady state is unique, and in the thermodynamic limit the mean-field description becomes exact for uncorrelated initial conditions, which implies that
\(
\lim_{S\to\infty}{|\langle \mathbf{S}\rangle|}/{S}=1.
\)
Therefore, approaching the critical line from the PT-broken phase, $\xi_{\rm KU}^{2}$ drops to zero, so \eqref{exxi} certifies the build-up of quantum entanglement near criticality.

For the DDM, other entanglement indicators (including both entanglement monotones and entanglement witnesses), such as negativity~\cite{Hannukainen} and QFI~\cite{Cabot, Montenegro}, have also been analyzed. In these works, quantum entanglement, as quantified by these indicators, exhibits pronounced enhancements at the transition point or throughout the PT-broken phase~\cite{Hannukainen, Cabot, Montenegro}.




An enhancement of the QFI in the vicinity of a critical point enables precision beyond the standard quantum limit, provided that the parameter generator, probe preparation, and measurement readout are suitably optimized.
These criticality-induced enhancements make quantum critical systems a resource for sensing and metrology~\cite{Cabot,Montenegro,cabot2025quantum} (the cited works do not discuss connections to the PT symmetry).

\subsection{Two-collective spin model with balanced gain and loss}
\label{sec72}
We analyze the PT symmetry, purity, and a quantum entanglement indicator in the two–collective-spin model with balanced gain and loss [Eq.~\eqref{2spindefinition}], and test whether these observables exhibit the same qualitative behavior as in the DDM.

\subsubsection{PT symmetry of TISS}
For DDM, the exact solution of TISS has been obtained~\cite{Puri}, but is not available in most cases. 
To investigate PT symmetry breaking in the TISS, the symmetry parameter~\cite{Huber2} and the PT symmetry parameter~\cite{Nakanishi2} have been investigated. Here, instead of these parameters, we investigate infidelity $1-F(\rho_{\rm ss},\rho_{\rm ref})$, defined as
\begin{align}
\label{QPTfide}
F(\rho_{\rm ss},\rho_{\rm ref})=\mathrm{Tr}\!\sqrt{\sqrt{\rho_{\rm ss}}\ \rho_{\rm ref}\sqrt{\rho_{\rm ss}}},
\end{align}
measuring how distinguishable two quantum states are, where we choose the reference state as $\rho_{\rm ref}=PT\rho_{ss}(PT)^{-1}$.


Fig.~$\ref{figfig}$ (d) shows the numerical analysis of the infidelity in the TISS for the two-collective spin model ($\ref{2spindefinition}$). This shows that the infidelity is close to 0 for $\Gamma_g=\Gamma_l<g$, while it is close to 1 for $\Gamma_g=\Gamma_l>g$. 
Moreover, as $S$ increases, the infidelity approaches 0 (1) in the PT-symmetric (PT-broken) regime, indicating sharpening in the thermodynamic limit.

\subsubsection{Purity and negativity}
Within the HP approximation, the TISS in the ordered phases is Gaussian and is fully characterized by the two-point covariance matrix. Therefore quantities such as the purity and the entanglement negativity, an entanglement indicator, $\mathcal{N}:=(||\bar{\rho}||_{1}-1)/2$, where $\bar{\rho}$ is the partial transpose of $\rho$ and $||X||_{1}=\textrm{Tr}\sqrt{X^{\dagger} X}$, can be evaluated \emph{exactly}~\cite{Huber1}. Here, the bipartition denotes the partial transpose with respect to subsystem \(B\).
In particular, on the PT-symmetric line $\Gamma_{g}=\Gamma_{l}=\Gamma$, these quantities are given by 
\begin{align}
\textrm{Pur}[\rho]=\begin{cases}
      1-(g/\Gamma)^{2}\ \ \ \ \ (\Gamma>g) \nonumber\\
      0\ \ \ \ \ (\Gamma<g),
   \end{cases}
 \end{align}
 and 
 \begin{align}
\mathcal{N}=\begin{cases}
      g/2\Gamma\ \ \ \ \ (\Gamma>g) \nonumber\\
      0\ \ \ \ \ (\Gamma<g),
   \end{cases}
 \end{align}
in the thermodynamic limit. As in the DDM, at the transition point, the TISS changes from a highly mixed phase to a more pure and entangled one. 

Thus, L-$\mathcal{PT}$-symmetric models exhibit common statistical and quantum signatures, which change dramatically at the transition point. It is worth highlighting that almost all of these signatures have not been directly derived from the definition of the L-$\mathcal{PT}$ symmetry ($\ref{HuberPT}$) itself. Determining whether these properties are intrinsic consequences of L-$\mathcal{PT}$ symmetry or instead arise from other ingredients (continuous time-translation symmetry breaking, permutational symmetry, or the presence of CEP) remains an important challenge for future work.




\section{Conclusion and Future outlook}
\label{sec8}
The L-$\mathcal{PT}$ phase transition is a genuinely nonequilibrium phase transition of the GKSL generator, characterized by changes in the steady state and the Lindbladian spectrum, and is, in general, beyond what effective non-Hermitian Hamiltonians capture.

\medskip
\noindent
The central findings are distilled as follows:
\begin{itemize}

\item \textbf{Mean-field theory}.  
When the microscopic GKSL generator satisfies the L-$\mathcal{PT}$ symmetry ($\ref{HuberPT}$), the resulting nonlinear mean-field equations inherit the n-PT symmetry ($\ref{nonlinearPT}$) for a broad class of models. Linear stability analysis then reveals DPTs from a dynamical phase with persistent periodic motion to a symmetry-broken gapped phase, associated with spontaneous breaking of the n-PT symmetry and typically occurring at a CEP, in single-collective--spin models, spin systems with long-range dissipation, and spatially extended bipartite bosonic lattices with a conserved $U(1)$ charge.

\item \textbf{Connection to non-reciprocal phase transitions}.
At the mean-field level, the symmetry structure underlying L-$\mathcal{PT}$ phase transitions and its spontaneous breaking is conceptually similar to that in standard continuous non-reciprocal phase transitions, yet the underlying physical contexts are essentially different.
In particular, L-$\mathcal{PT}$ transitions involve the breaking of the n-PT symmetry, whereas non-reciprocal transitions are characterized by the breaking of a nonlinear anti-PT symmetry (in our notation), leading to distinct dynamical consequences.

For L-$\mathcal{PT}$ phase transitions, PT-symmetric fixed points are centers, implying persistent closed orbits with periodic motion, and PT-broken fixed points appear as stable-unstable pairs rather than as a degeneracy of steady states.
By contrast, in non-reciprocal systems, the anti-PT-symmetric phase typically has a unique stable fixed point, while the broken phase often exhibits two stable limit cycles.

\item \textbf{DCTCs originating from the $\mathcal{PT}$ symmetry}. 
The oscillations arising in the mean-field dynamics induced by L-$\mathcal{PT}$ symmetry, when the frequencies of a stable PT-symmetric center are mutually commensurate, lie in the intersection of our mean-field and symmetry-protected DCTC classes.
Importantly, this behavior relies on symmetries beyond time-translation symmetry, namely, L-$\mathcal{PT}$ symmetry (n-PT symmetry), and it remains an open question whether it survives generic local perturbations that generate fluctuations beyond the mean-field description. 


\item \textbf{Statistical and quantum properties}.
Quantities such as purity and quantum entanglement indicators exhibit dramatic changes at L-\(\mathcal{PT}\) phase transition points. The TISS in the models studied here is highly mixed for the disordered (PT-unbroken) phase, and it is highly entangled throughout the ordered (PT-broken) phase and at the critical point in the thermodynamic limit.

\end{itemize}

\medskip
\noindent
Finally, we comment on the future prospects for L-$\mathcal{PT}$ phase transitions as follows:

\begin{itemize}
\item \textbf{Construction of spectral theory}.  
The development of a spectral theory for L-$\mathcal{PT}$ phase transitions remains a major challenge, serving as an analogue to the DPTs with unitary symmetry breaking discussed in Section~\ref{sec33}. A crucial step would be to derive, starting from the L-$\mathcal{PT}$ symmetry, general properties such as the emergence of purely imaginary Lindbladian eigenvalues and the PT-symmetry breaking of the TISS. If such spectral criteria can be fully established, they may offer a systematic way to characterize DPTs involving anti-unitary symmetry breaking at the microscopic level.

\item \textbf{Analysis of critical phenomena beyond mean-field theory}.
We have developed the mean-field theory of L-$\mathcal{PT}$ phase transitions.
The next frontier is to investigate their critical behavior beyond mean-field theory, fully accounting for fluctuations and spatial degrees of freedom. Employing the Keldysh functional formalism together with nonequilibrium renormalization group techniques would allow us to clarify universal properties and critical behavior~\cite{kamenev2023field, marino2016driven, sieberer2013dynamical, sieberer2016keldysh, sieberer2023universality}. 

\item \textbf{Exploration of non-trivial phenomena in the vicinity of CEPs}.
At the classical level, several nontrivial phenomena have already been reported in the vicinity of CEPs, including giant fluctuations~\cite{Hanai2}, a divergent entropy-production rate~\cite{suchanek2023entropy}, and fluctuation-induced first-order transitions~\cite{zelle2024universal}. A natural next question is whether CEPs also leave signatures in quantum observables; Sec.~\ref{sec7} reports critical behavior in entanglement indicators, motivating an examination of whether these features are directly tied to CEPs.

\item \textbf{Implementation}. 
Driven–dissipative collective-spin systems closely related to the DDM have already been realized in free-space cold-atom gases~\cite{Ferioli,goncalves2024driven} and in cavity-QED implementations of cooperative resonance fluorescence~\cite{Song2025SciAdv}. 
However, to our knowledge, the L-$\mathcal{PT}$ phase transition itself has not yet been observed experimentally. 
Beyond these DDM-related realizations, several other quantum simulation platforms provide complementary ingredients for its realization: all-to-all (or long-range) spin interactions have been demonstrated in trapped-ion chains~\cite{zhang2017observation}, Rydberg atom arrays~\cite{bernien2017probing,guardado2021quench}, and hybrid superconducting circuits~\cite{mlynek2014observation}. Collective decay has also been engineered in solid-state hybrid devices~\cite{angerer2018superradiant}.
These advances make dissipative collective-spin models experimentally accessible and suggest that observing L-$\mathcal{PT}$ phase transitions in such platforms may be within reach.

\end{itemize}

\noindent We hope that this review will facilitate further studies on nonequilibrium quantum phase transitions and phases of matter with anti-unitary symmetries and their symmetry breaking.




\section*{Acknowledgment}
The authors thank Shuji Sasamura, Ryo Hanai, Kazuya Fujimoto, and Kazuho Suzuki for fruitful discussions. Calculations of the magnetization and its dynamics were carried out using QuTip~\cite{johansson2012qutip}. Lindbladian eigenvalues were obtained using QuantumOptics.jl~\cite{kramer2018quantumoptics}. The work carried out by YN was supported by JSPS KAKENHI, Grant No. JP24K22850.
The work done by TS was supported by JSPS KAKENHI, Grants No. JP18H01141, No. JP18H03672, No. JP19L03665, No. JP21H04432, JP22H01143.


\section*{Appendix}

\renewcommand{\theequation}{A.\arabic{equation} }
\setcounter{equation}{0}
\renewcommand{\thesection}{\Alph{section}}
\setcounter{section}{0}

\renewcommand{\thefigure}{A.\arabic{figure} }
\setcounter{figure}{0}
\section{Example of Spontaneous Symmetry Breaking: Transverse Ising model}
\label{transising}
Consider a spin-$\tfrac12$ chain of length $N$ with periodic boundary conditions
$\sigma^{\alpha}_{N+1} = \sigma^{\alpha}_{1}$ $(\alpha = x,y,z)$. The Hamiltonian is given by
\begin{align}
H = -J\sum_{i=1}^N \sigma_i^z \sigma_{i+1}^z
      - h\sum_{i=1}^N \sigma_i^x,
\label{eq:A1}
\end{align}
for $J>0$, $h\ge 0$, and $g\equiv h/J$. This model has a global $Z_2$ parity symmetry,
\begin{align}
[H,P]=0,\qquad
P = \prod_{i=1}^N \sigma_i^x,\qquad
P\sigma_i^z P = -\sigma_i^z .
\label{eq:A2}
\end{align}

In the notation of Sec.~\ref{secSSB}, we take the lattice $\Lambda=\{1,\dots,N\}$ with volume
$V=|\Lambda|=N$, and choose the local observable
\begin{align}
o_x \equiv o_i := \sigma_i^z ,
\end{align}
so that the extensive operator
\begin{align}
O := \sum_{x\in\Lambda} o_x
    = \sum_{i=1}^N \sigma_i^z
\end{align}
plays the role of the order-parameter operator, and
$O^{(N)} := O/N$ is the intensive magnetization per spin. Because $O$ is parity-odd, its expectation value vanishes in any parity eigenstate. Indeed, for any eigenstate
$H|\psi_i\rangle = E_i|\psi_i\rangle$ with $P|\psi_i\rangle = p|\psi_i\rangle$, $p=\pm1$, one has
\begin{align}
\langle\psi_i|O|\psi_i\rangle
= \langle\psi_i|P O P|\psi_i\rangle
= -\langle\psi_i|O|\psi_i\rangle
\;\Rightarrow\;
\langle\psi_i|O|\psi_i\rangle = 0.
\label{eq:A3}
\end{align}

For any finite $N$ with $0<g\ll 1$, the ground state is unique and is well approximated by
the parity-even ``cat'' state
\begin{align}
|\psi_0\rangle \simeq \frac{1}{\sqrt2}
\bigl(|\!\uparrow\uparrow\cdots\uparrow\rangle
      +|\!\downarrow\downarrow\cdots\downarrow\rangle\bigr),
\label{eq:A4}
\end{align}
while the first excited state is
\begin{align}
|\psi_1\rangle \simeq \frac{1}{\sqrt2}
\bigl(|\!\uparrow\uparrow\cdots\uparrow\rangle
      -|\!\downarrow\downarrow\cdots\downarrow\rangle\bigr).
\label{eq:A5}
\end{align}
The energy splitting between them is exponentially small in $N$,
\begin{align}
E_1 - E_0 \sim J g^N .
\label{eq:A6}
\end{align}
In the thermodynamic limit $N\to\infty$, the parity-even and parity-odd ground states
become exactly degenerate, so that different symmetry sectors are separated only by an
infinitesimal splitting.

For $0<g\ll 1$ and any finite $N$, the unique parity-even ground state
$|\psi_0\rangle$ already exhibits long-range order in the sense of Eq.~\eqref{lro}: the normalized
two-point correlator of the order parameter is finite,
\begin{align}
\langle\psi_0|(O^{(N)})^2|\psi_0\rangle
= 1 + O(g^2),
\label{eq:A7}
\end{align}
while the one-point function vanishes due to the symmetry,
\begin{align}
\langle\psi_0|O^{(N)}|\psi_0\rangle = 0.
\label{eq:A8}
\end{align}

To realize spontaneous symmetry breaking as defined in Eq.~\eqref{eq:quasiavg}, we introduce a symmetry-breaking field $\epsilon$ coupled to $O$, following Eq.~\eqref{epshion}:
\begin{align}
H_\epsilon = H - \epsilon O.
\end{align}
Let $|\psi_0(\epsilon)\rangle$ be the unique ground state of $H_\epsilon$ for $\epsilon>0$.
In the ordered phase $0<g\ll 1$, this field selects the positively magnetized configuration,
and the intensive magnetization acquires a nonzero expectation value. More generally, if we
write $|\psi_0(\pm\epsilon)\rangle$ for the ground states of $H_\epsilon = H \mp \epsilon O$
with $\epsilon>0$, then
\begin{align}
\lim_{\epsilon\to0^+}
\lim_{N\to\infty}
\langle\psi_0(\pm\epsilon)|O^{(N)}|\psi_0(\pm\epsilon)\rangle
= \pm 1 + O(g^2),
\label{eq:A9}
\end{align}
so that the order parameter
\begin{align}
m
= \lim_{\epsilon\to0^+}\lim_{V\to\infty}\frac{1}{V}
  \langle O\rangle_{\epsilon,V}
\end{align}
defined in Eq.~\eqref{eq:quasiavg} takes the nonzero values $m=\pm 1 + O(g^2)$.
This explicit example illustrates the general statement in Sec.~\ref{secSSB} that a nonzero LRO
parameter $\sigma>0$ implies spontaneous breaking of the corresponding unitary symmetry in
the thermodynamic limit.

\renewcommand{\theequation}{B.\arabic{equation} }
\setcounter{equation}{0}

\renewcommand{\thefigure}{B.\arabic{figure} }
\setcounter{figure}{0}

\section{Phase transition and bifurcation}
\label{phasebifur}
\subsection{Near Equilibrium case}
Bifurcation theory formalizes qualitative changes in the behavior of nonlinear dynamical systems. A representative example appears in statistical physics: the relaxation of the Ising model toward equilibrium can be described, at a coarse-grained phenomenological level, by the time-dependent Ginzburg--Landau (TDGL) equation (Model A dynamics). Neglecting noise and setting the kinetic coefficient to unity, we write
\begin{equation}
\label{tGL}
\frac{\partial \phi}{\partial t} \;=\; -\,\frac{\delta \mathcal{F}[\phi]}{\delta \phi}\,
\end{equation}
where $\mathcal{F}$ is given by \eqref{free}.
For the spatially uniform case, Eq.~$(\rm\ref{tGL})$ reduces to the normal form
\begin{equation}
\label{eq:normal_form}
\dot \phi \;=\; - r\,\phi \;-\; u\,\phi^{3}+\mathcal{O}(\phi^5).
\end{equation}

Equation~$(\rm\ref{eq:normal_form})$ exhibits a \emph{supercritical pitchfork bifurcation} at \(r=0\). The stable fixed points are
\[
\phi_0 = 0 \quad\text{for } r>0, \qquad
\phi_0 = \pm \sqrt{-\,r/u} \quad\text{for } r<0.
\]
At the bifurcation point \(r=0\), the system transitions from a single symmetric fixed point to two degenerate fixed points. In the language of phase transitions, \(r=0\) is the critical point associated with a continuous transition accompanied by \(Z_2\) \textit{spontaneous} symmetry breaking.

In this way, once the correspondence between symmetry breaking in the dynamical system and the associated phase-transition behavior is clearly established, it is natural, by analogy with equilibrium statistical mechanics, to speak of the nonlinear symmetry as being \textit{spontaneously broken}.

\subsection{Open quantum system}
For a broad class of nonequilibrium systems, the order parameter \(\phi\) obeys a relaxational dynamics~$(\rm\ref{tGL})$. 
As a concrete open-quantum example, we consider the two-photon–driven Kerr oscillator (degenerate parametric oscillator) in a rotating frame, characterized by detuning \(\Delta\), Kerr nonlinearity \(U\), pump strength \(G\), and single-photon loss rate \(\gamma\). Its dynamics is governed by
\begin{align}
\hat H \;=\; -\,\Delta\,a^\dagger  a \;+\; \frac{U}{2N}\,a^{\dagger 2} a^{2}
\;+\; \frac{G}{2}\!\left( a^{\dagger 2}+ a^{2}\right),
\end{align}
\begin{align}
\partial_t\rho \;=\; -\,i[ H,\rho] \;+\; \frac{\gamma}{2}\,\mathcal{D}[ a]\rho,
\end{align}
where $N$ is a scaling parameter controlling the semiclassical limit. 
This model possesses the weak $\mathbb{Z}_2$ symmetry defined in Eq.~\eqref{weak}, implemented by the superoperator
$\hat{\mathcal{Z}}_2[\cdot] := e^{i\pi a^\dagger a} (\cdot), e^{-i\pi a^\dagger a}$, under which the field operator transforms as $a \mapsto -a$.

\subsubsection{Mean-field analysis}
The order parameter is the cavity amplitude \(\alpha:=\langle  a\rangle/\sqrt{N}\). A standard mean-field reduction yields
\begin{align}
\dot\alpha \;=\; \bigl(i\Delta-\tfrac{\gamma}{2}\bigr)\,\alpha \;-\; iG\,\alpha^\ast \;-\; iU\,|\alpha|^{2}\alpha .
\label{eq:SL}
\end{align}
which is invariant under \(\alpha\mapsto -\alpha\), reflecting a $Z_2$ symmetry. 
Writing \(\alpha = R e^{i\theta}\) gives
\begin{align}
\dot R=\Bigl(-\tfrac{\gamma}{2}-G\sin 2\theta\Bigr)R,\ \ 
\dot\theta=\Delta-G\cos 2\theta-U R^2 .
\label{eq:polar}
\end{align}
Near the bifurcation, the linear eigenvalues of the \((\alpha,\alpha^\ast)\) dynamics are
\(\lambda_\pm=-\tfrac{\gamma}{2}\pm s\) with \(s:=\sqrt{G^2-\Delta^2}\).
Hence the amplitude mode becomes slow as \(\lambda_+\to0\) at threshold, while the phase remains strongly damped with rate \(\sim\gamma\).
This time-scale separation justifies adiabatic elimination of \(\theta\) by imposing the slow manifold \(\dot\theta=0\), leading to the phase-locking constraint
\begin{align}
\cos(2\theta)&=\frac{\Delta-U R^2}{G},\nonumber\\
\sin(2\theta)&=-\frac{1}{G}\sqrt{\,G^2-(\Delta-U R^2)^2\,},
\label{eq:phaselock}
\end{align}
where the minus sign selects the stable branch.
At fixed points with \(R>0\), \(\dot R=0\) further gives \(\sin(2\theta_0)=-\gamma/(2G)\).

Expanding the square root in~$(\rm\ref{eq:phaselock})$ for small \(R\) (\(|2\Delta U R^2-U^2 R^4|\ll s^2\)) yields
\[
-\,G\sin 2\theta \;=\; s \;+\; \frac{\Delta U}{s}R^2  \;+\; \mathcal{O}(R^4),
\]
and substituting into \(\dot R\) gives
\begin{align}
\dot R \;=\; -\,r_{\mathrm{eff}}\,R \;-\; u_{\mathrm{eff}}\,R^{3} \;+\; \mathcal{O}(R^{5}),
\label{eq:normalform_full}
\end{align}
with
\begin{align}
r_{\mathrm{eff}} \;=\; \tfrac{\gamma}{2}-\sqrt{G^{2}-\Delta^{2}},\quad
u_{\mathrm{eff}} \;=\; -\,\frac{\Delta U}{\sqrt{G^{2}-\Delta^{2}}}.
\end{align}
Equation~$(\rm\ref{eq:normalform_full})$ is equivalently \(\dot R=-\partial_R V(R)\) with the Landau-type potential
\begin{align}
V(R) \;=\; \frac{r_{\mathrm{eff}}}{2}\,R^{2} \;+\; \frac{u_{\mathrm{eff}}}{4}\,R^{4} \;+\; \mathcal{O}(R^{6}).
\label{eq:landau}
\end{align}
For \(u_{\mathrm{eff}} > 0\) the transition at \(r_{\mathrm{eff}}=0\) is a supercritical pitchfork: the symmetric solution \(R=0\) loses stability and two symmetry-related attractors emerge at \(R_0=\sqrt{-r_{\mathrm{eff}}/u_{\mathrm{eff}}}\), giving
\begin{align}
\alpha_\pm \;=\; \pm\,R_0\,e^{i\theta_0},
\end{align}
\begin{align}
\sin(2\theta_0)=-\frac{\gamma}{2G},\quad
\cos(2\theta_0)=\frac{\Delta-U R_0^{2}}{G},
\end{align}
which differ by a global sign and thus spontaneously break the \(Z_2\) symmetry. The Landau description captures the universal content: the emergent Ising-like choice between \(\alpha_+\) and \(\alpha_-\) that organizes the phase transition in this driven–dissipative quantum system.


\begin{figure}[t]
   \vspace*{2.3cm}
     \hspace*{0.2cm}
\includegraphics[bb=0mm 0mm 90mm 150mm,width=0.54\linewidth]{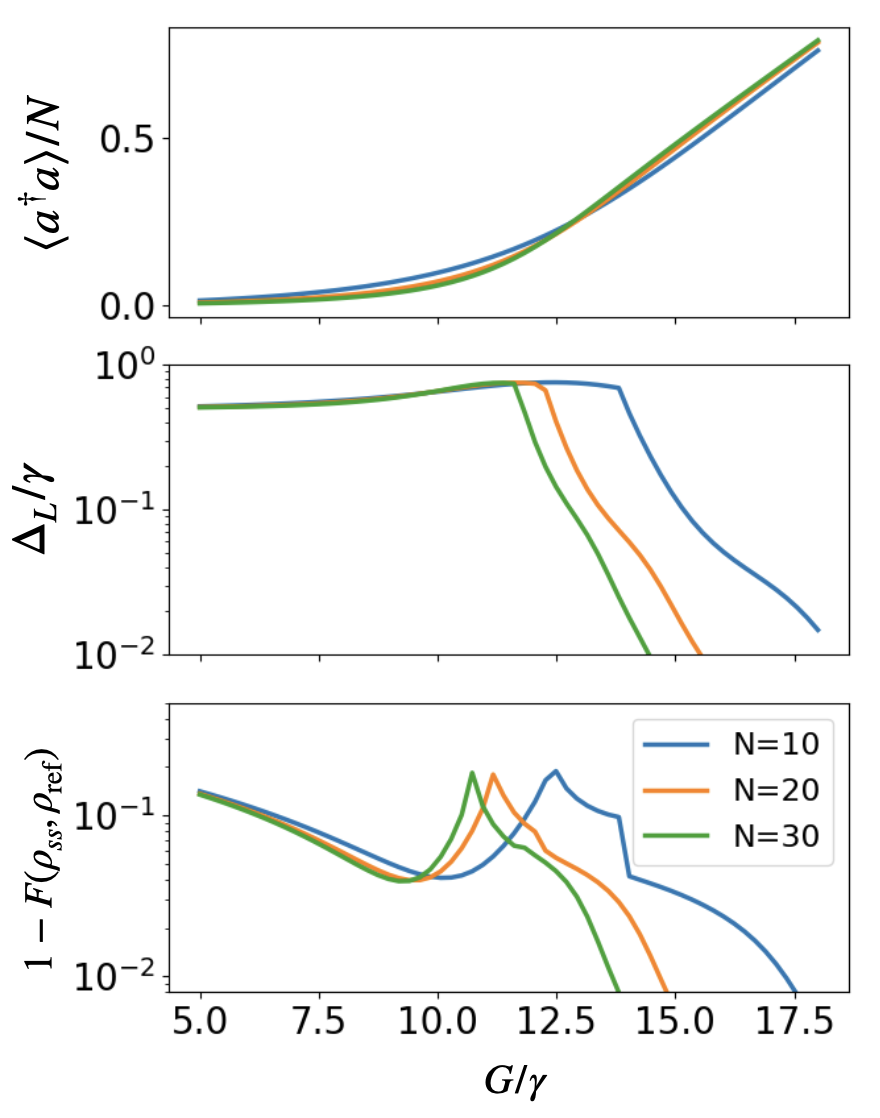}
\caption{\footnotesize\justifying Steady-state properties of a driven Kerr parametric oscillator versus the two-photon drive \(G/\gamma\) for different values of the scaling parameter \(N\) \((N=10,20,30)\).
Parameters: \(\gamma=1\), \(\ U=10\gamma\), \(\Delta=-10\gamma\); Fock-space cutoff \(n_{\mathrm{cut}}=60\) (i.e., \( G_c\simeq10\gamma\)).
  Top: rescaled photon number \(\langle a^\dagger a\rangle/N\).
  Middle: Lindbladian gap \(\Delta_L\) on a logarithmic scale.
  Bottom: infidelity \(1-F\) with respect to the spectral-ansatz state \(\rho_{\rm ref}=(\rho_{1}^{+}+\rho_{1}^{-})/2\) built from the eigenmatrix \(\rho_{1}\); the fidelity is \(F(\rho_{\rm ss},\rho_{\rm ref})\).
  Increasing \(N\) sharpens the crossover in \(\langle a^\dagger a\rangle/N\), steepens the drop of \(\Delta_L\) and suppresses the infidelity \(1-F(\rho_{\rm ss},\rho_{\rm ref})\) in the symmetry-broken phase.
}
  \label{fig:kpo_scaling}
\end{figure}

\subsubsection{Lindbladian spectral analysis}
The steady state \(\rho_{\rm ss}\) is obtained by exact diagonalization of the Lindbladian \(\mathcal{L}\) in a truncated Fock space (\(n_{\mathrm{cut}}=60\)).
We probe finite-size scaling by varying a dimensionless parameter \(N\).
Figure~\ref{fig:kpo_scaling} summarizes the steady-state properties as a function of \(G/\gamma\) for several values of \(N\).

The rescaled occupation \(\langle a^\dagger a\rangle/N\) shows a pronounced crossover as \(G/\gamma\) increases, which becomes sharper with increasing \(N\).
The Lindbladian gap \(\Delta_L/\gamma\) remains in order unity up to this crossover and then rapidly drops by several orders of magnitude.
 Infidelity \(1-F(\rho_{\rm ss},\rho_{\rm ref})\), where \(F\) is the fidelity defined in Eq.~\eqref{QPTfide} and \(\rho_{\rm ref}=(\rho_{1}^{+}+\rho_{1}^{-})/2\), decreases as \(N\) increases and is small in the symmetry-broken phase. Here, \(\rho_1\) denotes the eigenmatrix associated with the Lindbladian gap, and \(\rho_1^{\pm}\) are the positive operators obtained from the spectral decomposition \(\rho_1=\rho_1^{+}-\rho_1^{-}\).
This behavior is consistent with the spectral picture of DPTs, in which, in the symmetry-broken phase, the steady state approaches the symmetric mixture of components obtained from the spectral decomposition of \(\rho_{1}\)~\cite{Minganti}.


\renewcommand{\theequation}{C.\arabic{equation} }
\setcounter{equation}{0}

\renewcommand{\thefigure}{C.\arabic{figure} }
\setcounter{figure}{0}
\section{Example of the model with $\mathcal{PT}$ symmetry for shifted Lindbladian}
\label{appprosenpt}
\begin{figure}[t]
   \vspace*{0.9cm}
     \hspace*{0.1cm}
\includegraphics[bb=0mm 0mm 90mm 150mm,width=0.34\linewidth]{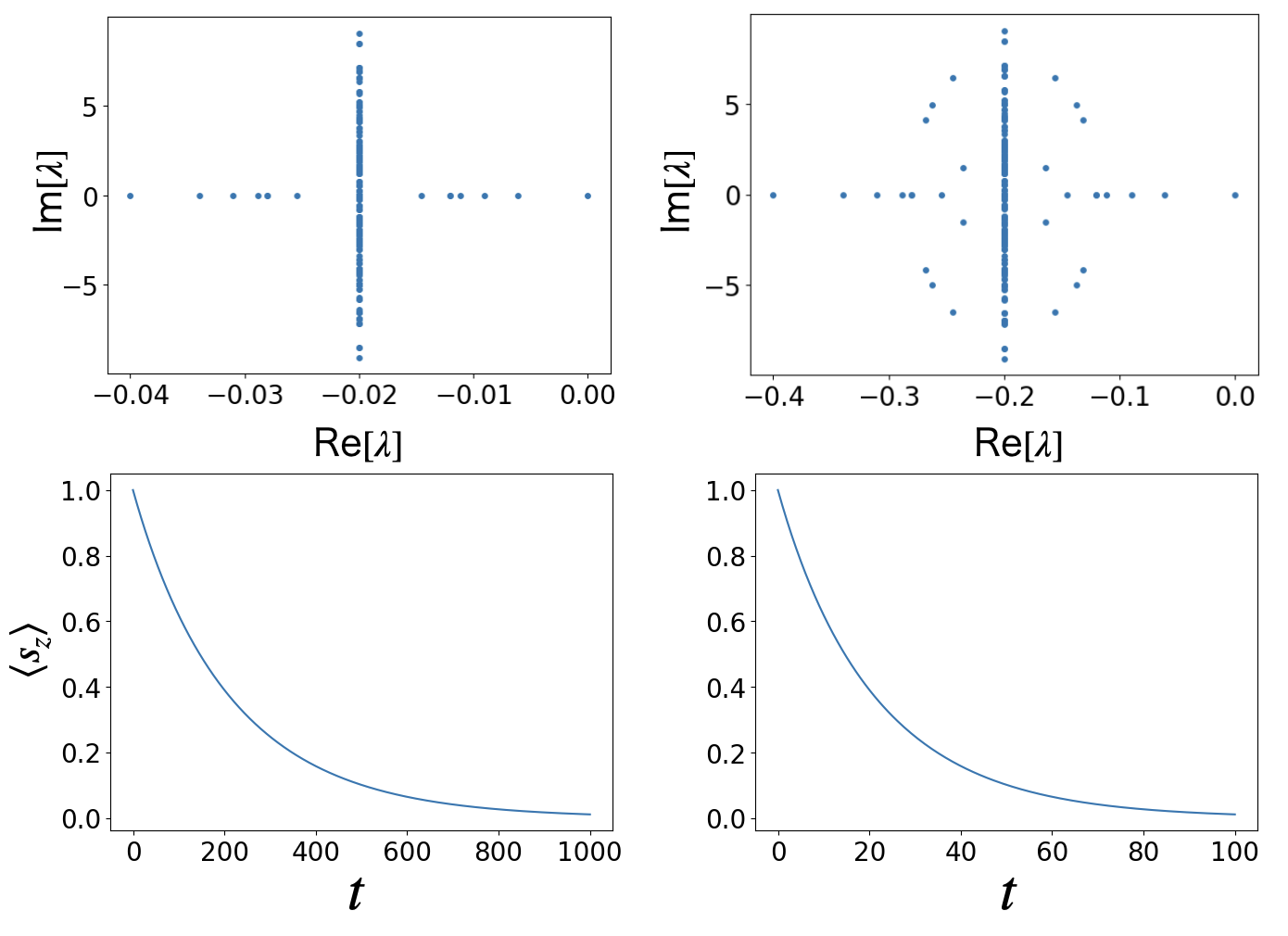}
\caption{\footnotesize\justifying Lindbladian eigenvalues and  dynamics of total magnetization $S_z:=\sum_j\sigma^z_j$ for dissipative XXZ spin chain that exhibits PT symmetry for a shifted Lindbladian ($\ref{ProzenPT2}$) with $\Delta=1/2$, $n=4$. (Left) $\gamma=0.02\ (<\gamma_{c})$ (Right) $\gamma=0.2\ (>\gamma_{c})$.}
    \label{figProsenPT}
\end{figure}

As an illustrative example for the model with $\mathcal{PT}$ symmetry for a shifted Lindbladian $\hat{\mathcal{L}}^\prime$ ($\ref{ProzenPT2}$), we consider the dissipative XXZ spin chain, 
where the Hamiltonian and Lindblad operators are defined as:
\begin{align}
H=\sum_{j=1}^{n-1}&(2\sigma_{j}^{+}\sigma_{j+1}^{-}+2\sigma_{j}^{-}\sigma_{j+1}^{+}+\Delta \sigma_{j}^{z}\sigma_{j+1}^{z})\nonumber\\
L_{1}&=\frac{\sqrt{\gamma}}{\sqrt{2}}\sigma_{1}^{+},\ \ \ \ L_{2}=\frac{\sqrt{\gamma}}{\sqrt{2}}\sigma_{n}^{-}.
\end{align}
Here, $\sigma_{j}^{\pm}=(\sigma_{j}^{x}\pm i\sigma_{j}^{y})/2$, and the parity operator is the reflection of space~\cite{Prosen1}. Fig.~$\rm\ref{figProsenPT}$ illustrates the eigenvalue structures above and below the spectral transition point $\gamma=\gamma_{c}$. For $\gamma<\gamma_{c}$, all the eigenvalues are on the lines of symmetry $l_{V}=-\gamma +i\mathbb{R} \ $and $l_{h}=\mathbb{R}$. In contrast, for $\gamma>\gamma_{c}$, some eigenvalues deviate from these lines. Although the spectral transition occurs at $\gamma=\gamma_{c}$, there is no significant difference in either the system's dynamics or its steady state. In particular, oscillatory dynamics at late times do not emerge, as there are no PIEs. 
We note that the relation between the level statistics of Lindbladian eigenvalues and $\mathcal{PT}$ symmetry breaking of the shifted Lindbladian has been investigated~\cite{Sa}.

\renewcommand{\theequation}{D.\arabic{equation} }
\setcounter{equation}{0}
\renewcommand{\thefigure}{D.\arabic{figure} }
\setcounter{figure}{0}
\section{Collective spin}
\label{colmath}
Here we briefly summarize some basic properties of the collective spin operators.  
They satisfy the standard SU(2) commutation relations
\begin{align}
\label{comDDM2}
[S_z, S_{\pm}] &= \pm S_{\pm}, \qquad
[S_{+}, S_{-}] = 2 S_z.
\end{align}
For a system of $N$ spin-$1/2$ particles with full permutation symmetry, the dynamics is restricted to the fully symmetric subspace with total spin $S = N/2$. This subspace is spanned by the eigenstates of $S_z$, $\ket{S=N/2,m}$ with $-N/2 \leq m \leq N/2$. On these states, the collective spin operators act as
\begin{align}
S^{2}\ket{S,m} &= S(S+1)\ket{S,m},\\
S_{+}\ket{S,m} &= \sqrt{S(S+1)-m(m+1)}\,\ket{S,m+1},\\
S_{z}\ket{S,m} &= m\ket{S,m},\\
S_{-}\ket{S,m} &= \sqrt{S(S+1)-m(m-1)}\,\ket{S,m-1}.
\end{align}

\renewcommand{\theequation}{E.\arabic{equation} }
\setcounter{equation}{0}
\renewcommand{\thefigure}{E.\arabic{figure} }
\setcounter{figure}{0}
\section{The derivation of the DDM in free space}
\label{deriddm}
\subsection{Born--Markov derivation of the collective decay term}

We consider $N$ identical two-level atoms in free space with transition frequency $\omega_0$ interacting with the quantized radiation field. In the electric dipole and
rotating-wave approximations, the total Hamiltonian can be written as
\begin{equation}
H = H_{\mathrm{S}} + H_{\mathrm{E}} + H_{\mathrm{I}},
\end{equation}
with
\begin{align}
H_{\mathrm{S}} &= \tfrac{\hbar\omega_0}{2} \sum_{j=1}^{N} \sigma_j^z,
\\
H_{\mathrm{E}} &= \sum_{\mathbf{k},\lambda} \hbar\omega_k
a_{\mathbf{k}\lambda}^\dagger a_{\mathbf{k}\lambda},
\\
H_{\mathrm{I}} &= \hbar \sum_{j,\mathbf{k},\lambda}
\Bigl(
g_{\mathbf{k}\lambda} e^{i\mathbf{k}\cdot\mathbf{r}_j}
\sigma_j^+ a_{\mathbf{k}\lambda}
+ g_{\mathbf{k}\lambda}^* e^{-i\mathbf{k}\cdot\mathbf{r}_j}
\sigma_j^- a_{\mathbf{k}\lambda}^\dagger
\Bigr).
\end{align}
Here 
$a_{\mathbf{k}\lambda}$ annihilates a photon in mode $(\mathbf{k},\lambda)$,
and $g_{\mathbf{k}\lambda}$ is the atom--field coupling constant. Below, we set $\hbar =1$.

We move to the interaction picture with respect to
$H_0 = H_{\mathrm{S}} + H_{\mathrm{E}}$.
The interaction-picture Hamiltonian is
\begin{equation}
H_{\mathrm{I}}(t)
=
\sum_{j,\mathbf{k},\lambda}
\Bigl(
g_{\mathbf{k}\lambda} e^{i\mathbf{k}\cdot\mathbf{r}_j}
\sigma_j^+ a_{\mathbf{k}\lambda}
e^{i(\omega_0-\omega_k)t}
+ h.c.
\Bigr).
\end{equation}
It is convenient to write this as
\begin{equation}
H_{\mathrm{I}}(t)
= \sum_{j} \Bigl(\sigma_j^+ F_j(t) + \sigma_j^- F_j^\dagger(t)\Bigr),
\end{equation}
where
\begin{equation}
F_j(t)
=
\sum_{\mathbf{k},\lambda}
g_{\mathbf{k}\lambda} e^{i\mathbf{k}\cdot\mathbf{r}_j}
a_{\mathbf{k}\lambda} e^{i(\omega_0-\omega_k)t}.
\end{equation}

Let $\chi(t)$ be the total density operator in the interaction picture.
The von Neumann equation reads
\begin{equation}
\frac{d\chi(t)}{dt}
=
-i\bigl[H_{\mathrm{I}}(t),\chi(t)\bigr].
\end{equation}
We formally integrate this equation and iterate once to obtain
\begin{align}
\frac{d\chi(t)}{dt}
=&
-i[H_{\mathrm{I}}(t),\chi(0)]\nonumber\\
&-\int_0^t d\tau\,
[H_{\mathrm{I}}(t),[H_{\mathrm{I}}(t-\tau),\chi(t-\tau)]].
\label{eq:BM-raw}  
\end{align}

We assume a factorized initial state
$\chi(0)=\rho(0)\otimes\rho_{\mathrm{E}}$ with the field in vacuum 
$\rho_{\mathrm{E}} = \ket{0}\bra{0}$, and assume that the field state remains
approximately unchanged (Born approximation). Tracing over the field
degrees of freedom we obtain the equation for the reduced density
operator $\rho(t)=\mathrm{Tr}_{\mathrm{E}}\chi(t)$:
\begin{equation}
\frac{d\rho(t)}{dt}
=
-\int_0^t d\tau\,
\mathrm{Tr}_{\mathrm{E}}
\Bigl(
[H_{\mathrm{I}}(t),[H_{\mathrm{I}}(t-\tau),\rho(t-\tau)\otimes\rho_{\mathrm{E}}]]
\Bigr),
\label{eq:BM}
\end{equation}
where the first term in Eq.~$\rm(\ref{eq:BM-raw})$ vanishes because
$\mathrm{Tr}_{\mathrm{E}}(F_j \rho_{\mathrm{E}})=0$.

\subsection{Evaluation of the double commutator}

Substituting
$H_{\mathrm{I}}(t) = \sum_j (\sigma_j^+ F_j(t) + \sigma_j^- F_j^\dagger(t))$
into ($\rm\ref{eq:BM}$) and expanding the double commutator, we encounter
terms containing products of two bath operators. For the vacuum state
$\rho_{\mathrm{E}}=\ket{0}\bra{0}$ the only non-vanishing two-point
correlation functions are
\begin{align}
\langle F_i(t) F_j^\dagger(t-\tau)\rangle
&=
\mathrm{Tr}_{\mathrm{E}}\bigl(F_i(t)F_j^\dagger(t-\tau)\rho_{\mathrm{E}}\bigr),
\nonumber \\
\langle F_i^\dagger(t)F_j(t-\tau)\rangle &= 0,
\end{align}
and similarly for correlators with two creation (or two annihilation)
operators. After some algebra one finds
\begin{align}
\frac{d\rho(t)}{dt}
=&
\sum_{i,j}\int_0^t d\tau\,
\Bigl\{
G_{ij}(\tau)
\bigl(
\sigma_j^- \rho(t-\tau) \sigma_i^+-\sigma_i^+\sigma_j^- \rho(t-\tau)
\bigr)
\nonumber\\
&
+ G_{ij}^*(\tau)
\bigl(
\sigma_i^- \rho(t-\tau)\sigma_j^+ - \rho(t-\tau)\sigma_j^+\sigma_i^-
\bigr)
\Bigr\},
\label{eq:BM-Gij}
\end{align}
where we have introduced the bath correlation functions,
\begin{equation}
G_{ij}(\tau)
=
\langle F_i(t) F_j^\dagger(t-\tau)\rangle.
\end{equation}

Explicitly, using the definition of $F_j(t)$,
\begin{equation}
G_{ij}(\tau)
=
\sum_{\mathbf{k},\lambda}
|g_{\mathbf{k}\lambda}|^2
e^{i\mathbf{k}\cdot(\mathbf{r}_i-\mathbf{r}_j)}
e^{-i(\omega_k-\omega_0)\tau}.
\label{eq:Gij}
\end{equation}

\subsection{Markov approximation and definition of the rates}

We now make the Markov approximation: the bath correlation time is
assumed much shorter than the characteristic relaxation time of the
atoms, so we can replace $\rho(t-\tau)\to\rho(t)$ and extend the upper
limit of the $\tau$-integral to infinity. This yields
\begin{equation}
\dot{\rho}
=
\sum_{i,j}\left\{
\Gamma_{ij} \bigl(
\sigma_j^- \rho \sigma_i^+ - \sigma_i^+\sigma_j^- \rho
\bigr)
+ \Gamma_{ij}^* \bigl(
\sigma_i^- \rho \sigma_j^+ - \rho \sigma_j^+\sigma_i^-
\bigr)
\right\},
\label{eq:BM-Gamma}
\end{equation}
where
\begin{equation}
\Gamma_{ij} = \int_0^\infty d\tau\, G_{ij}(\tau).
\end{equation}
Using ($\rm\ref{eq:Gij}$) we obtain
\begin{equation}
\Gamma_{ij}
=
\sum_{\mathbf{k},\lambda}
|g_{\mathbf{k}\lambda}|^2
e^{i\mathbf{k}\cdot(\mathbf{r}_i-\mathbf{r}_j)}
\int_0^\infty d\tau\, e^{-i(\omega_k-\omega_0)\tau}.
\end{equation}
The time integral is evaluated as
\begin{equation}
\int_0^\infty d\tau\, e^{-i(\omega_k-\omega_0)\tau}
=
\pi\delta(\omega_k-\omega_0)
- i\,\mathcal{P}\frac{1}{\omega_k-\omega_0},
\end{equation}
with the Cauchy principal value $\mathcal{P}$,
so that $\Gamma_{ij}$ naturally splits into a real and an imaginary
part,
\begin{equation}
\label{cauchy}
\Gamma_{ij}
=
\frac{1}{2}\gamma_{ij} + i S_{ij},
\end{equation}
where $\gamma_{ij}=2\,\mathrm{Re}\,\Gamma_{ij}$ gives decay rates and $S_{ij}$ give rise to a coherent
level shift (Lamb shift).

Substituting~$\rm(\ref{cauchy})$ into~$\rm(\ref{eq:BM-Gamma})$ and using $\gamma_{ij}=\gamma_{ji},\ S_{ij}=S_{ji}$, we can write the
equation in the standard GKSL
form:
\begin{align}
\frac{d\rho}{dt}
=
&-i[H_{\mathrm{LS}},\rho]
\nonumber\\
&+
\sum_{i,j}\frac{\gamma_{ij}}{2}
\Bigl(
2\sigma_j^- \rho \sigma_i^+
-
\sigma_i^+\sigma_j^- \rho
-
\rho \sigma_i^+\sigma_j^-
\Bigr),
\label{eq:Lehmberg}  
\end{align}
with
\begin{equation}
H_{\mathrm{LS}}
=\sum_{i,j} S_{ij}\sigma_i^+\sigma_j^-.
\end{equation}

\subsection{Small-sample regime and collective operators}
In the small-sample (subwavelength) regime, $|\mathbf{k}\cdot(\mathbf{r}_i-\mathbf{r}_j)|\ll1$, the phase factors
$e^{i\mathbf{k}\cdot(\mathbf{r}_i-\mathbf{r}_j)}$ can be approximated
by unity, and the decay tensor becomes essentially independent of
$i,j$:
\begin{equation}
\gamma_{ij} \simeq \gamma,
\qquad \forall i,j.
\end{equation}
Using the collective spin operators,
the dissipator reduces to
\begin{equation}
\mathcal{L}_\gamma[\rho]
=
\frac{\gamma}{2}
\Bigl(
2 S_- \rho S_+ - S_+S_- \rho - \rho S_+S_-
\Bigr).
\label{eq:dissipator}
\end{equation}

If we now add a coherent resonant drive at frequency $\omega_0$,
$H_{\rm drive}(t)=g\!\left(S_+ e^{-i\omega_0 t}+S_- e^{i\omega_0 t}\right)$, the
effective system Hamiltonian in the laser rotating frame is
\begin{equation}
H_{\mathrm{drive}} = g(S_+ + S_-)=2gS_x.
\end{equation}
Within the usual Born–Markov–secular treatment for a broadband vacuum in free
space, the full master equation in this frame takes the form
\begin{align}
\label{aaaa}
\frac{d\rho}{dt}
&=
-i[H_{\mathrm{drive}},\rho]
+
\mathcal{L}_\gamma[\rho]\nonumber\\
&=
-i\,2g\,[S_x,\rho]
+
\frac{\gamma}{2}
\Bigl(
2 S_- \rho S_+ - S_+S_- \rho - \rho S_+S_-
\Bigr).
\end{align}
Here, we neglected the Lamb shift term.
Eq.~$(\rm\ref{aaaa})$ is equivalent to equation~\eqref{DDM} with $\gamma=2\kappa/S$.

\renewcommand{\theequation}{F.\arabic{equation} }
\setcounter{equation}{0}
\renewcommand{\thefigure}{F.\arabic{figure} }
\setcounter{figure}{0}
\section{Third quantization}
\label{Thirdquantization}
\subsection{Third quantization for quadratic bosonic systems with linear baths}
The quadratic Hamiltonian and linear Lindblad operators for an arbitrary $n$ bosonic system can be written as
\begin{align}
\label{Hamiltonian}
H&=\underline{a}^{\dagger}\cdot\textbf{H}\underline{a}+\underline{a}\cdot\textbf{K}\underline{a}+\underline{a}^{\dagger}\cdot\bar{\textbf{K}}\underline{a}^{\dagger},\\
\label{dissipation11}L_{\mu}&=\underline{l}_{\mu}\cdot\underline{a}+\underline{k}_{\mu}\cdot\underline{a}^{\dagger}.
\end{align}
where $\textbf{H}=\textbf{H}^{\dagger}$ and $\textbf{K}=\textbf{K}^{T}$ are $n\times n$ matrices, $\underline{a}=(a_{1}, a_{2},..., a_{n})^{T}$ is a vector which consists of annihilation bosonic operators, $\underline{a}^{\dagger}=(a_{1}^{\dagger}, a_{2}^{\dagger},..., a_{n}^{\dagger})^{T}$ is a vector which consists of creation bosonic operators, and $\underline{l}_{\mu}$ and $\underline{k}_{\mu}$ are $n$ vectors for dissipation rates. Here, the underline means a vector. 

\begin{align}
\label{X}
\textbf{X}:=\frac{1}{2}\left(
   \begin{array}{cc}
     i\bar{\textbf{H}}-\bar{\textbf{N}}+\textbf{M} & -2i\textbf{K}-\textbf{L}+\textbf{L}^{T}  \\
      2i\bar{\textbf{K}} -\bar{\textbf{L}}+\bar{\textbf{L}}^{T} & -i\textbf{H}-\textbf{N}+\bar{\textbf{M}}
   \end{array}
  \right),
\end{align}
The matrices $\textbf{M}$, $\textbf{N}$ and $\textbf{L}$ are defined by
\begin{align}
\label{M}
\textbf{M}:&=\sum_{\mu}\underline{l}_{\mu}\otimes\underline{\bar{l}}_{\mu}=\textbf{M}^{\dagger},\ \ \ \ \textbf{N}:=\sum_{\mu}\underline{k}_{\mu}\otimes\underline{\bar{k}}_{\mu}=\textbf{N}^{\dagger},\nonumber\\
\textbf{L}:&=\sum_{\mu}\underline{l}_{\mu}\otimes\underline{\bar{k}}_{\mu}.
\end{align}
Assuming that the matrix $\textbf{X}$ ($\rm\ref{X}$) is diagonalizable, it can be written as
\begin{align}
\label{20}
\textbf{X}=\textbf{P}{\boldsymbol{\Delta}}\textbf{P}^{-1},\ \ \ \ \ {\boldsymbol{\Delta}}=\textrm{diag}\{\beta_{1},\beta_{2},...,\beta_{2n} \},
\end{align}
where $\textbf{P}$ is a $2n$ $\times$ $2n$ matrix.
Then, the following theorems hold~\cite{Prosen2}.\\

\noindent\textbf{\textit{Theorem F.1}}: If the matrix $\textbf{X}$ ($\rm\ref{X}$) is diagonalizable and the real parts of all the eigenvalues are positive, i.e. $\forall j,\rm{Re} [\beta_{j}]>0$, the unique nonequilibrium steady state $\rho_{ss}$ exists, satisfying
\begin{align}
\label{327}
\hat{\mathcal{L}}\rho_{ss}=0.
\end{align}

\noindent\textbf{\textit{Theorem F.2}}: If the matrix $\textbf{X}$ ($\rm\ref{X}$) is diagonalizable and the real parts of all the eigenvalues are positive, i.e. $\forall j,\rm{Re} [\beta_{j}]>0$, the full spectrum of the Lindbladian is given by a $2n$ component multi-index of super-quantum numbers $\underline{m}\in\mathbb{Z}_{+}^{2n}$,
\begin{align}
\label{newbasis}
\lambda_{\underline{m}}=-2\sum_{r=1}^{2n}m_{r}\beta_{r}.
\end{align}

\subsection{Third quantization for the two-collective spin model with gain and loss}
We apply the third quantization for the AM phase with $S=\infty$. Its GKSL equation can be rewritten as 
\begin{align}
\dot{\rho}=-i[g(a_Aa_B+{\rm{H.c.}}),\rho]+\Gamma_g\mathcal{D}[a_A]\rho+\Gamma_l\mathcal{D}[a_B]\rho.
\end{align}
The matrices $\textbf{K}$ and $\textbf{M}$ in Eqs.($\rm\ref{X}$), ($\rm\ref{M}$) are given by
\begin{align}
\textbf{K}=\frac{g}{2}\left(
   \begin{array}{rr}
      0 & 1  \\
      1 & 0 
    \end{array}
  \right),\ \ \textbf{M}=\left(
   \begin{array}{rr}
      \Gamma_{g} & 0  \\
      0 & \Gamma_{l}
    \end{array}
  \right),
\end{align}
and $\textbf{H}=\textbf{N}=\textbf{L}=\textbf{0}$. Thus, the matrix $\textbf{X}$ ($\rm\ref{X}$) is given by Eq.~($\ref{XX}$),
which is Hermitian.
Similarly, the matrix $\textbf{X}$ for the FM $\ket{\Uparrow\Uparrow}$ phase is given by Eq.~($\ref{FMX}$).

\renewcommand{\theequation}{G.\arabic{equation} }
\setcounter{equation}{0}
\renewcommand{\thefigure}{G.\arabic{figure} }
\setcounter{figure}{0}
\section{PT symmetry of eigenmodes with a nonzero imaginary part for the DDM}
\label{imagptpt}
\begin{figure}[t]
   \vspace*{-0.6cm}
     \hspace*{-0.2cm}
\includegraphics[bb=0mm 0mm 90mm 150mm,width=0.48\linewidth]{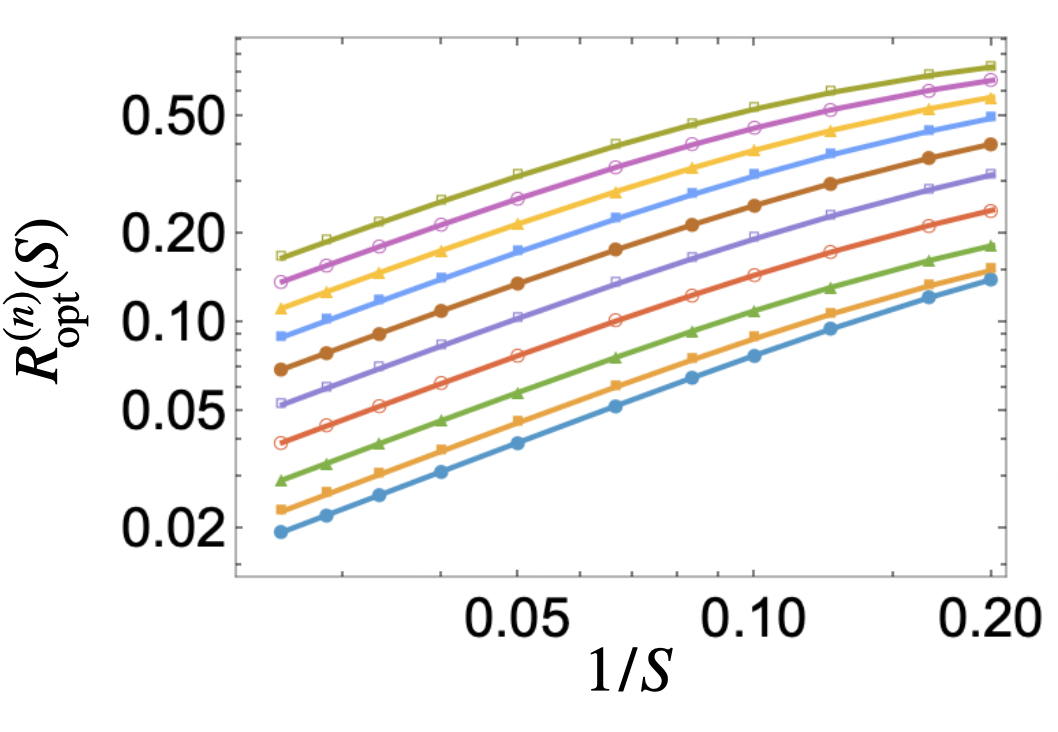}
\caption{\footnotesize\justifying The $S$-dependence of the PT-symmetry measure $R_{\mathrm{opt}}^{(n)}(S)$ at $\kappa/g=0.5$ in the DDM~\eqref{DDM}.
For each total spin $S$ we evaluate $R_{\mathrm{opt}}^{(n)}(S)$ for the first ten right eigenmodes of the Lindbladian whose eigenvalues have a nonzero imaginary part, focusing on those with $\mathrm{Im}\,\lambda_n \approx 2$, ordered by increasing $|\mathrm{Re}\,\lambda_n|$.
Both axes are plotted on a logarithmic scale.}
  \label{fig:k}
\end{figure}

To quantify how close a right Lindbladian eigenmode $\rho_n$ is to being PT-symmetric, we introduce the dimensionless Frobenius–norm measure
\begin{equation}
  R_{\mathrm{opt}}^{(n)}(S)
  :=
  \frac{\bigl\|\rho_n - \alpha_n\,PT\rho_n(PT)^{-1}\bigr\|_{F}}
       {\bigl\|\rho_n\bigr\|_{F}
        + \bigl|\alpha_n\bigr|\,\bigl\|PT\rho_n(PT)^{-1}\bigr\|_{F}},
  \label{eq:Ropt-def}
\end{equation}
where $\|\cdot\|_{F}$ denotes the Frobenius norm.
The complex coefficient $\alpha_n$ is chosen as,
\begin{equation}
  \alpha_n
  =
  \frac{\mathrm{Tr}\ \!\bigl[(PT\rho_n(PT)^{-1})^{\dagger}\rho_n\bigr]}
       {\mathrm{Tr}\ \!\bigl[(PT\rho_n(PT)^{-1})^{\dagger}PT\rho_n(PT)^{-1}\bigr]}.
  \label{eq:alphaopt-def}
\end{equation}
By construction $0 \le R_{\mathrm{opt}}^{(n)}(S) \le 1$, with $R_{\mathrm{opt}}^{(n)}(S)=0$ if and only if
$\rho_n$ is exactly proportional to its PT transform, $\rho_n \propto PT\rho_n(PT)^{-1}$.

Fig.~\ref{fig:k} shows $R_{\mathrm{opt}}^{(n)}(S)$ for the DDM at $\kappa/g=0.5$.
For each spin length $S$ we focus on right eigenmodes whose Lindbladian eigenvalues $\lambda_n$ lie in the band $\mathrm{Im[\lambda_n]}\sim2$, and we plot the first ten modes ordered by $|\mathrm{Re}\,\lambda_n|$.
On the double-logarithmic scale all curves exhibit a clear monotonic decrease with $S$ and are consistent with an algebraic decay of $R_{\mathrm{opt}}^{(n)}(S)$ towards zero.
This demonstrates that, in the large-$S$ limit, these oscillatory modes with finite imaginary parts of $\lambda_n$ become asymptotically PT-symmetric,
\begin{equation}
  R_{\mathrm{opt}}^{(n)}(S) \xrightarrow[S\to\infty]{} 0,
  \qquad
  \rho_n \propto PT\rho_n(PT)^{-1}.
\end{equation}
This result is consistent with the mean-field expectation that the oscillations around the centers are n-PT symmetric.

\vskip\baselineskip
\section*{Reference}
\bibliography{ref_fixed_patched_unified_v2}

\end{document}